\def\cleanVersion{1}
\documentclass[12pt,oneside]{article}

\usepackage{ifthen}
\newboolean{showfigures}
\newboolean{linenumbers}
\newboolean{showcomments}
\newboolean{printversion}
\setboolean{showfigures}{true}
\setboolean{linenumbers}{false}
\ifdefined\cleanVersion\setboolean{showcomments}{false}\else\setboolean{showcomments}{true}\fi
\setboolean{printversion}{false}

\usepackage{fullpage}
\usepackage{setspace}
\usepackage{amsmath,amssymb,amsthm,mathtools}
\usepackage{microtype}
\usepackage{bm,dsfont,mathrsfs,braket}

\usepackage{adjustbox,array,booktabs,graphicx,multicol,multirow}
\usepackage{changepage,comment,enumitem,etoolbox,pifont,relsize,textcomp,xpatch}
\usepackage{pythonhighlight}
\usepackage[table]{xcolor}
\usepackage{xurl}

\definecolor{accent}{HTML}{244653}
\definecolor{rulecolor}{HTML}{B4C0C5}
\definecolor{lightgreen}{HTML}{E8F1EA}
\definecolor{lightgray}{gray}{0.95}
\definecolor{falcocolor}{HTML}{A33A32}
\definecolor{omarcolor}{HTML}{795392}

\ifthenelse{\boolean{printversion}}{%
  \colorlet{accent}{black}
  \colorlet{rulecolor}{black!45}
  \colorlet{falcocolor}{black}
  \colorlet{omarcolor}{black}
  \colorlet{lightgreen}{black!8}
}{}

\setlist[enumerate]{leftmargin=*,itemsep=0.3ex,topsep=0.5ex,parsep=0pt}
\setlist[itemize]{leftmargin=*,itemsep=0.2ex,topsep=0.5ex,parsep=0pt}

\usepackage{caption,subcaption}
\ifthenelse{\boolean{showfigures}}{}{%
  \renewenvironment{figure}{}{}
  \renewenvironment{figure*}{}{}
  \excludecomment{figure}
  \excludecomment{figure*}
}

\usepackage{algorithm,algpseudocode}
\algrenewcommand\algorithmicrequire{\textbf{Input:}}
\algrenewcommand\algorithmicensure{\textbf{Output:}}

\usepackage{tikz,tikz-cd}
\usetikzlibrary{arrows.meta,positioning,shapes.geometric,decorations.pathreplacing}

\makeatletter
\newcommand{\paperAuthors}{}
\newcommand{\paperAffiliations}{}
\newcommand{\paperEmail}{}
\renewcommand{\author}[1]{\gappto\paperAuthors{{#1}\quad}}
\newcommand{\address}[1]{\gappto\paperAffiliations{#1\par}}
\newcommand{\email}[1]{\gdef\paperEmail{#1}}
\renewcommand{\maketitle}{%
  \begin{center}
    {\LARGE\bfseries\color{accent}\@title\par}
    \medskip
    {\normalsize\paperAuthors\par}
  \end{center}
  \ifdefined\anonymousVersion\else
  \begingroup
    \renewcommand{\thefootnote}{}
    \footnotetext{\small\paperAffiliations\textit{Email: }\paperEmail}
  \endgroup
  \fi
  \setcounter{footnote}{0}
}
\makeatother

\usepackage{titlesec}

\titleformat{\section}[block]
  {\normalfont\large\scshape\color{accent}}
  {\thesection.}{0.7em}{}
  [\vspace{0.3ex}{\color{rulecolor}\titlerule[0.4pt]}]

\titlespacing*{\section}
  {0pt}
  {2.2ex plus 0.7ex minus 0.3ex}
  {1.0ex plus 0.2ex}

\titleformat{\subsection}[block]
  {\normalfont\normalsize\bfseries}
  {\thesubsection.}{0.6em}{}

\titlespacing*{\subsection}
  {0pt}
  {1.5ex plus 0.5ex minus 0.2ex}
  {0.3ex}

\titleformat{\subsubsection}[runin]
  {\normalfont\normalsize\itshape}
  {\thesubsubsection.}{0.6em}{}[.]

\titlespacing*{\subsubsection}
  {0pt}
  {1.0ex plus 0.3ex minus 0.2ex}
  {0.6em}

\usepackage{needspace}
\pretocmd{\section}{\Needspace{4\baselineskip}}{}{}
\pretocmd{\subsection}{\Needspace{3\baselineskip}}{}{}

\usepackage[authoryear,round]{natbib}
\usepackage{multibib}
\newcites{App}{Supplementary References}
\usepackage{hyperref}
\hypersetup{
  colorlinks=true,
  linkcolor=accent,
  citecolor=accent,
  urlcolor=accent,
  bookmarksnumbered=true
}

\usepackage[noabbrev,capitalise,nameinlink]{cleveref}

\creflabelformat{equation}{#2\textup{(#1)}#3}
\numberwithin{equation}{section}

\newtheoremstyle{paperresult}
  {0.5\baselineskip}
  {0.2\baselineskip}
  {\itshape}
  {}
  {\bfseries}
  {.}
  {0.5em}
  {}

\newtheoremstyle{paperstatement}
  {0.5\baselineskip}
  {0.2\baselineskip}
  {\normalfont}
  {}
  {\bfseries}
  {.}
  {0.5em}
  {}

\theoremstyle{paperresult}
\newtheorem{theorem}{Theorem}
\newtheorem*{theorem*}{Theorem}
\newtheorem{lemma}{Lemma}
\newtheorem*{lemma*}{Lemma}
\newtheorem{prop}{Proposition}
\newtheorem*{prop*}{Proposition}
\newtheorem{corollary}{Corollary}

\theoremstyle{paperstatement}
\newtheorem{assumption}{Assumption}
\newtheorem{condition}{Condition}
\newtheorem{definition}{Definition}
\newtheorem*{definition*}{Definition}

\newtheorem{remark}{Remark}
\newtheorem*{problem*}{Problem Statement}
\crefname{assumption}{Assumption}{Assumptions}
\crefname{condition}{Condition}{Conditions}
\crefname{definition}{Definition}{Definitions}
\crefname{example}{Example}{Examples}
\crefname{remark}{Remark}{Remarks}
\crefname{lemma}{Lemma}{Lemmas}
\crefname{prop}{Proposition}{Propositions}
\crefname{corollary}{Corollary}{Corollaries}
\crefname{cor}{Corollary}{Corollaries}
\crefname{fig}{Figure}{Figures}

\usepackage[left,mathlines]{lineno}

\AtBeginDocument{\ifthenelse{\boolean{linenumbers}}{\linenumbers}{}}

\newcommand{\1}{\mathds{1}}
\newcommand{\bv}{\bm{v}}
\newcommand{\bV}{\bm{V}}

\newcommand{\bW}{\bm{W}}

\newcommand{\indep}{\perp\!\!\!\perp}

\newcommand{\Var}{\operatorname{Var}}
\newcommand{\op}{o_p}
\newcommand{\Op}{\mathcal{O}_p}

\DeclareMathOperator*{\argmax}{arg\,max}
\DeclareMathOperator*{\argmin}{arg\,min}

\newcommand{\bx}{\bm{x}}
\newcommand{\bX}{\bm{X}}

\newcommand{\E}{\mathbb{E}}
\newcommand{\Prob}{\mathbb{P}}

\ifthenelse{\boolean{showcomments}}{
\newcommand{\falco}[1]{{\color{red} (FB: #1)}}
\newcommand{\falcoblock}[1]{{\color{red}\bfseries [BLOCKING] }{\color{red} #1}}
\newcommand{\duy}[1]{{\color{magenta} (DP: #1)}}
\newcommand{\kevin}[1]{{\color{teal} (KJ: #1)}}
}{
\newcommand{\falco}[1]{}
\newcommand{\falcoblock}[1]{}
\newcommand{\duy}[1]{}
\newcommand{\kevin}[1]{}
}

\title{Interpretable Discovery of Heterogeneous Causal Dose--Response Curves}

\ifdefined\anonymousVersion\else
\author{Falco J. Bargagli-Stoffi\textsuperscript{1}}
\address{\textsuperscript{1}Department of Biostatistics, University of California, Los Angeles}
\email{falco@ucla.edu (Corresponding author)}

\author{Duy Pham\textsuperscript{2}}
\address{\textsuperscript{2}Department of Statistics and Data Science, University of California, Los Angeles}

\author{Edward H. Kennedy\textsuperscript{3}}
\address{\textsuperscript{3}Department of Statistics \& Data Science, Carnegie Mellon University}

\author{Heejun Shin\textsuperscript{4}}
\address{\textsuperscript{4}Department of Applied Statistics, Konkuk University}

\author{Noah Greifer\textsuperscript{5}}
\address{\textsuperscript{5}Institute for Quantitative Social Science, Harvard University}

\author{Kevin P. Josey\textsuperscript{6}}
\address{\textsuperscript{6}Department of Biostatistics and Informatics, Colorado School of Public Health}

\author{Rachel C. Nethery\textsuperscript{7}}
\address{\textsuperscript{7}Department of Biostatistics, Harvard T.H. Chan School of Public Health}

\fi

\date{}

\begin{document}

\ifthenelse{\boolean{linenumbers}}{\linenumbers}{}
\doublespacing

\maketitle

\begin{abstract}
\singlespacing
Discovering subgroups with heterogeneous treatment effects is a central task in causal inference.
While methods for heterogeneous effect discovery have proliferated for discrete treatments, their counterparts for continuous treatments remain underdeveloped.
In this paper, we introduce a novel, data-adaptive algorithm for discovering interpretable subgroups with heterogeneous causal dose-response curves.
This algorithm combines doubly-robust dose-response curve estimation with recursive partitioning, using flexible objective functions to quantify heterogeneity across potential subgroups' dose-response curves.
First, we show that the standard doubly-robust pseudo-outcome for continuous treatments cannot serve this purpose.
We therefore construct the pseudo-outcome separately for each candidate subgroup, using subgroup-specific g-computation and a subgroup-specific stabilizing density.
Finally, we characterize the differences between dose-response curves detected by each objective function.
We establish uniform convergence of the splitting criteria and give conditions for recovery of the population greedy tree.
%Monte Carlo simulations characterize discovery performance of the algorithm across signal strengths and confounding scenarios.
We apply it to county-level data on fine particulate matter pollution (PM$_{2.5}$) and mortality in the contiguous United States over the period 2000--2019. 
The discovered heterogeneous subgroups vary by education, income, and race.
\ifdefined\anonymousVersion The proposed algorithm is implemented in an \textsf{R} package.\else
The proposed algorithm is implemented in the \textsf{R} package \href{https://github.com/fbargaglistoffi/discoverHTE}{\texttt{discoverHTE}}.
\fi
\end{abstract}


\vspace{0.5cm}

\section{Introduction}
\label{s:intro}

Discovering how causal effects vary across populations is central to precision medicine, targeted interventions, and public health.
Population averages can obscure variation relevant to these decisions \citep{ding2016randomization, ding2019decomposing, dominici2020controlled}.
Because the heterogeneous subgroups are often unknown, discovery entails estimating their response functions and representing their differences through interpretable rules \citep{kosorok2019precision}.

Substantial methodological progress has been made for heterogeneous effects discovery and estimation.
On the discovery side, tree-based methods provide interpretable and flexible tools for heterogeneous subgroup discovery \citep{athey2016recursive,wager2018estimation,hahn2020bayesian, bargagli2022heterogeneous,bargagli2025heterogeneous, bargagli2024causal,huang2025distilling}.
On the estimation side, meta-learners offer complementary approaches for conditional average treatment effects (CATE) estimation \citep{Nie2017, kunzel2019metalearners}, while doubly robust methods ensure robustness to model misspecification \citep{chernozhukov2018double, semenova2021debiased,kennedy2023towards}.

The bulk of these methods, however, was developed in the context of discrete treatment. Yet, in many applications the treatment is inherently continuous: prominent examples are pollutant concentrations in environmental epidemiology, drug dosages in pharmacology, and policy intensity or duration in the social sciences.
For such continuous treatments, causal effects can be characterized by \textit{dose-response curves} rather than scalar contrasts, and summarizing heterogeneity often requires comparing these functions across subgroups: subgroups may differ in average response levels, in the effects of changing dose, or both.
Recent methodological advances have improved dose-response estimation for continuous treatments \citep{Kennedy2017,ren2021bayesian,  giffin2023generalized,wu2024matching,colangelo2026double}. 

In the context of heterogeneous effects with a continuous exposure, \cite{Woody2020} extend Bayesian tree ensembles to continuous treatments, focusing on effect estimation through sum-of-trees models that do not naturally yield interpretable subgroups. 
\cite{shin2025treatment} address multivariate continuous treatments with an emphasis on variable importance rather than subgroup discovery.\footnote{See also \cite{boileau2025nonparametric,hines2025variable,bargagli2026causal} for general methods for variable importance in the context of heterogeneous effects.} 
\cite{jin2023change} focus on testing for differential effects between two subgroups under a continuous treatment using a semiparametric change-plane model, assuming a two-group structure separated by a linear boundary rather than learning a flexible subgroup partition.\footnote{We note that population-average curve estimation, variable importance, prespecified subgroups, and restricted change-plane or treatment-trigger models provide complementary outputs.
\Cref{app:related} compares these approaches, including recent cluster-based methods and alternative policy estimands.} 
\cite{thomas2018dose} use model-based recursive partitioning in randomized dose-finding trials.
None of these methods, however, compares doubly robust estimates of subgroup curves for continuous treatments in observational studies.

\subsection{Contributions}

To fill this gap in the literature, we develop an algorithm for interpretable discovery of heterogeneity in causal dose-response curves, with three main methodological contributions.

First, we show that the population-marginal pseudo-outcome of \cite{Kennedy2017}, evaluated with the true nuisance functions, returns the population-average curve in every subgroup (\Cref{sec:pseudo}).
Reweighting its residual alone cannot change that target.
Second, we construct the pseudo-outcome separately for each candidate subgroup in the tree search (\Cref{sec:new_pseudo}).
The subgroup-specific g-computation component determines the appropriate target, while the subgroup stabilizing density supplies the correction needed for double robustness under outcome-model misspecification.
Third, we characterize differences between dose-response curves through objective functions (e.g., functional ANOVA, integrated $L^2$ distance, and area under the curve [AUC]\footnote{As illustrated in \Cref{sec:of}, AUC measures weighted average response differences, the centered shape component measures variation in additive dose contrasts, and $L^2$ and FANOVA respond to both, with FANOVA additionally favoring balanced splits.}) and use them to select interpretable subgroup rules, with optional density-based weights that emphasize common exposure support (\Cref{sec:recursive} and \Cref{sec:of}).

We establish tree consistency under uniform estimation bounds, stable admissibility, and a split-gain margin.
Finally, we apply the method to environmental epidemiology, analyzing county-level data on fine particulate matter (PM$_{2.5}$) and mortality in the United States between 2000 and 2019.
The fitted curves suggest differential vulnerability to air pollution by educational attainment.
The estimated mortality increase from 5 to 9 $\mu$g/m$^3$ is larger in county-years with lower educational attainment.

%\vspace{-0.25cm}
%\subsection{Paper Organization}
%\vspace{-0.25cm}
%\Cref{s:data} introduces the data and research questions. \Cref{s:methods,s:application,s:conclusions} present the method, application, and discussion.

%\vspace{-0.25cm}
\section{Data Description and Motivation}
\label{s:data}

Our central research questions concern the discovery of subgroups with different causal dose--response relationships under a continuous treatment and the characteristics that distinguish them.
As discussed in the previous section, environmental epidemiology provides a prominent example: how does the relationship between air pollution and mortality vary across communities?
Specifically, fine particulate matter (PM$_{2.5}$) exposure is a leading environmental risk factor for mortality, contributing to cardiovascular disease, respiratory illness, and premature death \citep{zorzetto2026bayesian}.
To investigate these questions, we construct a county-level dataset linking long-term exposure to PM$_{2.5}$ with mortality across the contiguous United States.
Our goal is to characterize the causal PM$_{2.5}$--mortality relationship and discover, in a data-adaptive manner, interpretable subgroups of counties with different dose--response functions.
The county scale is relevant to air-quality planning, where designated areas are often defined using counties or portions thereof \citep{cfr_40_part81}.

\subsection{Data and Preprocessing}\label{sec:data}

\begin{figure}%[htbp]
\centering
\includegraphics[width=0.86\textwidth]{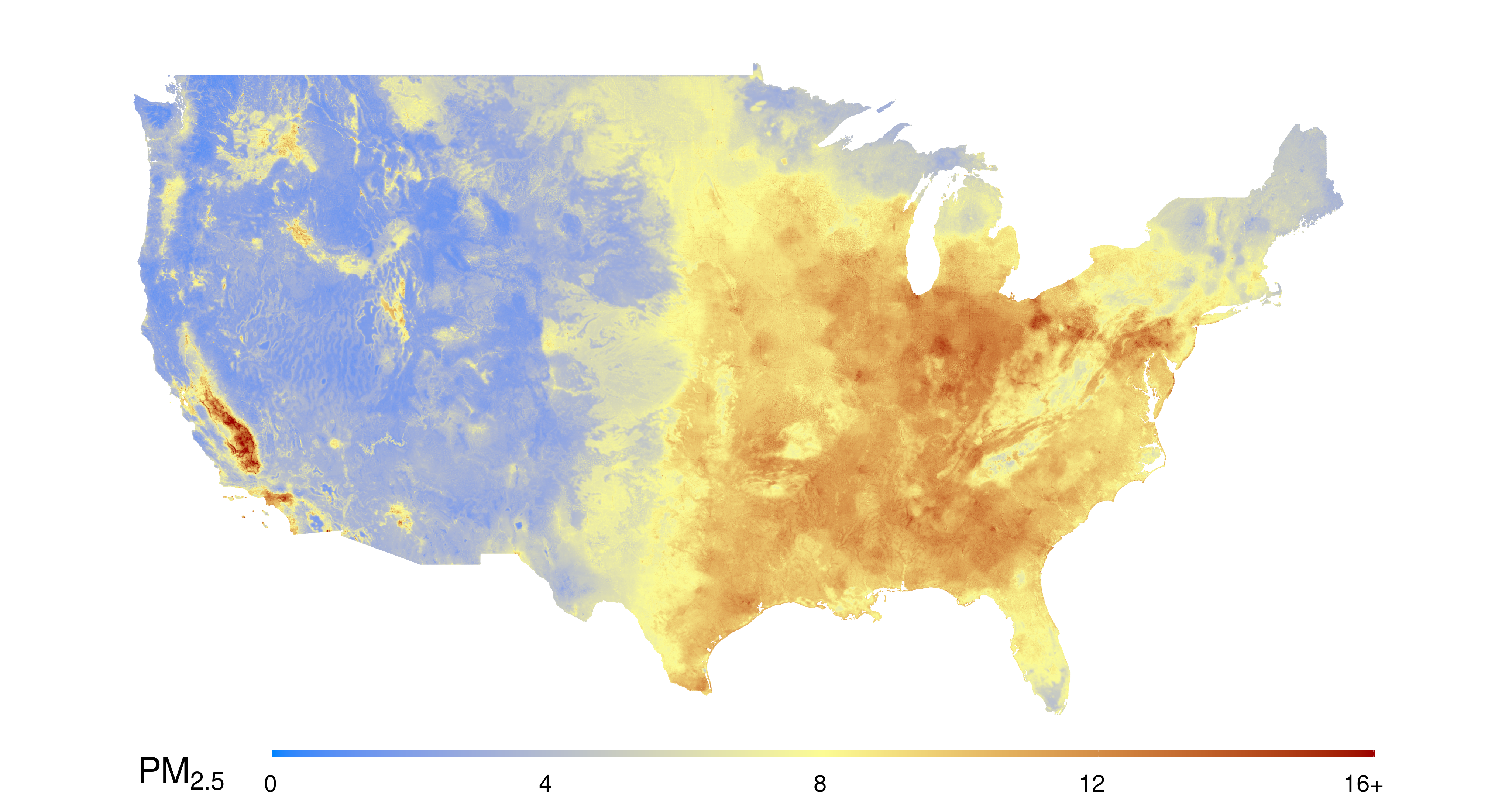}
\caption{\footnotesize Illustrative spatial distribution of annual average PM$_{2.5}$ concentrations ($\mu$g/m$^3$) across the contiguous United States, based on satellite-derived estimates from \protect\cite{van2021monthly}.}
\label{fig:map_pm25}
\end{figure}

\textit{Exposure.}
Our analysis covers the 20-year period 2000--2019.
We restrict the analysis to this pre-pandemic period to avoid pandemic-related changes in mortality patterns, particularly given evidence associating long-term PM$_{2.5}$ exposure with COVID-19 mortality \citep{Wu2020COVID}.
Annual average PM$_{2.5}$ concentrations ($\mu$g/m$^3$) were derived from satellite-based estimates \citep{van2021monthly}, available at a spatial resolution of $0.01^{\circ} \times 0.01^{\circ}$ and visualized in \cref{fig:map_pm25}.
County-level exposures were obtained by averaging all grid-cell values falling within each county.
We use one-year-lagged concentrations as the exposure of interest, with a mean of $8.71\ \mu$g/m$^3$, a standard deviation of $2.71\ \mu$g/m$^3$, and a range of $2.86$ to $18.65\ \mu$g/m$^3$ across county-years.\footnote{Contemporaneous concentrations have a similar distribution, with a mean of $8.59\ \mu$g/m$^3$ and a range of $2.81$ to $18.65\ \mu$g/m$^3$.}
In 2024, the U.S. Environmental Protection Agency lowered the primary annual National Ambient Air Quality Standard (NAAQS) for PM$_{2.5}$ from $12$ to $9\ \mu$g/m$^3$ \citep{usepa_pm_naaqs_2024}.
We use the $9\ \mu$g/m$^3$ level as a policy-relevant benchmark in the subsequent analyses: approximately one-third of county-year exposures in our sample exceed this level.

\textit{Outcome.}
Death counts were obtained from the CDC WONDER database \citep{cdcwonder2024} for the period 2000--2019, by county, year, and age stratum, for the three strata 65--74, 75--84, and 85 years and older.\footnote{Since our exposure of interest is the previous year's PM$_{2.5}$ concentration, the analysis excludes mortality records from 2000.} We restrict to ages 65 and above, as older adults represent one of the groups at the highest risk from exposure to air pollution \citep{Wu2020}.
The analytic sample comprises 2{,}815 counties in the contiguous United States with linked exposure data, 53{,}479 county-years, and 160{,}329 county-year-age cells, covering 29.4 million deaths over 652 million person-years.
Because all three strata are aged 65 and over, the estimand throughout is age-standardized mortality among adults aged 65 and older, expressed per 100{,}000 person-years.

\textit{Covariates.}
County-level socioeconomic and demographic characteristics were obtained from the American Community Survey \citep{acs2024formal}: median home value, median household income, percent of the population identifying as non-Hispanic Black, percent identifying as Hispanic, percent with higher educational attainment, percent below the poverty level, and percent of housing units owner-occupied.
Similar socioeconomic and demographic covariates are commonly included in confounding adjustment in studies of long-term PM$_{2.5}$ exposure and mortality \citep{Wu2020,Wei2021DoseResponse,Bennett2019CountyMortality}.
We additionally adjust for an indicator of urbanicity based on the CDC's National Center for Health Statistics urban--rural classification scheme \citep{ingram20142013}, and for K\"oppen climate type to capture broad climatic differences across regions.\footnote{Due to missing covariate values, we exclude ten county-years from Williamsburg, VA, spanning 2010 to 2019, from the analysis, leaving 53{,}469 county-years.}

\subsection{The Population-Average Relationship}
\label{sec:motivation}

\begin{figure}%[htbp]
\centering
\includegraphics[width=0.75\textwidth]{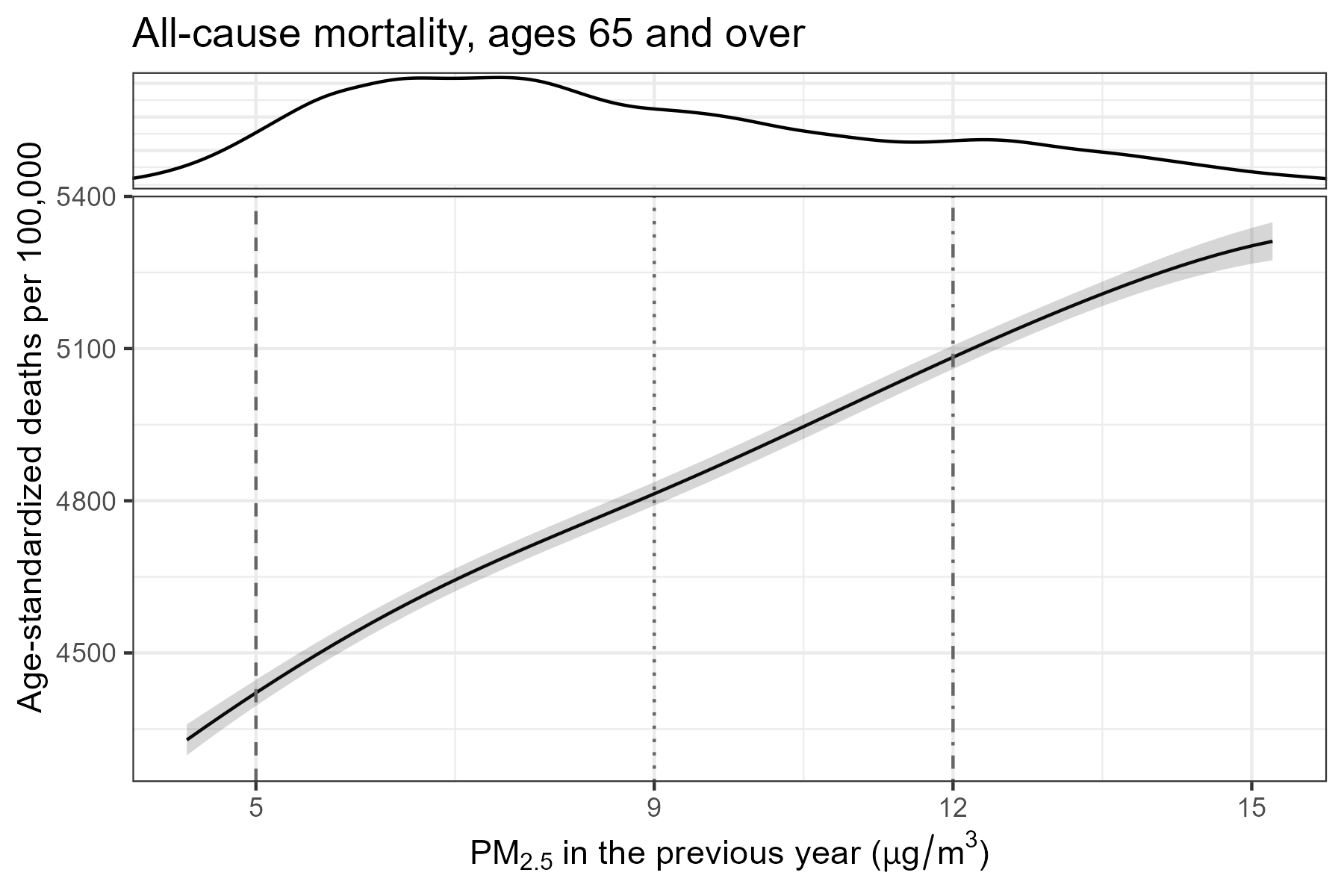}
\caption{\footnotesize Estimated population-average dose-response curve relating one-year-lagged fine particulate matter (PM$_{2.5}$) exposure, on the horizontal axis, to county-level mortality (deaths per 100,000 person-years), on the vertical axis. 
Exposure ranges from 4.31 to 15.2 $\mu$g/m$^3$, corresponding to the 1st and 99th percentiles across all 53{,}479 county-years.
The solid black line represents the estimated exposure-response function, with the shaded region indicating the reported pointwise 95\% interval.
The density plot shows the observed distribution of PM$_{2.5}$ exposure across counties.
The vertical lines mark 5, 9, and 12 $\mu$g/m$^3$. The latter two correspond to the current and previous EPA annual standards, respectively.}
\label{fig:erf_pm25}
\end{figure}

Before examining how the relationship between air pollution and mortality varies across communities, it is important to evaluate the population-level PM$_{2.5}$--mortality relationship.
\Cref{fig:erf_pm25} presents the population-average dose-response curve, estimated by adapting the pseudo-outcome procedure of \cite{Kennedy2017} to count outcomes as described in \Cref{app:estimation}.
The fitted population curve is positive and approximately linear across the estimation range, consistent with previous findings \citep{Wu2020}.%\footnote{\Cref{sec:q2} compares its estimated contrasts below and above the regulatory reference point.}
The pointwise interval widens toward both ends of the exposure distribution, consistent with the reduced availability of observations at particularly low and high PM$_{2.5}$ concentrations.

While this curve is informative about the population-level health response to air pollution, it is uninformative about whether the health harm induced by exposure to different levels of PM$_{2.5}$ affects all communities equally.
Existing studies of heterogeneity in this relationship fix the subgroups ex ante, typically subgroups defined by socioeconomic status, race, or urbanicity \citep{josey2023air}.
Such analyses evaluate prespecified subgroup contrasts but do not themselves search for new partitions or unanticipated combinations of modifiers.

%Furthermore, as \Cref{fig:map_pm25} makes clear, exposure is spatially structured, and it is also socially structured: \cite{jbaily2022air} document systematic disparities in PM$_{2.5}$ exposure across income and demographic groups in the United States, with the disparity persisting even as absolute concentrations have fallen.
%Candidate partitions of the county set along urbanicity or socioeconomic composition therefore tend also to partition the exposure range, and comparing dose-response curves across subgroups whose exposure distributions barely overlap requires extrapolation into regions where one subgroup has almost no data.
%This is the practical form that violations of the positivity assumption take in continuous-treatment settings, and it motivates the weighting function introduced in \Cref{sec:weighting}.

\subsection{Research Questions}
\label{sec:rq}

These considerations, together with the gaps in the literature identified above, motivate three research questions:

\begin{itemize}
\item[(Q1)] Which combinations of county characteristics define interpretable subgroups with systematically different PM$_{2.5}$ dose--response relationships?
\item[(Q2)] To what extent do the discovered subgroup curves differ in mortality levels and in their causal responses to changes in PM$_{2.5}$ exposure, particularly below the $9\ \mu$g/m$^3$ standard?
\item[(Q3)] How do the discovered subgroups and their interpretation depend on the choice of heterogeneity objective and on weighting by common exposure support?
\end{itemize}

Questions Q1 and Q2 address substantive questions about which communities have different dose--response relationships and whether these differences imply different effects of changes in exposure, with potential implications for environmental monitoring and mitigation.
Question Q3 examines how the findings depend on the definition of heterogeneity and the relative weight assigned to different exposure levels when comparing subgroup curves.
We address these questions using the methodological framework developed in the following section.

%We note at the outset that NAAQS enforcement and monitoring investment are geographic decisions taken at the county level or above.
%The county is the unit of analysis and of the reported potential-outcome estimands, while policies changing county pollution concentrations may operate over wider geographic areas.
%The causal interpretation therefore requires that the hypothetical exposure intervention be well defined at the county level, and the analysis does not establish independent manipulability of pollution in neighboring counties or individual-level effects.

\section{Methods}
\label{s:methods}

\subsection{Problem Formulation}

Consider $n$ i.i.d. units indexed by $i = 1, \ldots, n$.
Each unit has covariates $\bX_i \in \mathbb{R}^p$, potential effect modifiers $\bV_i \in \mathbb{R}^q$ (a subvector of $\bX_i$), continuous treatment $Z_i \in \mathbb{R}$, and outcome $Y_i$.
Effect modifiers may be continuous (e.g., income) or discrete (e.g., sex).
We compare curves on a nondegenerate compact interval $\mathcal{Z} = [z_{\min}, z_{\max}]$ contained in the interior of the treatment support.
In our motivating application, $Z_i$ represents pollutant concentration.
Under the potential outcomes framework \citep{rubin1974estimating}, each unit has potential outcomes $Y_i(z)$ for all $z \in \mathcal{Z}$, with $Y_i = Y_i(Z_i)$ observed.
We assume that the potential-outcome process is jointly measurable and we adopt standard identification assumptions for causal dose-response curves \cite[see, e.g., ][]{wu2024matching}:

\begin{assumption}[Consistency]\label{ass:consistency}
$Y_i = Y_i(z)$ if $Z_i = z$.
For fixed-dose identification, assume also that $\lim_{r\downarrow 0}\E[Y_i-Y_i(z)\mid |Z_i-z|<r,\bX_i]=0$ almost surely for each $z\in\mathcal Z$.
\end{assumption}

\begin{assumption}[Unconfoundedness]
\label{ass:unconfoundedness}
$\{Y_i(z): z \in \mathcal{Z}\} \indep Z_i \mid \bX_i$.
\end{assumption}

\begin{assumption}[Overlap]\label{ass:overlap}
For the conditional treatment density $\pi(z\mid\bx)$, $\pi(z\mid\bx)\geq\epsilon>0$ for $z\in\mathcal Z$ and $P_{\bX}$-almost every $\bx$.
\end{assumption}

Under these assumptions, the population average \textit{dose-response curve} is identified as
\begin{equation}
\theta(z) = \E[Y_i(z)] = \E \big[\E[Y_i \mid Z_i = z, \bX_i]\big] = \int \mu(z, \bx) \, d\Prob(\bx),\footnote{For contrasts at specified doses, we additionally need to assume that the relevant causal mean curves are continuous at those doses.}
\end{equation}
where $\mu(z,\bx) = \E[Y_i \mid Z_i = z, \bX_i = \bx]$ is the outcome regression.
%At a fixed dose, \Cref{ass:consistency,ass:unconfoundedness} give the version $\mu(z,\bx)=\lim_{r\downarrow0}\E[Y_i\mid |Z_i-z|<r,\bX_i=\bx]=\E[Y_i(z)\mid\bX_i=\bx]$.
For a subgroup defined by $\bV \in \mathcal{L}$, a subset of the effect modifier space, the conditional dose-response curve is
\begin{equation}\label{eq:subgroup_drc}
\theta(z \mid \mathcal{L}) = \E[Y_i(z) \mid \bV_i \in \mathcal{L}] = \int \mu(z, \bx) \, d\Prob(\bx \mid \bV \in \mathcal{L}).
\end{equation}

Our first goal is to discover subgroups of the population that exhibit meaningfully different dose-response relationships.
What constitutes a ``meaningful difference'' depends on the policy or clinical context: one may be interested in detecting differences in dose-response shape, average outcome level, or both.\footnote{Other functional contrasts are possible, under this general construction, but left for future work.}
For two subgroups $L$ and $R$, let $d(z)=\theta_L(z)-\theta_R(z)$ denote the difference in their mean potential outcomes at dose $z$.
The difference between the subgroups in the additive causal effect of changing the dose from $z_0$ to $z$ is $d(z)-d(z_0)$.\footnote{Our objective is to characterize heterogeneity across entire dose--response functions without restricting attention to a
prespecified intervention.
This allows the discovered subgroups to be examined under different policy-relevant exposure contrasts.
A complementary approach considers stochastic interventions that modify the exposure distribution, providing a framework for evaluating the effects of particular intervention regimes
\citep{diaz2012population,schindl2024incremental}.}
This distinction is essential because parallel curves may differ in level without implying heterogeneous causal effects of changing dose.
We formalize these goals through the choice of objective function (\Cref{sec:of}), which quantifies heterogeneity in a manner aligned with the scientific question.

Beyond quantifying heterogeneity, our second goal is to discover subgroups in a way that retains \textit{interpretability}.
Here, interpretability refers to describing each subgroup by a short sequence of \textit{if-then} decision rules. Such decision rules are highly interpretable as they resemble human decision-making processes \citep{bargagli2024causal}. 
In what follows, we obtain such rules via recursive partitioning, which systematically explores complex heterogeneity patterns to produce simple, actionable and mutually-exclusive subgroup definitions in the form of binary decision trees.

\subsection{Pseudo-Outcomes for Dose-Response Curves}
\label{sec:pseudo}

Recursive partitioning requires, at every candidate split, an estimate of the dose-response curve within each subgroup.
The doubly-robust approach of \cite{Kennedy2017} constructs a \emph{pseudo-outcome} whose conditional mean given treatment equals the curve of interest, and then regresses that pseudo-outcome on treatment using off-the-shelf nonparametric methods.
Throughout, we write $\pi(z \mid \bx)$ for the generalized propensity score, $\varpi(z) = \int \pi(z \mid \bx)\, d\Prob(\bx)$ for the marginal treatment density, and, for a subgroup $\mathcal{L}$,
$\varpi(z \mid \mathcal{L}) = \int \pi(z \mid \bx)\, d\Prob(\bx \mid \bV \in \mathcal{L})$
for the treatment density within that subgroup.
The nuisance quantities are the outcome regression $\mu(z,\bx)$ and generalized propensity density $\pi(z\mid\bx)$, together with the stabilizing densities $\varpi(z)$ and $\varpi(z\mid\mathcal L)$ obtained by averaging $\pi$ over the corresponding covariate distributions.
These distributions are also estimated in practice when constructing the g-computation components.
The pseudo-outcome can be read as a g-computation estimate of the dose-response curve, debiased by a weighted outcome residual \citep{austin2018generalized,zhang2016quantitative,Kennedy2017}.
The construction of \cite{Kennedy2017} is
\begin{equation}\label{eq:pseudo_marginal}
\varphi_i \;=\;
\underbrace{\frac{\varpi(Z_i)}{\pi(Z_i \mid \bX_i)}}_{\substack{\text{\normalfont stabilizer}\\
%\text{\normalfont $w(Z_i,\bX_i)$}
}}
\;\cdot\;
\underbrace{\big\{Y_i - \mu(Z_i, \bX_i)\big\}}_{\substack{\text{\normalfont residual}\\
%\text{\normalfont mean zero given } (Z_i,\bX_i)
}}
\;+\;
\underbrace{\int \mu(Z_i, \bx)\, d\Prob(\bx)}_{\substack{\text{\normalfont g-computation}\\ \text{\normalfont component}}}.
\end{equation}
The three factors play distinct roles.
The \emph{stabilizer} $w(z, \bx) = \varpi(z)/\pi(z \mid \bx)$ depends on the treatment and the covariates alone and is bounded under \Cref{ass:overlap} (together with an upper bound on the marginal treatment density, as required by \Cref{con:density}).
The \emph{residual} $Y_i - \mu(Z_i, \bX_i)$ has conditional mean zero given $(Z_i, \bX_i)$, by the definition of $\mu$.
The \emph{g-computation component} $b(z)=\int\mu(z,\bx)\,d\Prob(\bx)$ is the outcome-regression-based representation of the population dose-response curve at $z$.
Although $b(z)=\theta(z)$ when $\mu$ and the covariate distribution are evaluated at their population values, neither is known in practice, so $b(z)$ must be estimated and its model error corrected by the weighted residual term.

Compactly, $\varphi_i = w(Z_i, \bX_i)\{Y_i - \mu(Z_i, \bX_i)\} + b(Z_i)$.
The construction satisfies $\E[\varphi_i \mid Z_i = z] = \theta(z)$ \citep{Kennedy2017}, and the stabilizer is what delivers double robustness: it is the factor that allows a misspecified outcome regression to be corrected by a correctly specified treatment model, and vice versa.

The g-computation component and the stabilizer play distinct roles when the same construction is applied within a subgroup.
For fixed population quantities, $b(z)$ depends on treatment alone and determines which curve the pseudo-outcome targets, whereas the stabilizer corrects misspecification of the estimated outcome regression.
Neither component is supplied for free: the identity below concerns their population values, and the estimated versions require nuisance and empirical-process analysis.
\Cref{lem:offset} in the Supplementary Material formalizes the role of the additive component.
We assume the weighted residual and additive component are integrable.
The score identities below hold for almost every dose in each subgroup of positive probability.
Applying \Cref{lem:offset} to \eqref{eq:pseudo_marginal}, whose g-computation component is $b(z) = \theta(z)$, gives the following Proposition.

\begin{prop}[Insufficiency of the marginal pseudo-outcome]\label{prop:insufficient}
Under \Crefrange{ass:consistency}{ass:overlap}, with $\mu$, $\pi$, and $\varpi$ evaluated at their true population values, the pseudo-outcome \eqref{eq:pseudo_marginal} satisfies
$\E[\varphi_i \mid Z_i = z, \bV_i \in \mathcal{L}] = \theta(z) \: \text{for every subgroup } \mathcal{L}$.
%Consequently, for any split $s$ of any subgroup, the two child curves obtained from \eqref{eq:pseudo_marginal} coincide, and the population objective satisfies $\mathcal{G}(s) = 0$ for every $\mathcal{G}$ in \Cref{sec:of}.
\end{prop}

%\Cref{prop:insufficient} is stronger than a statement about bias.
\Cref{prop:insufficient} shows that, at the population level, conditioning the marginal pseudo-outcome on subgroup membership returns the same curve in every subgroup, so this pseudo-outcome cannot serve as the foundation for an algorithm aiming to discover heterogeneous causal dose--response curves.\footnote{Estimated curves can nevertheless differ through sampling and nuisance-estimation error; finite-sample stopping requires a separate rule.}
Because \Cref{lem:offset} holds for \emph{any} stabilizer $w$, no reweighting of the residual can repair this: the pseudo-outcome must change to target conditional effects.\footnote{The failure persists with estimated nuisances when the outcome regression is uniformly consistent and stabilizing ratios remain bounded.
\Cref{app:marginal_details} gives the precise expansion, explains the empirical g-computation error, and relates the construction to orthogonal statistical learning.
The distinction from regular conditional influence-function targets is discussed in \Cref{rem:cif}.}

\subsection{A Pseudo-Outcome for Subgroup Dose-Response Curves}\label{sec:new_pseudo}

\begin{definition}[A corrected pseudo-outcome for subgroup dose-response curves]
For a fixed subgroup, extending \cite{doss2024nonparametric} with subgroup membership as the modifier, for each candidate subgroup $\mathcal{L}$, we define \vspace{-0.25cm}
\begin{equation}\label{eq:pseudo_subgroup}
\varphi^{\mathcal{L}}_i \;=\;
\underbrace{\frac{\varpi(Z_i \mid \mathcal{L})}{\pi(Z_i \mid \bX_i)}}_{\substack{\text{\normalfont subgroup stabilizer}}}
\;\cdot\;
\underbrace{\big\{Y_i - \mu(Z_i, \bX_i)\big\}}_{\text{\normalfont residual}}
\;+\;
\underbrace{\int \mu(Z_i, \bx)\, d\Prob(\bx \mid \bV \in \mathcal{L})}_{\substack{\text{\normalfont subgroup g-computation} \\ \text{\normalfont $= \theta(Z_i \mid \mathcal{L})$}}},
\end{equation}
computed on units with $\bV_i \in \mathcal{L}$.
\end{definition}

Both modifications are with respect to the same object, the covariate distribution restricted to the subgroup, and both are needed as shown in the following Proposition.\footnote{\Cref{prop:subgroup} is stated for the population version \eqref{eq:pseudo_subgroup}; the fold-indexed estimator in \eqref{eq:pseudo_hat} uses within-subgroup covariate averages and a training-only subgroup stabilizing density.
\Cref{app:estimation} distinguishes averaging fitted propensity densities from direct density smoothing; their errors must be controlled for the construction actually used.
The resulting additional error is carried explicitly through \Cref{lem:erf_uniform_sub}.}

\begin{prop}[Identification and double robustness]\label{prop:subgroup}
Under \Crefrange{ass:consistency}{ass:overlap}, the pseudo-outcome \eqref{eq:pseudo_subgroup} satisfies
$\E[\varphi^{\mathcal{L}}_i \mid Z_i = z, \bV_i \in \mathcal{L}] = \theta(z \mid \mathcal{L})$.
Moreover, for a fresh evaluation observation independent of the nuisance-training and empirical g-computation samples, write $\hat\varphi^{\mathcal{L}}_i$ for the version constructed from $(\hat\mu,\hat\pi,\hat\varpi)$ and $\hat\Prob(\cdot\mid\bV\in\mathcal L)$ for the estimated within-subgroup covariate distribution,
\begin{align}\label{eq:dr_remainder}
&\E[\hat\varphi^{\mathcal{L}}_i \mid Z_i = z, \bV_i \in \mathcal{L}] - \theta(z \mid \mathcal{L})\nonumber\\
&= \int \big\{\mu(z,\bx) - \hat\mu(z,\bx)\big\}
\left\{ \frac{\pi(z \mid \bx) \,/\, \varpi(z \mid \mathcal{L})}{\hat\pi(z \mid \bx) \,/\, \hat\varpi(z \mid \mathcal{L})} - 1 \right\} dP_{\mathcal L}(\bx) \nonumber \\
&\quad + \int \hat\mu(z, \bx)\, d(\hat P_{\mathcal L}-P_{\mathcal L})(\bx).
\end{align}
The expectation is conditional on the fitted quantities; $P_{\mathcal L}$ and $\hat P_{\mathcal L}$ denote the true and estimated within-subgroup covariate laws.
The product term vanishes if $\hat\mu=\mu$ or the normalized treatment-density ratio is correct; the latter holds if both $\hat\pi=\pi$ and $\hat\varpi(\cdot\mid\mathcal L)=\varpi(\cdot\mid\mathcal L)$.
The empirical g-computation term remains in either case and is $O_p(n_{\mathcal L}^{-1/2})$ for an independent covariate-averaging sample of comparable size at a fixed subgroup under a uniform second-moment bound; adaptive subgroups require the separate bound in \Cref{con:nuisance}.
\end{prop}

The subgroup g-computation component identifies the target when the outcome regression is correct.
The subgroup stabilizer also permits correction of outcome-model error when the treatment-density components are correct. 
However, retaining a marginal stabilizer does not generally provide that cancellation.
\Cref{rem:stabilizer,app:po_sim} give the exact bias and an illustrative example in which the apparent direction of a subgroup curve reverses.

\subsection{Recursive Partitioning for Continuous Treatments}\label{sec:recursive}

To discover subgroups, we seek a partition of the effect modifier space into subspaces $\{\mathcal{L}_1, \ldots, \mathcal{L}_K\}$ such that the conditional dose-response curves $\theta(z \mid \mathcal{L}_k)$ differ substantially across regions while remaining relatively homogeneous within each region.
We pursue this through recursive partitioning: starting from the full population (what is often referred to as the root node), we iteratively split the sample into subgroups (i.e., nodes) that maximize heterogeneity in dose-response.
This yields a tree structure on $\mathbb{R}^q$, where each terminal node corresponds to a subgroup defined by a conjunction of simple rules (e.g., ``\texttt{age > 65} \& \texttt{sex = female}'').
The tree is grown via a sequence of binary splits \citep[for a review, see][]{loh2014fifty}.
%Throughout, the population reference tree is the \emph{population greedy tree} that maximizes the chosen splitting objective locally at each node subject to the same admissibility and depth restrictions as the algorithm.
%It need not be the globally optimal partition among all trees of the permitted size; our consistency claims concern the greedy population target.

Formally, a tree $\psi$ induces a partition $\Pi(\psi) = \{\mathcal{L}_1, \ldots, \mathcal{L}_{|\psi|}\}$ into $|\psi|$ terminal nodes, where each $\mathcal{L}_k$ is defined by intersections of halfspaces of the form $\{V_j \leq c\}$ or $\{V_j > c\}$ where $c \in \mathbb{R}$ is a selected thresholding value.
Each split $s = (j, c, \mathcal{L})$ divides a node $\mathcal{L}$ on variable $V_j$ at threshold $c$ into children: $\mathcal{L}_L = \{v \in \mathcal{L} : v_j \leq c\}$ and $\mathcal{L}_R = \{v \in \mathcal{L} : v_j > c\}.$

Once a node has been split into two child nodes, further splits of the child nodes are considered.
Splitting continues until a stopping criterion is reached.
After the conclusion of splitting, the final set of child nodes are referred to as ``terminal nodes'' (or leaves). %which in our setting define the mutually exclusive subspaces of the modifier space with heterogeneous dose-response curves.

To select splits, including both the variable to split on and the splitting threshold, we require an \textit{objective function}, $\mathcal{G}(s)$, that quantifies the heterogeneity in dose-response relationships between the two child nodes of any candidate split.
Given an objective function, the optimal split at node $\mathcal{L}$ is
\begin{equation}
s^*(\mathcal{L}) = \argmax_{s \in \mathcal{S}(\mathcal{L})} \mathcal{G}(s),
\end{equation}
where $\mathcal{S}(\mathcal{L})$ is the set of candidate splits at $\mathcal{L}$, and the tree is grown by recursively selecting splits.
The choice of objective function determines what type of heterogeneity the method prioritizes.
We now present three objective functions with different interpretations and split rankings.
All these objective functions can incorporate a weighting function $w_s(z)$, illustrated in \Cref{sec:weighting}, which allows the comparison to be focused on the area of common support.

\subsection{Objective Functions}\label{sec:of}

\subsubsection{Functional ANOVA}
The functional ANOVA objective measures weighted deviations of child dose-response curves from the parent dose-response curve, analogous to between-group variability in classical ANOVA.
Write $p_s=\Prob(\bV\in\mathcal L_L\mid\bV\in\mathcal L)$ for the population proportion in the left child; the empirical criterion replaces $p_s$ by $n_L/n_P$:
\begin{align}\label{eq:fanova}
\mathcal{G}_{\text{FANOVA}}(s) &= p_s \int_{\mathcal{Z}} [\theta(z \mid \mathcal{L}_L) - \theta(z \mid \mathcal{L})]^2 w_s(z) \, dz \nonumber\\
&\quad + (1-p_s) \int_{\mathcal{Z}} [\theta(z \mid \mathcal{L}_R) - \theta(z \mid \mathcal{L})]^2 w_s(z) \, dz,
\end{align}
where $n_P = n_L + n_R$ is the parent node size.

The FANOVA objective quantifies between-subgroup functional variability, with an additional preference for balanced child nodes.
Large/small values indicate substantial/irrelevant between-subgroup curve differences under the chosen weight.% small values do not establish that a uniform treatment policy is optimal.

\subsubsection{Integrated $L^2$ Distance}

The $L^2$ objective measures the weighted integrated squared distance between child dose-response curves:
\begin{align}\label{eq:l2}
\mathcal{G}_{L^2}(s) &= \int_{\mathcal{Z}} [\theta(z \mid \mathcal{L}_L) - \theta(z \mid \mathcal{L}_R)]^2 w_s(z) \, dz.
\end{align}
This is, possibly, the most direct measure of heterogeneity: it is large when the dose-response curves differ substantially across the treatment range.
The $L^2$ objective is symmetric in the two children and does not depend on the parent node's dose-response curve.

These two objectives are not independent.
Writing $p = \Prob(\bV \in \mathcal{L}_L \mid \bV \in \mathcal{L})$ and noting that $\theta(z \mid \mathcal{L}) = p\,\theta(z \mid \mathcal{L}_L) + (1-p)\,\theta(z \mid \mathcal{L}_R)$ by \eqref{eq:subgroup_drc}, substitution into \eqref{eq:fanova} gives the following.

\begin{remark}[FANOVA is a balance-penalized $L^2$]\label{prop:fanova_l2}
\textit{At the population level, for any measurable weighting function $w_s$ for which the weighted squared differences are absolutely integrable,}
\begin{equation}
\mathcal{G}_{\text{\normalfont FANOVA}}(s) = p\,(1-p)\; \mathcal{G}_{L^2}(s).
\end{equation}
\end{remark}

Two consequences are worth stating.
First, FANOVA multiplies the $L^2$ gain by $p(1-p)$, which is maximized at $p=1/2$; rankings agree when this factor is common across candidates, but need not agree otherwise.
Without the $n_L/n_P$ weights, the corresponding factor is $2p^2 - 2p + 1$, which is maximized as $p \to 0$ or $p \to 1$ and hence favors unbalanced splits.
Second, this identity concerns population curves.
With a separately fitted parent, the empirical FANOVA criterion also includes
$\int\{\hat\theta_P-\hat p\hat\theta_L-(1-\hat p)\hat\theta_R\}^2\hat w_s$;
the exact identity is given in \Cref{eq:empirical_fanova}.

\vspace{-0.5cm}
\subsubsection{Area Under the Curve}
The AUC objective compares the weighted average response across treatment levels:
\begin{align}
\mathcal{G}_{\text{AUC}}(s) &= \left| \int_{\mathcal{Z}} \theta(z \mid \mathcal{L}_L) w_s(z) \, dz - \int_{\mathcal{Z}} \theta(z \mid \mathcal{L}_R) w_s(z) \, dz \right|.
\end{align}
The absolute value ensures the objective is invariant to child labeling, though in practice the direction of the level shift is readily identified from the estimated dose-response curves.
This objective measures weighted mean differences, including but not limited to uniform vertical shifts.
Oppositely signed differences can cancel, as characterized in \Cref{prop:detection}.

The AUC objective identifies differences in weighted average outcome levels; it does not require one curve to remain above the other throughout the treatment range.
A difference in level is not by itself a difference in the effect of changing the treatment.
Two subgroups with parallel curves have the same contrast $\theta(z \mid \mathcal{L}) - \theta(z' \mid \mathcal{L})$ for every pair of doses $(z, z')$.
The AUC objective can therefore prioritize prognostic level differences even when causal dose contrasts are identical across subgroups.
An AUC-selected subgroup therefore need not have differential causal susceptibility; \Cref{prop:l2_decomp} separates level and shape.

The $L^2$ and AUC objectives are also related.
Write $d_s(z) = \theta(z \mid \mathcal{L}_L) - \theta(z \mid \mathcal{L}_R)$ for the difference curve and $\bar d_s = \int_{\mathcal{Z}} d_s(z)\, w_s(z)\, dz$ for its weighted mean.

\begin{remark}[$L^2$ decomposes into level and shape]\label{prop:l2_decomp}
\textit{For any measurable $w_s\geq0$ with $\int_{\mathcal Z}w_s(z)\,dz=1$ and $\int_{\mathcal Z}d_s(z)^2w_s(z)\,dz<\infty$,}
\begin{equation}\label{eq:l2_decomp}
\mathcal{G}_{L^2}(s) = \mathcal{G}_{\text{\normalfont AUC}}(s)^2 + \mathcal{G}_{\text{\normalfont shape}}(s), \qquad
\mathcal{G}_{\text{\normalfont shape}}(s) = \int_{\mathcal{Z}} \{d_s(z) - \bar d_s\}^2 w_s(z)\, dz.
\end{equation}
\textit{For uniform weighting,} $w_s \equiv 1$, \textit{the identity becomes} $\mathcal{G}_{L^2}(s) = \mathcal{G}_{\text{\normalfont AUC}}(s)^2/|\mathcal{Z}| + \int_{\mathcal{Z}} \{d_s(z) - \bar d_s\}^2\, dz$ \textit{with} $\bar d_s = |\mathcal{Z}|^{-1}\int_{\mathcal{Z}} d_s(z)\, dz$.
\end{remark}

The decomposition yields an exact characterization of the heterogeneity visible to each criterion.
Here the shape component is a diagnostic and a possible additional objective function; the empirical analyses below compare AUC, $L^2$, and FANOVA.

\begin{remark}[Heterogeneity detected by each objective]\label{prop:detection}
{\em Fix a split with $0<p_s<1$ and a nonnegative weight satisfying $\int_{\mathcal Z}w_s=1$ and $\int_{\mathcal Z}d_s^2w_s<\infty$.
Write $\nu_s(dz)=w_s(z)\,dz$ and $m_s=\int d_s\,d\nu_s$.
Then AUC is zero exactly when $m_s=0$, whereas $L^2$ and FANOVA are zero exactly when $d_s=0$ $\nu_s$-almost everywhere.
The shape component is zero exactly when $d_s=m_s$ $\nu_s$-almost everywhere, and
\begin{equation}\label{eq:shape_contrasts}
\mathcal G_{\mathrm{shape}}(s)=\frac12\iint
\{d_s(z)-d_s(t)\}^2\,d\nu_s(z)\,d\nu_s(t).
\end{equation}
Consequently, AUC has zero gain and shape has strictly positive gain if and only if $m_s=0$ and $d_s$ is nonzero on a set of positive $\nu_s$-measure.
In that case $\mathcal G_{L^2}=\mathcal G_{\mathrm{shape}}>0$ and $\mathcal G_{\mathrm{FANOVA}}=p_s(1-p_s)\mathcal G_{\mathrm{shape}}>0$.}
\end{remark}

\Cref{eq:shape_contrasts} averages squared differences between the children's additive causal dose contrasts.
It distinguishes such heterogeneity from a constant vertical separation, which gives positive AUC, $L^2$, and FANOVA gains but zero shape gain.
For example, on $[0,1]$ with $w_s=1$, the increasing curves $\theta_L(z)=3/2+3z$ and $\theta_R(z)=5/2+z$ both average to $3$.
Their AUC gain is zero, but their shape and $L^2$ gains are $1/3$, and their FANOVA gain is $p_s(1-p_s)/3$.
Thus AUC can miss nonzero dose-contrast heterogeneity through exact cancellation.
Curve crossing alone does not imply cancellation: the weighted mean difference must be zero.
The characterization is relative to the chosen weight, and $L^2$ and FANOVA have the same population zero-gain splits but can rank nonzero splits differently.
\Cref{app:objective_detection} gives the proofs, a level--shape example family, and the role of weighting.

\vspace{-0.25cm}
\subsection{Weighting}\label{sec:weighting}

A challenge in comparing dose-response curves across subgroups is that treatment distributions may differ: one subgroup may have observations concentrated at low doses while another may have observations at high doses.
However, comparing dose-response curves in regions where one subgroup has little data leads to unreliable inference due to strong reliance on extrapolation.
This is a concrete feature of the application in \Cref{s:application}, where urban and rural counties have markedly different PM$_{2.5}$ support (see, e.g., \Cref{fig:map_pm25}).

To mitigate this, we propose a density-based weighting function that emphasizes regions of common support:
\begin{equation}\label{eq:weight}
w_s(z) = \frac{f_L(z) \cdot f_R(z)}{\int_{\mathcal{Z}} f_L(t) \cdot f_R(t) \, dt},
\end{equation}
where $f_L(z) = f(Z = z \mid \bV \in \mathcal{L}_L)$ and $f_R(z) = f(Z = z \mid \bV \in \mathcal{L}_R)$ are the treatment densities in the left and right child nodes.

The weight $w_s(z)$ is small wherever either $f_L(z)$ or $f_R(z)$ is small, emphasizing regions of common support.
This is analogous to trimming or overlap weighting in binary treatment settings \citep{crump2009dealing, li2018balancing}.
Weighting changes the scientific comparison and does not establish validity where positivity fails (\Cref{rem:density_tension}).
The application therefore reports both uniform and density-weighted objectives.
Alternative weights and the scope of the convergence bound are discussed in \Cref{rem:density_tension}.

\subsection{Algorithm}

The method proceeds in two phases, depicted in \Cref{alg:general}.

The discovery sample determines the partition using cross-fitted subgroup-specific pseudo-outcomes and the selected objective.

The independent inference sample is used to re-estimate nuisance functions and subgroup curves within the selected terminal nodes.
The reported analyses use fixed-depth trees without minimum-gain stopping.
The tree-recovery and modifier-selection guarantees concern the optional variant that also stops when the best estimated gain does not exceed $\lambda_n$.
Further estimator and inference details appear in \Cref{app:estimation,app:honest}.

\vspace{0.25cm}
{\begin{algorithm} \singlespacing
\caption{Continuous treatment HTE discovery with optional minimum-gain stopping}\label{alg:general}
\begin{algorithmic}[1]
    \Require Data $\{(Y_i, Z_i, \bX_i, \bV_i)\}_{i=1}^n$, objective $\mathcal{G}$, max depth $d_{\max}$, min leaf size $n_{\min}$; optional minimum gain $\lambda_n$
    \State Randomly split into $\mathcal{D}_{\text{dis}}$ and $\mathcal{D}_{\text{inf}}$ independently of the observations
    \State \textbf{Discovery phase} (on $\mathcal{D}_{\text{dis}}$):
    \State \quad Fit outcome, propensity, and node-density nuisances outside evaluation folds
    \State \quad \textbf{Tree construction:} starting from the root, recursively:
    \State \quad\quad For each eligible split $s$: form $\hat\varphi^{\mathcal{L}_L}, \hat\varphi^{\mathcal{L}_R}$ per \eqref{eq:pseudo_hat}, estimate $\hat{w}_s(z)$, compute $\hat{\mathcal{G}}(s)$
    \State \quad\quad Stop if no eligible split remains; otherwise maximize $\hat{\mathcal G}(s)$ over eligible splits
    \State \quad\quad Stop at depth $d_{\max}$ or, if minimum-gain stopping is enabled, when $\hat{\mathcal G}(\hat s)\leq\lambda_n$
    \State \quad\quad Otherwise split at the maximizer $\hat s$
    \State \textbf{Inference phase} (on $\mathcal{D}_{\text{inf}}$):
    \State \quad Cross-fit all score nuisances within $\mathcal D_{\text{inf}}$; recompute pseudo-outcomes per terminal node
    \State \quad Estimate dose-response curves $\hat{\theta}_h(z)$ within each terminal node $h$ of $\hat{\psi}$
    \Ensure Tree $\hat\psi$, subgroup curves, and pointwise intervals under the conditions of \Cref{prop:honest}
\end{algorithmic}
\end{algorithm}}

\subsection{Theoretical Properties}
\label{sec:theory}

We derive uniform objective convergence and greedy-tree consistency conditional on the estimation, admissibility, separation, and stopping conditions below.

\subsubsection{Regularity Conditions}

In addition to \Crefrange{ass:consistency}{ass:overlap}, the theory requires smoothness, density, and moment regularity, nuisance-error bounds, and uniform control of the local-linear design and score processes.
\Cref{app:conditions} states these requirements explicitly, including the high-level smoothing bounds in \Cref{con:process} and stable admissibility in \Cref{con:admissibility}.
The rate and tree results are conditional on these bounds; polynomial node complexity alone does not establish them.

For consistent stopping at population zero-gain nodes, a minimum-gain rule is needed in addition to maximum-depth and leaf-size constraints, since the estimated objectives are nonnegative even at a node where all population child curves agree.
For this theoretical variant, we use a threshold $\lambda_n$ that vanishes more slowly than the error in estimating the splitting gains.

\begin{condition}[Stopping threshold]\label{con:stopping}
Let $\varepsilon_n$ denote the uniform estimation-error rate in \Cref{prop:gain_uniform}.
The splitting threshold satisfies $\lambda_n\to0$ and $\varepsilon_n/\lambda_n\to0$.
For nodes of a fixed population target tree at which a split is admissible and desirable, the maximal population gain is bounded below by a constant $\gamma>0$.
At nodes declared terminal by the gain criterion, the maximal admissible population gain is zero; nodes terminated by maximum depth or leaf-size constraints are not required to have zero population gain.
\end{condition}

The density-weight estimation bound is stated in \Cref{lem:weight}; uniform weighting requires no density estimation.

\begin{prop}[Uniform objective convergence]\label{prop:gain_uniform}
Let $\mathcal S_n$ denote all eligible candidate splits at nodes reachable within depth $d_{\max}$.
Under \Crefrange{ass:consistency}{ass:overlap} and \Cref{app:conditions}, for any of the three objectives in \Cref{sec:of},
\begin{equation}\label{eq:gain_rate}
\sup_{s\in\mathcal S_n}|\hat{\mathcal G}(s)-\mathcal G(s)|=O_p(\varepsilon_n),\qquad
\varepsilon_n=\varepsilon_{\theta,n}+d_{w,n}.
\end{equation}
The terms $\varepsilon_{\theta,n}$ and $d_{w,n}$ are defined in \eqref{eq:sn} and \Cref{lem:weight}, respectively, and vanish under the stated bandwidth and nuisance-product conditions.
\end{prop}

\begin{condition}[Margin condition]\label{con:margin}
For exact tree recovery on a fixed finite candidate grid, each nonterminal node of the population greedy tree has a unique optimal split $s^*$ satisfying $\mathcal G(s^*)-\max_{s\neq s^*}\mathcal G(s)\geq\delta$ for some $\delta>0$.
\end{condition}

A fixed positive margin between adjacent thresholds is generally untenable when candidates become dense along a continuous modifier.
Accordingly, we distinguish consistency of splitting gains and stopping from exact recovery on a fixed grid.

\begin{theorem}[Tree consistency]\label{thm:tree_consistency}
Under \Crefrange{ass:consistency}{ass:overlap}, the conditions of \Cref{app:conditions}, and \Cref{con:stopping}, with empirical and population candidate sets agreeing on the event in \Cref{con:admissibility}, at an evaluated node below the maximum depth with a nonempty admissible split set, a zero maximal population gain leads to splitting with probability tending to zero, whereas a maximal population gain of at least $\gamma$ is split with probability approaching one.
For every fixed node and candidate set, if $s^*$ and $\hat s$ maximize the population and estimated gains over that same set, respectively, then
\begin{equation}\label{eq:split_regret}
0\leq\mathcal G(s^*)-\mathcal G(\hat s)\leq 2\sup_{s\in\mathcal S_n}|\hat{\mathcal G}(s)-\mathcal G(s)|=O_p(\varepsilon_n).
\end{equation}
If candidate thresholds are restricted to a fixed finite grid,\footnote{Finite grids of candidate cutpoints are used in common tree-based methods. For instance, Bayesian additive regression trees also restrict splits to a discrete set of candidate cutpoints \citep[Section~2.2.2]{chipman2010bart}.}
\Cref{con:margin} holds, and the depth is fixed, then the estimated greedy tree agrees with its population counterpart with probability approaching one:
\begin{equation}
\Prob(\hat\psi=\psi^*)\to1.
\end{equation}
For increasingly dense continuous threshold sets the result instead yields vanishing split-gain regret; convergence of thresholds follows from additional uniqueness, compactness, and continuity conditions, not exact equality of the trees.
\end{theorem}

%The separation and fixed-grid assumptions in the exact-recovery result are deliberately stronger than those needed to recover high-scoring splits.
In applications, the selected variables may be more stable than the ordering of splits or their exact cut points, motivating a separate variable-selection result.

Let $\mathcal M^*$ denote the designated set of relevant coordinates for a chosen representation of the conditional dose-response surface, and let $\mathcal M(\psi)$ denote the variables used by a tree $\psi$.
Here relevance includes response-level modifiers, as distinguished from heterogeneity in additive dose contrasts in \Cref{prop:l2_decomp}.
Under \Cref{con:modifier_sep}, at each eligible node reached using relevant variables and having positive maximal gain, the best relevant split exceeds every irrelevant split in population gain by a fixed positive margin.

\begin{theorem}[Modifier set consistency]\label{thm:modifier_consistency}
Under \Crefrange{ass:consistency}{ass:overlap}, \Cref{app:conditions}, \Cref{con:stopping,con:modifier_sep}, and fixed $d_{\max}$,
\begin{equation}
\Prob\{\mathcal M(\hat\psi)\subseteq\mathcal M^*\}\to1.
\end{equation}
If additionally a fixed finite split grid satisfies \Cref{con:margin} and the population greedy tree uses every variable in $\mathcal M^*$, then $\Prob\{\mathcal M(\hat\psi)=\mathcal M^*\}\to1$.
\end{theorem}

The first conclusion allows ties among relevant variables and does not require exact threshold recovery.
The modifier separation condition and proof are given in \Cref{app:modifier_selection}.

The sample-splitting structure additionally enables inference for dose-response curves within terminal nodes conditional on the discovered tree.
\Cref{prop:honest} establishes conditional pointwise inference for a node selected from a fixed finite family under additional variance, moment, bandwidth, and nuisance-remainder conditions.
Independence of discovery and inference alone does not establish these conditions or justify intervals for clustered observations.

Additional proofs are provided in \Cref{app:proofs}, and the pointwise inference result and its proof appear in \Cref{app:honest}.
An extensive Monte Carlo evaluation of finite-sample discovery performance, across five data-generating settings and two sample sizes, is reported in \Cref{app:simulations}.

\section{Application: Air Pollution and Mortality}
\label{s:application}

We now apply the method of \Cref{s:methods} to the data of \Cref{s:data} and answer (Q1)--(Q3) in turn.
We report pointwise 95\% intervals. The i.i.d.\ result in \Cref{prop:honest} does not establish their coverage for repeated county observations.
We use the $L^2$ objective with uniform weighting and binarized covariates unless otherwise indicated (\Cref{app:app_results}).

We randomly split the counties into a discovery sample (1{,}407 counties, 26{,}718 county-years) and an inference sample (1{,}408 counties, 26{,}751 county-years).
We grow the tree in the discovery sample and estimate its node-specific curves in the inference sample, cross-fitting the nuisance models over two folds within each sample (\Cref{app:estimation}).
We dichotomize continuous covariates at their medians across the 53{,}469 county-years and use the same bandwidth of 1.5 as in \Cref{fig:erf_pm25}.
The 1st and 99th percentiles of exposure in the inference sample are 4.30 and 15.1 $\mu$g/m$^3$.

\subsection{Which Communities Have Different Dose-Response Relationships (Q1)}\label{sec:q1}

\begin{figure}[htbp]
\centering
\includegraphics[width=1\textwidth]{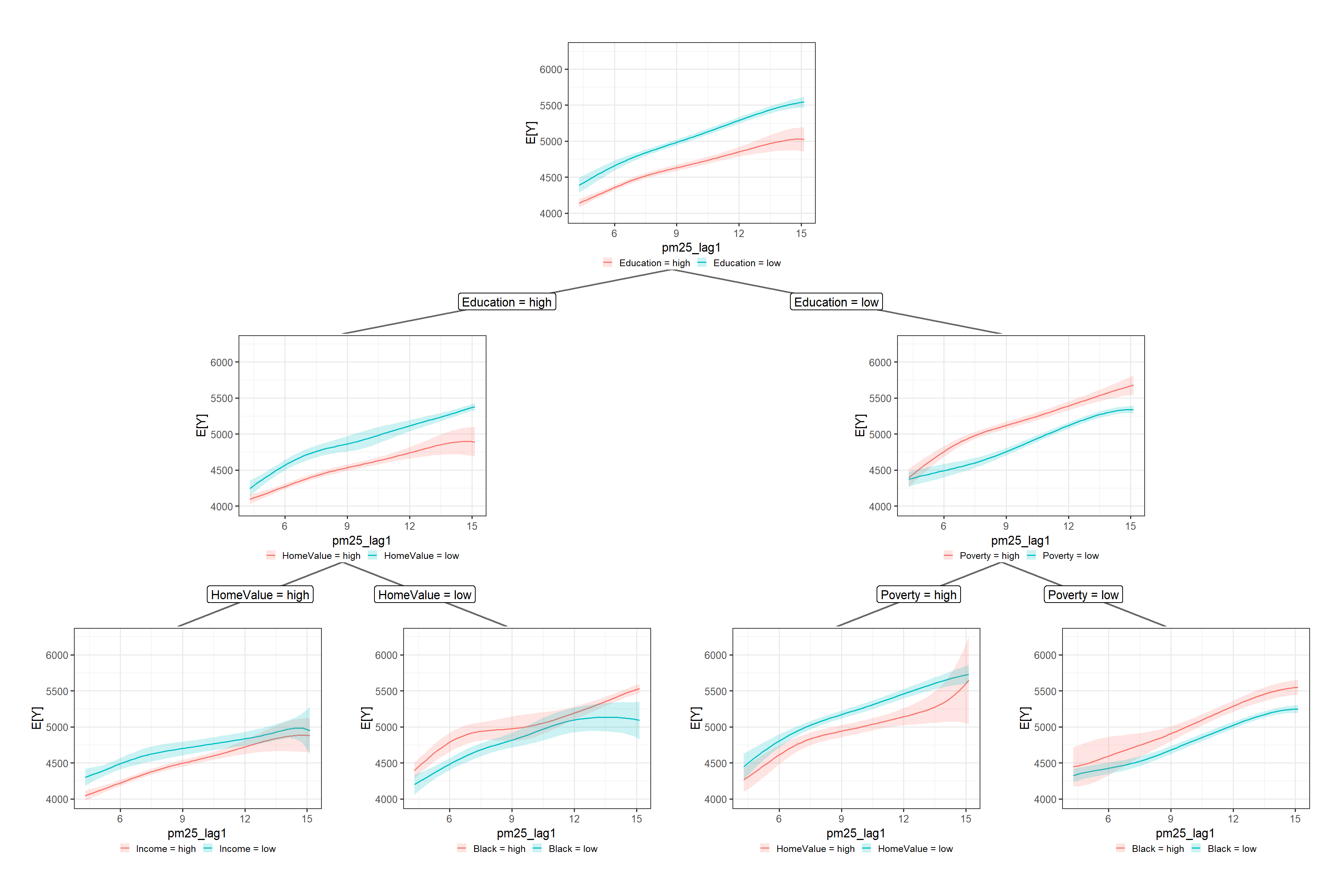}
\caption{Discovered honest tree under the $L^2$ and AUC objectives with uniform weighting.
Each panel shows the estimated dose-response curves of the two children of the corresponding split, with pointwise confidence bands based on cluster-robust standard errors.}
\label{fig:tree_l2_unweighted_pm25}
\end{figure}

\Cref{fig:tree_l2_unweighted_pm25} displays the tree discovered under the $L^2$ objective with uniform weighting and binarized covariates.
The complete node-by-node estimates are provided in \Cref{table:tab_tree_l2_unweighted} of the Supplementary Material. Both cluster-robust and Conley heteroskedasticity- and autocorrelation-consistent (HAC) standard errors are reported \citep{conley1999gmm, conley2010spatial}; the latter use a Bartlett kernel with a cutoff of 200 kilometers.

The first split is on educational attainment. The second split is on home value among counties with higher educational attainment and on poverty among those with lower attainment. The third level splits on income, Black population share, and home value once more. 

\textit{The estimated educational attainment gap varies with dose.}
At the root, counties with higher and lower educational attainment have reported mortality estimates of 4{,}229 and 4{,}506 at $5\ \mu$g/m$^3$, 4{,}632 and 4{,}984 at $9\ \mu$g/m$^3$, and 4{,}852 and 5{,}289 at $12\ \mu$g/m$^3$, respectively. The corresponding gaps are 277, 352, and 437 deaths per 100{,}000 person-years.
From 5 to 9 $\mu$g/m$^3$, estimated mortality increases by 478 deaths per 100{,}000 person-years in the lower-education group and by 403 in the higher-education group, a difference of 75.

\textit{Some fitted curves flatten while others continue to rise.}
Among counties with higher educational attainment and lower home values, estimated mortality from 12 to 15 $\mu$g/m$^3$ rises from 5{,}095 to 5{,}098 in the subgroup with a lower Black population share and from 5{,}190 to 5{,}519 in the subgroup with a higher Black population share.
The pointwise intervals at 12 $\mu$g/m$^3$ overlap under both variance estimators.

\textit{Mortality levels and dose contrasts vary with socioeconomic conditions.}
Counties with higher educational attainment, home values, and income have the lowest estimated mortality at each of the three doses: 4{,}116, 4{,}495, and 4{,}720 deaths per 100{,}000 person-years at 5, 9, and 12 $\mu$g/m$^3$, respectively.
Counties with lower educational attainment, higher poverty, and lower home values have the largest estimated increase between 5 and 9 $\mu$g/m$^3$, from 4{,}608 to 5{,}170, followed by a further increase to 5{,}458 at 12 $\mu$g/m$^3$.

\textit{Fitted mortality varies with Black population share in two branches.}
Black population share splits two branches of the tree: counties with higher educational attainment and lower home values, and counties with lower educational attainment and lower poverty.
In both branches, the subgroup with a higher Black population share has higher fitted mortality.
In the second branch, this subgroup has the largest estimated increase between 9 and 12 $\mu$g/m$^3$, from 4{,}904 to 5{,}286 deaths per 100{,}000 person-years.

\subsection{Mortality Levels and Dose Contrasts Below the Current Standard (Q2)}\label{sec:q2}

On the population curve of \Cref{fig:erf_pm25}, mortality rises by 392 deaths per 100{,}000 person-years from 4{,}421 to 4{,}813 between 5 and $9\ \mu$g/m$^3$, or roughly 98 deaths per $\mu$g/m$^3$, and by a further 269 from 4{,}813 to 5{,}082 between 9 and $12\ \mu$g/m$^3$, or roughly 90 deaths per $\mu$g/m$^3$. 
No estimated curve flattens below 12 $\mu$g/m$^3$. 
Below this level, subgroups differ mainly in how steeply mortality rises, with some curves rising more slowly between 9 and 12 $\mu$g/m$^3$ than between 5 and 9 $\mu$g/m$^3$. 
Flattening appears only at the upper end of the exposure range, where, however, support is more sparse.
As such, these fitted contrasts provide no visual indication of flattening near the regulatory reference point, although a formal comparison of slopes or contrasts would be needed to assess their difference. 

The subgroups differ in how much of that increase they accrue and where.
Even among counties with higher educational attainment, home values, and income, which have the lowest mortality rates across all dosage levels, the point estimates correspond to average slopes of about 94 and 126 deaths per 100{,}000 person-years per $\mu$g/m$^3$ below and above $9\ \mu$g/m$^3$, respectively, and therefore do not support a claim that the curve flattens immediately above the standard.
Among counties with lower educational attainment, higher poverty, and lower home values, which have the steepest dosage response curve, the corresponding average slopes are about 187 and 72 deaths per 100{,}000 person-years per $\mu$g/m$^3$ below and above 9 $\mu$g/m$^3$.

These comparisons have two main implications.
First, when subgroup exposure-response curves differ in shape, a common reduction in PM$_{2.5}$ can correspond to different estimated changes in county mortality rates depending on the starting exposure level and subgroup.
Second, differences between subgroup mortality levels should not be interpreted as differences in the causal effect of exposure reductions unless the corresponding dose contrasts also differ.
The dose-response approach makes both dimensions visible.

\subsection{Sensitivity to Objective and Weighting (Q3)}
\label{sec:q3}

With uniform weighting, AUC and $L^2$ recover the same tree (\Cref{fig:tree_l2_unweighted_pm25}).
FANOVA changes only the third-level split among counties with higher educational attainment and home values, selecting poverty in place of income (\Cref{fig:tree_fanova_unweighted_pm25}).

With density weighting, all three objective functions discover the same tree. 
The density-weighted tree and its subgroup estimates appear in \Cref{fig:tree_l2_density_pm25,table:tab_tree_l2_density} of the Supplementary Material.
Educational attainment remains the root split.
In the higher-education branch, the density-weighted tree splits on income and then on home value within each income group.

Density weighting emphasizes doses where both child densities are high, which can change split ordering and attenuate tail differences. 
The choice of weighting therefore affects the comparison, particularly at the extremes of the exposure range.

All three trees select educational attainment, home value, poverty, and Black population share.
The lower-education branch is identical across trees: it splits on poverty, then on home value among higher-poverty counties and on Black population share among lower-poverty counties.
\Cref{app:evalue} reports E-values for the 5 to 9 $\mu$g/m$^3$ contrast in the population and terminal nodes of the uniformly weighted $L^2$ tree \citep{vanderweele2017sensitivity}.

Taken together, these results emphasize that the main structure of the tree was robust across splitting objectives and weighting schemes.
Furthermore, in every terminal subgroup of the primary tree, the estimated mortality increase from 5 to 9 $\mu$g/m$^3$ was also robust to weak unmeasured confounding, with E-values ranging from 1.35 to 1.49.

\section{Discussion}
\label{s:conclusions}

Two subgroups can have different mean outcomes at every dose and still experience the same effect of a change in dose.
We developed a recursive partitioning method that discovers subgroups with different causal dose--response curves and characterizes how those curves differ.
Even with known nuisance functions, regressing the population-marginal pseudo-outcome on dose within each subgroup yields the same population-average curve in every subgroup.
We therefore construct the pseudo-outcome separately for each candidate subgroup, averaging the outcome regression over its covariate distribution and using its exposure density to stabilize the residual correction.
This gives a doubly robust estimator of each subgroup curve and allows us to use differences between curves to decide where the tree should split.

The choice of splitting objective determines which differences the tree can discover.
When two curves cross, differences of opposite sign can cancel in the AUC objective even when the curves differ substantially over part of the dose range.
The $L^2$ objective detects these differences, and its decomposition separates differences in average response levels from differences in curve shape.
A difference in shape means that, for at least one pair of doses, the additive effect of moving from one dose to the other differs between subgroups.

Our theory distinguishes identifying the variables that define heterogeneity from recovering the full decision rules.
Under the stated regularity and separation conditions, a tree with minimum-gain stopping asymptotically excludes irrelevant variables even when several relevant splits are equally good.
Exact recovery of the population greedy tree holds under stronger conditions, including a fixed grid and a unique, separated best split at each internal node (\Cref{thm:tree_consistency,thm:modifier_consistency}).
In the main simulation setting, the relevant variables were easier to recover than the exact rules when the signal was weak.

In our analysis of 2{,}815 U.S. counties, all fixed-depth trees selected educational attainment first and recovered the same lower-education branch, although some later splits in the higher-education branch varied with the objective and weighting scheme.
From 5 to 9 $\mu$g/m$^3$ of PM$_{2.5}$, estimated age-standardized mortality among adults aged 65 and older rose by 478 deaths per 100{,}000 person-years in counties with lower educational attainment and by 403 in counties with higher attainment.
A causal interpretation of these county-level contrasts depends on adequate exposure overlap and control of confounding within subgroups.
The estimated mortality gap thus widened by 75 deaths per 100{,}000 person-years over this range, suggesting differential vulnerability to fine particulate matter in communities that already faced higher mortality at the lower dose.

\section*{Data and Code Availability}

Mortality data are publicly available from the CDC WONDER database (\url{https://wonder.cdc.gov/}).
PM$_{2.5}$ concentration estimates are available from the Atmospheric Composition Analysis Group at Washington University in St.\ Louis (\url{https://sites.wustl.edu/acag/datasets/}).
County-level socioeconomic and demographic variables are available from the American Community Survey (\url{https://www.census.gov/programs-surveys/acs}).
\ifdefined\anonymousVersion
The method is implemented in a development \textsf{R} package; identifying software details are omitted for anonymous review.
\else
The method is implemented in the \textsf{R} package \texttt{discoverHTE} \citep{discoverhte2026}.
\fi

\section*{Supplementary Materials}

The Supplementary Material accompanying this manuscript contains estimator details, proofs, analytical examples, simulations, and additional application results.

\ifdefined\anonymousVersion\else
\section*{Funding}
FBS acknowledges funding from Amazon Web Services (AWS) grant on ``AI/ML for Identifying Social Determinants of Health'' and from the Society of Hellman Fellows.

\fi

\section*{Conflict of Interest}
The authors declare no conflict of interest.

\section*{Declaration of Generative AI Use}
LLMs were used to assist with manuscript copy-editing, consistency checks, code review, mathematical arguments checking and supporting formal-verification materials.
The authors are responsible for the manuscript's claims, analyses, and conclusions.

\bibliographystyle{abbrvnat}
\bibliography{biblio}

\clearpage

\ifthenelse{\boolean{showcomments}}{
\typeout{JASA DRAFT: Main manuscript page limit is not enforced while editorial comments are visible.}
}{
\typeout{JASA MAIN MANUSCRIPT PAGES (INCLUDING REFERENCES): \the\numexpr\value{page}-1\relax}
\ifnum\value{page}>36
\PackageError{JASA}{Main manuscript exceeds 35 pages including references}{Move supporting tables and figures to the separately paginated Supplementary Material or shorten the main text.}
\fi
}
\singlespacing
\appendix
% supplement is paginated independently of the main paper
\pagenumbering{arabic}
\renewcommand{\thepage}{S\arabic{page}}
\setcounter{page}{1}
\numberwithin{equation}{section}
\counterwithin{figure}{section}
\counterwithin{table}{section}

\begin{center}
\textbf{Supplementary Materials to\\ Interpretable Discovery of Heterogeneous Causal Dose--Response Curves}
\medskip

\ifdefined\anonymousVersion\else
Falco J. Bargagli-Stoffi, Duy Pham, Edward Kennedy, Heejun Shin,\\
Noah Greifer, Kevin P. Josey, and Rachel C. Nethery
\fi
\end{center}

\section{Dose-Response Curve Estimation Details}
\label{app:estimation}

\subsection{Estimator}

Within a node $\mathcal{L}$ we estimate $\theta(z \mid \mathcal{L})$ by kernel-weighted local linear regression of the subgroup-conditional pseudo-outcomes \eqref{eq:pseudo_subgroup} on treatment,
\begin{equation}\label{eq:locpoly}
\hat\theta(z \mid \mathcal{L}) = \hat\beta_0, \quad (\hat\beta_0, \hat\beta_1) = \argmin_{\beta_0, \beta_1} \sum_{i : \bV_i \in \mathcal{L}} K_h(Z_i - z)\left(\hat\varphi^{\mathcal{L}}_i - \beta_0 - \beta_1 \frac{Z_i - z}{h}\right)^2,
\end{equation}
with $K_h(\cdot) = h^{-1}K(\cdot/h)$ and $K$ Gaussian.
We emphasize that \eqref{eq:locpoly} is applied to $\hat\varphi^{\mathcal{L}}$ and not to the marginal pseudo-outcome \eqref{eq:pseudo_marginal}; by \Cref{prop:insufficient} the latter would return the population-average curve in every node.
Explicitly, for a node $\mathcal{L}$,
\begin{equation}\label{eq:pseudo_hat}
\hat\varphi^{\mathcal L}_i=
\frac{\hat\varpi_{k(i)}(Z_i\mid\mathcal L)}{\hat\pi_{k(i)}(Z_i\mid\bX_i)}
\{Y_i-\hat\mu_{k(i)}(Z_i,\bX_i)\}
+\frac1{n_{\mathcal L}}\sum_{j:\bV_j\in\mathcal L}\hat\mu_{k(i)}(Z_i,\bX_j),
\end{equation}
where $k(i)$ denotes the evaluation fold. In the theoretical construction, each subscript-$k$ nuisance is fitted outside that fold.
Both the stabilizer and g-computation component are node-specific.
The displayed g-computation component reuses node covariates, including evaluation covariates; its effect is controlled by the uniform empirical g-computation bound rather than an exact conditional-independence claim.

\subsection{Nuisance Estimation}

\textit{Generalized propensity score.}
We use a Gaussian treatment density with fitted conditional mean $\hat m(\bx)$ and residual standard deviation $\hat\sigma(\bx)$:
\begin{equation}
\hat\pi(z\mid\bx)=\frac{1}{\hat\sigma(\bx)}\phi\!\left(\frac{z-\hat m(\bx)}{\hat\sigma(\bx)}\right),
\end{equation}
where $\phi$ is the standard normal density.
The count-outcome analysis uses a constant residual standard deviation. With SuperLearner \citepApp{van2007super}, a second fit to the log-squared residuals estimates the conditional scale.
For evaluation fold $I_k$, a training-only stabilizer is
\[
\hat\varpi_k(z\mid\mathcal L)=\frac1{n_{\mathcal L,-k}}
\sum_{j\notin I_k:\bV_j\in\mathcal L}\hat\pi_k(z\mid\bX_j),
\]
where $n_{\mathcal L,-k}$ is the corresponding training-node count.
Direct density smoothing using only these training exposures is another option with a different error decomposition.
A split requires a positive training-node count in each fold.
In the application we average the fitted propensity densities over the evaluation rows in each subgroup.
\Cref{con:nuisance} and the convergence result use the training-fold average displayed above.

\textit{Outcome regression.}
For continuous outcomes in the simulations, $\mu(z, \bx)$ is fitted by SuperLearner.
For the application, the outcome is a death count in a county-year-age cell. The working outcome model is a negative-binomial generalized additive model with a log link, comprising a smooth term in the exposure; for each continuous covariate in $\bX_c$, a smooth main effect and a tensor-product interaction with the exposure; for the discrete covariates $\bX_d$ and the age stratum, linear main effects and linear interactions with the exposure; and a log-population offset:
\begin{equation}\label{eq:negbinom}
\begin{split}
\log \E[\text{deaths} \mid Z, \bX, \text{age}]
  &= s_0(Z) + \sum_{c} \bigl\{ s_c(X_c) + t_c(Z, X_c) \bigr\}
     + \gamma^\top \bX_d + Z\,\eta^\top \bX_d \\
  &\quad + \alpha_{\text{age}} + Z\,\delta_{\text{age}} + \log(\text{population}),
\end{split}
\end{equation}
where $s_0$ and $s_c$ are penalized cubic regression splines (basis dimensions 20 and 10), and $t_c$ is a tensor-product interaction smooth that excludes the corresponding main effects (basis dimension $5\times5$). The model is fitted with the \texttt{bam} function of the \texttt{mgcv} package \citepApp{wood2017generalized,wood2015bam}, using the negative-binomial family with estimated dispersion \citepApp{wood2016smoothing} and tensor-product interaction
smooths \citepApp{wood2006tensor}.

Age stratum enters as a factor, and predictions are combined across strata using the 2000 U.S. standard population to produce age-standardized rates: 18{,}215, 12{,}360, and 4{,}259 out of 34{,}834 for 65--74, 75--84, and 85 years and older, respectively. Fitted and
g-computation predictions are converted to mortality rates per 100{,}000 and age-standardized within each county-year before entering and forming the residual in the pseudo-outcome on \Cref{eq:pseudo_hat}.

\subsection{Interval-Censored Counts}

Under the CDC WONDER suppression rule for counts of one to nine deaths, a suppressed cell indicates that the count lies in $\{1, \ldots, 9\}$ rather than that it is missing.
Writing $Y_i$ for the count in cell $i$, $c_i = 1$ for a suppressed cell, and $f(\cdot;\lambda_i,\phi)$ and $F(\cdot;\lambda_i,\phi)$ for the negative-binomial probability mass and distribution functions with mean $\lambda_i$ given by the working model in \Cref{eq:negbinom} and dispersion $\phi$, the log-likelihood contribution of cell $i$ is
\begin{equation}
\ell_i = (1 - c_i)\, \log f(Y_i; \lambda_i, \phi) \;+\; c_i\, \log\big\{F(9; \lambda_i, \phi) - F(0; \lambda_i, \phi)\big\}.
\end{equation}
Discarding suppressed observations can induce exposure-dependent selection, particularly in smaller counties.
The penalized likelihood $\sum_i \ell_i$ is maximized by an expectation--conditional-maximization procedure. Suppressed counts are initialized
at the midpoints of their intervals; at each iteration, every suppressed count is replaced by its conditional expectation $\E[Y_i \mid l_i \le Y_i \le u_i; \hat\lambda_i, \hat\phi]$ under the current fit,
the model is refitted to the completed data with $\phi$ held fixed, and $\phi$ is then updated by maximizing $\sum_i \ell_i$ given the fitted means. In the pseudo-outcome, the outcome of a suppressed cell is its conditional
expectation under the model fitted on the training folds,
$\tilde Y_i = \E[Y_i \mid l_i \le Y_i \le u_i; \hat\lambda_i, \hat\phi]$, and
$\tilde Y_i = Y_i$ otherwise; observed and imputed counts and fitted means are converted to rates per 100{,}000 and age-standardized within each county-year before forming the residual in \Cref{eq:pseudo_hat}. For a suppressed count the imputation error satisfies $|\tilde Y_i-Y_i|\leq u_i-l_i$.
Its contribution to the pseudo-outcome is weighted by the stabilizing ratio, so the usual double-robustness argument requires additional assumptions for these cells.

\subsection{Bandwidth and Variance}

For a linear smoother fitted to fixed pseudo-outcomes, leave-one-out cross-validation can be computed from the full-fit residuals as
$\mathrm{CV}(h)=n^{-1}\sum_i\{(\hat\varphi_i-\hat\theta_h(Z_i))/(1-H_{h,ii})\}^2$,
where $\hat\theta_h$ denotes the full-sample smoother and $H_{h,ii}$ its diagonal leverage.
In the application, we use a fixed bandwidth of 1.5 and give each county-year equal weight.
\Cref{prop:honest} uses a deterministic bandwidth; extending its conclusion to a selected bandwidth requires a suitable uniform or stochastic-equivalence argument.

Under i.i.d.\ sampling, the local-linear asymptotic variance has the sandwich form stated in \Cref{prop:honest}, involving both the local-linear design and the conditional variance of the subgroup pseudo-outcome.
For the application, we include the empirical subgroup g-computation term in the local-linear influence function and aggregate contributions by county.
Fix a node $\mathcal L$ with inference units $i = 1,\ldots,n_{\mathcal L}$, observation
weights $w_i$, an anchor $z$, and bandwidth $h$. Let $u_i = (Z_i - z)/h$, $\mathbf g(u) = (1,u)^\top$ and $\hat\eta_i$ residuals at $z$ of the weighted local-linear fit described in \Cref{eq:locpoly}. The local-linear design matrix is
$$
\hat S(z) = \frac{1}{n_{\mathcal L}}\sum_{i=1}^{n_{\mathcal L}}
  w_i K_h(Z_i - z)\,\mathbf g(u_i)\mathbf g(u_i)^\top.
$$
Write $\bar\mu_{\mathcal L}(t)$ and $\bar\pi_{\mathcal L}(t)$ for the weighted node
averages of $\hat\mu(t,\bX_j)$ and $\hat\pi(t\mid\bX_j)$ over $j\in\mathcal L$. The subgroup g-computation component contributes
\[
\widehat{\mathrm{int}}_i(z) = \int \bigl\{\hat\mu(t,\bX_i) - \bar\mu_{\mathcal L}(t)\bigr\}
  \bar\pi_{\mathcal L}(t)\,\mathbf g\!\left(\frac{t - z}{h}\right) K_h(t - z)\,dt,
\]
which is evaluated by the trapezoidal rule on the exposure grid used to fit the nuisance functions. The estimated influence function of $\hat\theta(z\mid\mathcal L)$ is then
\begin{equation}\label{eq:psi_hat}
\hat\psi_i(z) = \mathbf e_1^\top \hat S(z)^{-1}\, w_i
  \bigl\{K_h(Z_i - z)\,\mathbf g(u_i)\,\hat\eta_i + \widehat{\mathrm{int}}_i(z)\bigr\},
  \qquad i = 1,\ldots,n_{\mathcal L}.
\end{equation}
For the application in \Cref{s:application}, let $c = 1,\ldots, C$ index the counties represented in $\mathcal L$. The county-clustered robust standard errors sum the influences over the county-years of county $c$ and produce the covariance of the estimates $\hat\theta(z\mid\mathcal L)$ and $\hat\theta(z'\mid\mathcal L)$ at anchors $z$ and $z'$
\begin{equation}\label{eq:V_cr}
\hat V_{\mathrm{CR}}(z,z')
= \frac{C}{C-1}\cdot\frac{1}{n_{\mathcal L}^2}\sum_{c=1}^{C}\Bigl(\sum_{i\in c}\hat\psi_i(z)\Bigr)\Bigl(\sum_{i\in c}\hat\psi_i(z')\Bigr),
\end{equation}
with $C/(C - 1)$ being the standard small-sample correction for a finite number of clusters.

The implementation of Conley spatial HAC standard errors \citep{conley1999gmm,conley2010spatial} additionally includes products of the county
totals of distinct counties within a cutoff distance $\kappa$:
\begin{equation}\label{eq:V_conley}
\hat V_{\mathrm{C}}(z,z')
= \frac{C}{C-1}\cdot\frac{1}{n_{\mathcal L}^2}
\sum_{c=1}^{C}\sum_{c'=1}^{C} W_{cc'}\Bigl(\sum_{i\in c}\hat\psi_i(z)\Bigr)\Bigl(\sum_{i\in c'}\hat\psi_i(z')\Bigr),
\end{equation}
where $W_{cc'} = (1 - d_{cc'}/\kappa)\,\mathbf 1\{d_{cc'} < \kappa\}$ is the Bartlett kernel and $d_{cc'}$ the
great-circle distance in kilometers between the 2010 population centroids of counties $c$ and $c'$. The cutoff used for the standard errors in tables and figures in \Cref{app:additional_disc,app:additional_cont,app:additional_conley} is 200 kilometers.

%The pointwise inference result in \Cref{prop:honest} assumes independent observations.

\subsection{Relation to Conditional Influence Functions}

\begin{remark}[Why the general theory of conditional influence functions does not apply]\label{rem:cif}
The failure established above is an exception to a general pattern, which is worth making explicit since it is the reason the exception is easy to miss.
\citeApp{chernozhukov2024conditional} study conditional objects of the form $\theta_0(\bV) = \E[m(W, \gamma_0) \mid \bV]$, where $\gamma_0$ is a regression function and $m$ is a linear functional of it, and show that whenever a conditional mean-square continuity condition holds (that is, whenever there exists a square-integrable conditional Riesz representer $\alpha_0(Z,\bX)$ with $\E[m(W,\gamma) \mid \bV] = \E[\alpha_0(Z,\bX)\gamma(Z,\bX) \mid \bV]$ for all square-integrable $\gamma$) the conditional influence function is $m(W, \gamma_0) - \theta_0(\bV) + \alpha_0(Z, \bX)\{Y - \gamma_0(Z, \bX)\}$.
This is the influence function of the corresponding \emph{unconditional} parameter with $\theta_0$ replaced by $\theta_0(\bV)$, so within that class the same pseudo-outcome serves the marginal and the conditional object, and no new construction is required.
The binary-treatment case illustrates the pattern: there $m(W, \gamma) = \gamma(1, \bX) - \gamma(0, \bX)$ is a function of the covariates, and conditioning the augmented inverse probability weighted pseudo-outcome on $\bV$ returns the conditional average treatment effect, which is what licenses the doubly-robust learner of \citeApp{kennedy2023towards}.

The conditional dose-response curve lies outside this class.
Taking $m(W, \gamma) = \gamma(z, \bX)$ at a fixed treatment level $z$, the defining property of the conditional Riesz representer would require $\alpha_0(\zeta, \bx) \pi(\zeta \mid \bx) = \delta(\zeta - z)$ for almost every $\bx$, so $\alpha_0$ is a Dirac delta rather than a square-integrable function and the continuity condition fails.
This is the same irregularity that prevents $\theta(z)$ itself from being pathwise differentiable and that motivates the smoothed construction of \citeApp{Kennedy2017}.
It is instructive that the conditional average derivative $\E[\partial\gamma_0(Z,\bX)/\partial z\mid\bV]$ can admit the representer $-\partial_z\log\pi(Z\mid\bX)$ under appropriate differentiability and vanishing boundary terms, even though the pointwise curve level remains irregular.
The pattern above therefore offers no guidance here, and \Cref{prop:insufficient} shows that following it would be actively misleading.
\end{remark}

\section{The Cost of a Marginal Pseudo-Outcome}
\label{app:po_sim}

\Cref{prop:insufficient,prop:subgroup} are statements about population limits.
This appendix illustrates both in a design with closed-form nuisance functions and conditional means for each pseudo-outcome.
Nuisance functions are supplied exactly or deliberately misspecified as described below, rather than estimated.
The closed-form conditional means isolate the pseudo-outcome construction; the fitted Monte Carlo curves additionally reflect smoothing and sampling variation.

\subsection{Design}

Let $V \sim \text{Bernoulli}(0.5)$ be a single binary effect modifier, let $X_1, X_2 \sim \mathcal{N}(0,1)$ independently of $V$, and let
\begin{equation}\label{eq:po_dgp}
Z \mid X, V \sim \mathcal{N}\big(1 + 0.5X_1 - 0.5X_2 + 0.5V,\; 1\big), \qquad
Y = 1 + X_1 - X_2 + Z\,(0.5 + 1.5V) + \varepsilon,
\end{equation}
with $\varepsilon \sim \mathcal{N}(0,1)$ independent of $(Z, X, V)$.
The modifier enters both the treatment assignment and the outcome model, so that the stabilizing density and the g-computation component each differ between the marginal and subgroup-conditional constructions and neither difference is switched off by the design.

Marginalizing $X$ analytically, $Z \mid V = v \sim \mathcal{N}(1 + 0.5v,\, 3/2)$, so the marginal treatment density is the equal mixture $\varpi(z) = \tfrac12 \phi(z; 1, 3/2) + \tfrac12 \phi(z; 1.5, 3/2)$, which is not itself normal.
Since $X \indep V$ and $\E[X_1 - X_2] = 0$, the g-computation components and the target curves are
\begin{equation}\label{eq:po_truth}
\theta(z \mid V = v) = 1 + z(0.5 + 1.5v), \qquad \theta(z) = 1 + 1.25\,z,
\end{equation}
so the true subgroup curves have slopes $0.5$ and $2$ and the population-average curve lies midway between them.

We compare four estimators, all formed by local linear kernel regression of the relevant pseudo-outcome on treatment within each subgroup:

\begin{center}
\begin{tabular}{llll}
\toprule
Arm & G-computation & Stabilizer & Outcome regression \\
\midrule
A1 & marginal & marginal & exact \\
A2 & subgroup & subgroup & exact \\
B1 & subgroup & subgroup & misspecified \\
B2 & subgroup & marginal & misspecified \\
\bottomrule
\end{tabular}
\end{center}

In the B arms the outcome regression is misspecified as $\hat\mu(z, \bx, v) = 1 + 1.25z$.
This choice is deliberate: $\hat\mu$ has no dependence on $\bx$ or $v$, so both g-computation components collapse to the same function of $z$ and the two B arms are distinguished by the stabilizer alone.

We use $n = 5{,}000$, $1{,}000$ replicates, a Gaussian kernel, and an evaluation grid of $100$ equally spaced points spanning the fifth to ninety-fifth percentiles of the marginal treatment distribution, $[-0.806, 3.306]$.

\subsection{The G-Computation Component Determines the Target}

With nuisances exact, the residual in \eqref{eq:pseudo_marginal} has conditional mean zero and \Cref{lem:offset} applies, so A1 targets $\theta(z)$ in both subgroups.
The simulation reproduces this.
Averaged over the grid, the estimated vertical separation between the two subgroup curves is $0.0063$ under A1, against a true separation of $1.8750$ and an exact limit of zero; under A2 it is $1.8805$.
The mean signed bias of A1 is $+0.9355$ for $V = 0$ and $-0.9332$ for $V = 1$, against exact values of $\pm 0.9375$: the two estimates are displaced in opposite directions by equal amounts, onto the common midpoint $\theta(z)$.
Their grid-average separation is approximately $0.006$, and the two curves are drawn at different stroke widths in \Cref{fig:po_loss} so that both remain visible.

For A2 the mean signed bias is $-0.0013$ and $+0.0043$, against a mean absolute bias of $0.066$ and $0.068$.
The absolute figures reflect finite-sample variability as well as any bias and should not be interpreted as systematic bias alone.

\subsection{The Stabilizer Determines Double Robustness}

In the B arms the residual no longer has conditional mean zero and the stabilizer is no longer inert.
Retaining the subgroup g-computation component but pairing it with the marginal stabilizer leaves, by \Cref{rem:stabilizer}, the bias
\begin{equation}\label{eq:po_bias}
\{r_v(z) - 1\}\, z\, (1.5v - 0.75), \qquad r_v(z) = \frac{\varpi(z)}{\varpi(z \mid V = v)},
\end{equation}
which under \eqref{eq:po_dgp} admits the closed form $r_0(z) = \tfrac12 + \tfrac12 e^{(z - 1.25)/3}$ and $r_1(z) = \tfrac12 + \tfrac12 e^{-(z-1.25)/3}$.
The bias vanishes at $z = 1.25$, where the marginal and subgroup treatment densities coincide, and again at $z = 0$ through the explicit factor of $z$, and grows away from both points.
It is of the same sign in both subgroups within each region, so the failure is a common shift rather than a symmetric spreading.

Because the pointwise separation between subgroups has Monte Carlo standard deviation of $6.2$ and $9.5$ at the grid edges, neither B arm's separation is informative on its own.
The two arms are evaluated on common replicates, however, so their paired difference isolates the stabilizer with far less noise.
At the upper grid endpoint the paired difference is $+0.6906$ with Monte Carlo standard error $0.1034$, against an exact value of $+0.6055$; at $z = 3.015$ it is $+0.4530$ with standard error $0.0397$, against $+0.4030$.

The most visible consequence is at high treatment levels.
Under the marginal stabilizer the fitted curve for $V = 0$ turns over at $z = 2.10$, against a closed-form turning point of $2.16$, and falls to $1.21$ at the upper endpoint where the exact limit is $1.43$ and the truth is $2.65$.
The estimator does not merely attenuate the dose-response in that subgroup; it reverses its apparent direction.

\ifthenelse{\boolean{showfigures}}{
\begin{figure}[htbp]
\centering
\includegraphics[width=\textwidth]{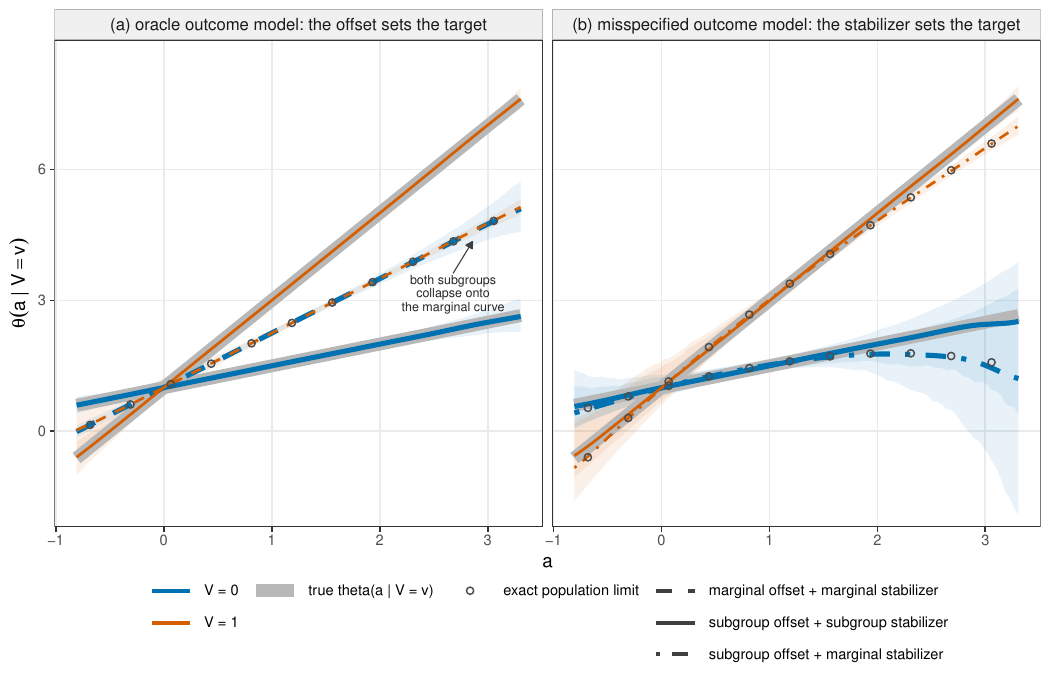}
\caption{Two failure modes of the marginal pseudo-outcome under design \eqref{eq:po_dgp}.
Pale halos are the true subgroup curves $\theta(z \mid V = v)$; bands are the fifth to ninety-fifth percentile across $1{,}000$ replicates; open circles are exact population limits computed in closed form.
\emph{Panel (a)}, outcome regression exact: the marginal g-computation component with marginal stabilizer returns $\theta(z)$ in both subgroups, so the two estimates coincide and all heterogeneity disappears; their grid-average separation is approximately $0.006$ and the two curves are drawn at different stroke widths so that both are visible.
The subgroup-conditional construction recovers each curve, and has no separate overlay because its exact limit is the true curve.
\emph{Panel (b)}, outcome regression misspecified as $\hat\mu(z,\bx,v) = 1 + 1.25z$: with the subgroup g-computation component held fixed, the stabilizer determines the target.
The marginal stabilizer leaves the bias \eqref{eq:po_bias}, which vanishes at $z = 0$ and $z = 1.25$ and grows toward both grid edges, reversing the apparent direction of the dose-response for $V = 0$ beyond $z = 2.10$.
The subgroup-conditional construction remains correct despite the misspecified outcome model.
Widening bands toward the grid edges reflect the unbounded stabilized weights of this illustrative Gaussian design.}
\label{fig:po_loss}
\end{figure}
}{}

\subsection{Numerical Detail}

\textit{Bandwidth.}
A fixed bandwidth $h = 1.06\,\hat\sigma_{Z \mid V} n_V^{-1/5} = 0.2751$ is used throughout rather than a cross-validated one.
With exact nuisances, the regression functions in panel (a) are affine, so local-linear smoothing reproduces their conditional means whenever the design is nonsingular.
This does not imply that realized leave-one-out cross-validation loss decreases monotonically with bandwidth.
The fixed bandwidth makes the comparison reproducible.
The misspecified marginal-stabilizer regression in panel (b) is nonlinear, so its fixed-bandwidth fitted limit can differ from the unsmoothed conditional-mean formula.
Repeating the study at $h \in \{0.15, 0.2751, 0.5\}$ leaves every conclusion intact: the A1 grid-average separation is $-0.005$, $-0.003$ and $+0.003$, the A2 separation is $1.871$, $1.872$ and $1.875$, and the paired difference in panel (b) is $+0.751$, $+0.691$ and $+0.645$, positive at every bandwidth.

\textit{Overlap and the stabilized weights.}
Because \eqref{eq:po_dgp} gives $Z$ a Gaussian conditional distribution, $\pi(z \mid \bx)$ is not bounded away from zero and the design does not satisfy \Cref{ass:overlap}.
The stabilized weight $\varpi(z)/\pi(z \mid \bx, v)$ has $\E[w] = 1$ but infinite second moment, with an empirical ninety-ninth percentile of $5.21$ and maximum of $88.6$ at $n = 5{,}000$.
This design deliberately separates mean identification from moment regularity.
The conditional-mean argument in \Cref{lem:offset} requires integrability rather than a uniform overlap bound, but correct centering does not eliminate the variance of the weighted residual.
Heavy weights can affect precision in both panels.
Unconditional weight moments and the local kernel-weighted moments relevant to smoothing on a compact interval must be distinguished; neither the exact mean identity nor the plotted bands verifies the assumptions of the asymptotic theory.

\textit{Symbolic and numerical validation.}
Beyond the simulation, the algebraic and identification results of \Cref{sec:pseudo} were checked symbolically and by numerical quadrature against the design \eqref{eq:po_dgp}, in which every conditional expectation is computable to machine precision.
Numerical integration recovered the population and subgroup curves in \Cref{prop:insufficient,prop:subgroup}.
With a misspecified outcome regression, the discrepancy was at most $2.3\times10^{-14}$ when the two density components and the subgroup covariate law were supplied at their population values.
The expression in \Cref{rem:stabilizer} agreed with the numerical bias to within $3.5\times10^{-14}$. We also checked the FANOVA identity symbolically.

We checked the closed forms against $2\times10^6$ Monte Carlo draws. Across 116 checks, the largest discrepancy was 3.89 standard errors, at the left edge of the exposure grid.
The mean and median across replicates differed by at most 0.014 in panel (a) and 0.124 in panel (b).

\section{Sensitivity to Unmeasured Confounding}
\label{app:evalue}

\Cref{ass:unconfoundedness} requires that the measured covariates suffice to adjust for confounding of the exposure-mortality relationship at the county-year level, which is not verifiable.
We therefore report E-values \citepApp{vanderweele2017sensitivity}: the minimum association, on the rate ratio scale, that an unmeasured confounder would need with both exposure and mortality, conditional on the measured covariates, to explain away an estimated contrast.

We use the 5 to 9 $\mu$g/m$^3$ contrast from (Q2) throughout. Holding the discovery tree fixed, we draw 250 bootstrap samples by resampling inference counties with replacement.

\begin{table}[htbp]
\centering
\caption{E-values for the mortality rate ratio (RR) between 5 and 9 $\mu$g/m$^3$ by subgroup; the E-value is the minimum association, on the rate ratio scale, that an unmeasured confounder would need with both exposure and mortality, conditional on the measured covariates, to explain away the estimated contrast}
\label{tab:evalues}
\footnotesize
\setlength{\tabcolsep}{3pt}
%\begin{tabular}{>{\raggedright\arraybackslash}p{0.45\textwidth}rrrrr}
\begin{tabular}{lrrrr}
Subgroup & RR & 95\% CI & E-value & E-value (CI) \\
\midrule
Education = low \& Poverty = high \& HomeValue = high & 1.122 & (1.077, 1.190) & 1.49 & 1.36 \\
Education = low \& Poverty = high \& HomeValue = low & 1.120 & (1.091, 1.161) & 1.49 & 1.41 \\
Education = high \& HomeValue = low \& Black = low & 1.116 & (1.083, 1.153) & 1.48 & 1.38 \\
Population & 1.099 & (1.089, 1.112) & 1.43 & 1.40 \\
Education = low \& Poverty = low \& Black = high & 1.093 & (1.048, 1.299) & 1.41 & 1.27 \\
Education = high \& HomeValue = high \& Income = high & 1.091 & (1.074, 1.106) & 1.41 & 1.35 \\
Education = high \& HomeValue = low \& Black = high & 1.089 & (1.023, 1.148) & 1.40 & 1.18 \\
Education = high \& HomeValue = high \& Income = low & 1.073 & (1.050, 1.105) & 1.35 & 1.28 \\
Education = low \& Poverty = low \& Black = low & 1.071 & (1.051, 1.096) & 1.35 & 1.28 \\
\bottomrule
\end{tabular}
\end{table}

On the population curve in \Cref{fig:erf_pm25}, the rate ratio is 1.089 and the E-value is 1.40.
For the root of the tree, which uses only the inference sample, the corresponding values are 1.099 and 1.43 (\Cref{tab:evalues}).
All subgroup intervals exclude one. E-values range from 1.35 to 1.49 for the subgroup estimates and from 1.18 to 1.41 for the confidence limits closest to one.
These values are comparable to those reported in other observational studies \citepApp{blum2020use}.

\section{Regularity Conditions}
\label{app:conditions}

We state the formal regularity conditions required for the theoretical results.

\begin{condition}[Smoothness]\label{con:smoothness}
There is a fixed open interval $J$ containing $\mathcal Z$, with positive distance between $\mathcal Z$ and the complement of $J$, on which $z\mapsto\mu(z,\bx)$ is twice continuously differentiable for $P_{\bX}$-almost every $\bx$.
For some finite $M$, $\sup_{z\in J,\bx}|\partial_z^k\mu(z,\bx)|\leq M$ for $k=0,1,2$, using versions with a common null set in $\bx$.
\end{condition}

\begin{condition}[Nuisance Estimation and Empirical Processes]\label{con:nuisance}
The outcome regression, propensity, and subgroup stabilizing density are fitted outside each evaluation fold, using a fixed number of folds.
Write $\mathcal T_k$ for the training information for fold $k$.
In particular, a stabilizer estimated from evaluation observations is not independent training information and requires a separate dependence argument.
The empirical g-computation component may reuse evaluation covariates only subject to the uniform g-computation component bound below.
Estimated propensities and subgroup stabilizing densities are truncated to uniformly positive and bounded values, and the fitted outcome functions and their evaluation classes are uniformly bounded and have polynomial covering numbers.
Writing $P_{\mathcal L}$ for the conditional distribution given $\bV\in\mathcal L$, suppose there are nonrandom, nonnegative, bounded sequences $a_{\mu,n}$, $a_{\pi,n}$, and $a_{\varpi,n}$ such that uniformly over folds, admissible nodes, and $z\in J$,
\begin{align*}
\|\hat\mu(z,\cdot)-\mu(z,\cdot)\|_{L^2(P_{\mathcal L})} &= O_p(a_{\mu,n}),\\
\|\hat\pi(z\mid\cdot)-\pi(z\mid\cdot)\|_{L^2(P_{\mathcal L})} &= O_p(a_{\pi,n}),\\
|\hat\varpi(z\mid\mathcal L)-\varpi(z\mid\mathcal L)| &= O_p(a_{\varpi,n}).
\end{align*}
The within-node empirical g-computation components satisfy $\sup_{k,\mathcal L,z\in J}|(\hat P_{\mathcal L,k}-P_{\mathcal L})\hat\mu_k(z,\cdot)|=O_p(t_n)$, where $t_n=\sqrt{\log n/n_{\min}}$.
The classes obtained by multiplying cross-fitted pseudo-outcome score functions by candidate-node indicators and treatment-kernel translates have uniformly polynomial conditional covering numbers, with complexity constants independent of $n$.
The deterministic product $a_{\mu,n}(a_{\pi,n}+a_{\varpi,n})\to0$.
\end{condition}

\begin{condition}[Kernels and Bandwidths]\label{con:kernel}
The kernels for local-linear regression and density estimation are bounded, symmetric probability densities with bounded variation, finite second moments, and finite squared integrals; Gaussian kernels are allowed.
The comparison range $\mathcal Z$ is a compact interior interval of treatment support, so ordinary density smoothing is not evaluated at support boundaries.
The bandwidths $h$, $h_\pi$, and $h_w$ used for curve estimation, subgroup density estimation (if used), and density weighting (if used) satisfy $h,h_\pi,h_w\to0$, $(\log n)^3/(n_{\min}h)\to0$, $\log n/(n_{\min}h_\pi)\to0$, and $\log n/(n_{\min}h_w)\to0$.
The stronger restriction on $h$ permits an envelope term of order $(\log n)^2/(n_{\min}h)$ to be absorbed into $\sqrt{\log n/(n_{\min}h)}$; the required process bounds are stated separately in \Cref{con:process}.
For estimated bandwidths the stated bounds are required to hold uniformly over the bandwidth-selection ranges.
\end{condition}

\begin{condition}[Bounded Moments]\label{con:moments}
The conditional outcome residual has uniformly bounded sub-exponential tails given $(Z,\bX)$, and the conditional response and treatment-density estimators satisfy the bounded-envelope conditions in \Cref{con:smoothness,con:nuisance}.
\end{condition}

\begin{condition}[Treatment Density Regularity]\label{con:density}
For all admissible subgroups $\mathcal{L}$ with $\Prob(\bV \in \mathcal{L}) \geq \rho_n$, where $\rho_n=n_{\min}/(2n)$ and $n_{\min}/\log n\to\infty$, the conditional treatment density $f(z \mid \bV \in \mathcal{L})$ satisfies:
\begin{enumerate}
    \item[(i)] $f(z \mid \bV \in \mathcal{L}) \geq f_{\min} > 0$ for all $z \in \mathcal{Z}$ (bounded away from zero);
    \item[(ii)] $f(z \mid \bV \in \mathcal{L}) \leq f_{\max} < \infty$ for all $z \in \mathcal{Z}$ (bounded above);
    \item[(iii)] $f(z \mid \bV \in \mathcal{L})$ is twice continuously differentiable on $J$, with derivatives uniformly bounded there.
\end{enumerate}
The density bounds in (i)--(ii) hold on $J$.
The true propensity density $\pi(z\mid\bx)$ is uniformly bounded above on $J$; estimated denominators have the lower bounds in \Cref{con:nuisance}.
\end{condition}

\begin{condition}[Finite Candidate Splits]\label{con:splits}
The maximum depth $d_{\max}$ and the dimension $q$ are fixed, and each candidate split variable has at most $C_n=O(n)$ thresholds, yielding a polynomial-size collection of attainable nodes for fixed depth.
For exact tree recovery we additionally require a fixed finite candidate grid; the data-dependent grids used in practice are instead covered by uniform gain convergence and appropriate argmax consistency statements.
\end{condition}

\begin{condition}[Uniform Local-Linear Design and Score Bounds]\label{con:process}
Let $a_n=\sqrt{\log n/(n_{\min}h)}$, $g(u)=(1,u)^\top$, and $u_i=(Z_i-z)/h$.
For an eligible node define
\begin{align*}
S_{n,\mathcal L}(z)&=\frac1{n_{\mathcal L}}\sum_{i:\bV_i\in\mathcal L}K_h(Z_i-z)g(u_i)g(u_i)^\top,\\
\ell_{i,\mathcal L}(z)&=\frac{K_h(Z_i-z)}{n_{\mathcal L}}e_1^\top S_{n,\mathcal L}(z)^{-1}g(u_i).
\end{align*}
The following bounds hold simultaneously over all eligible candidate rectangles, $z\in\mathcal Z$, and the allowed bandwidths; sums below are over observations in the node.
\begin{enumerate}[label=(\roman*)]
\item With probability tending to one, all these matrices have smallest eigenvalue at least a constant $c>0$, and $\sup_{\mathcal L,z}\sum_i|\ell_{i,\mathcal L}(z)|=O_p(1)$.
Writing $\theta_{\mathcal L}(t)=\theta(t\mid\mathcal L)$, the weighted Taylor remainder satisfies
\[
\sup_{\mathcal L,z}\left|\sum_i\ell_{i,\mathcal L}(z)
\{\theta_{\mathcal L}(Z_i)-\theta_{\mathcal L}(z)-\theta'_{\mathcal L}(z)(Z_i-z)\}\right|=O_p(h^2).
\]
\item Let $\eta_i^{\mathcal L}=\varphi_i^{\mathcal L}-\theta_{\mathcal L}(Z_i)$.
Let $R_i^{\mathcal L}$ be the estimated-minus-oracle pseudo-outcome difference with the estimated g-computation component replaced by $P_{\mathcal L}\hat\mu_{k(i)}$, where $k(i)$ is the evaluation fold, and define
$B_{k,\mathcal L}(t)=\E[R^{\mathcal L}(W)\mid Z=t,\bV\in\mathcal L,\mathcal T_k]$ for a fresh evaluation observation.
The two centered processes satisfy
\[
\sup_{\mathcal L,z}\left|\sum_i\ell_{i,\mathcal L}(z)\eta_i^{\mathcal L}\right|=O_p(a_n),\qquad
\sup_{\mathcal L,z}\left|\sum_i\ell_{i,\mathcal L}(z)
\{R_i^{\mathcal L}-B_{k(i),\mathcal L}(Z_i)\}\right|=O_p(a_n).
\]
\item For a noncompact kernel, contributions outside $J$ also obey
\[
\sup_{\mathcal L,z}\sum_{i:Z_i\notin J}|\ell_{i,\mathcal L}(z)|
\{|B_{k(i),\mathcal L}(Z_i)|+|e_{k(i),\mathcal L}(Z_i)|\}
=O_p(h^2+t_n+r_n+s_n),
\]
where $e_{k,\mathcal L}(t)=(\hat P_{\mathcal L,k}-P_{\mathcal L})\hat\mu_k(t,\cdot)$,
$r_n=a_{\mu,n}a_{\pi,n}$, and $s_n=a_{\mu,n}a_{\varpi,n}$.
This tail condition is automatic for compactly supported kernels once $h$ is small enough.
For Gaussian kernels it requires a tail-envelope argument, as does the Taylor-remainder bound in (i).
\item With probability tending to one every eligible node has population probability at least $\rho_n=n_{\min}/(2n)$; empirical child proportions differ from their population counterparts by $O_p(t_n)$ uniformly over eligible splits.
Any fold-specific count bounds needed for (i)--(iii) hold on this same uniform event.
For density weighting, the simultaneous child-density rate assumed in \Cref{lem:weight} also holds.
\end{enumerate}
\end{condition}

These are explicit high-level assumptions on the actual smoother and fold construction, not conclusions inferred solely from the number of realized thresholds.
For compact kernels, affine reproduction and a uniform bound on the sum of absolute equivalent weights give the Taylor bound from \Cref{con:smoothness}.
Establishing the matrix, score, tail, and eligible-node bounds from primitive entropy and moment conditions requires additional empirical-process arguments.
The results below are conditional on \Cref{con:process}; their current proofs do not supply that additional verification.

\begin{condition}[Stable Admissibility]\label{con:admissibility}
For the tree and modifier-consistency claims, there are population admissible sets $\mathcal S_n^0(\mathcal L)$ and events $A_n$ with $\Prob(A_n)\to1$ on which the empirical admissible sets equal $\mathcal S_n^0(\mathcal L)$ simultaneously at all candidate nodes under consideration.
Population and empirical trees use the same child regions and depth restriction on this event.
The sets in the separation and margin conditions refer to $\mathcal S_n^0$.
For exact recovery these population sets and the candidate grid are fixed.
\end{condition}

On a fixed finite grid, one sufficient leaf-size rule is $n_{\min}/n\to0$, with every population-admissible child having positive probability and every population-zero child excluded.
The law of large numbers simultaneously retains all positive-probability children and excludes zero-probability children.
If $n_{\min}/n\to\alpha>0$, strict separation of each relevant child probability from $\alpha$ is needed instead.
In particular, at a probability-$1/2$ split with $n_{\min}=n/2$, empirical eligibility requires exactly equal counts and need not stabilize.
A split-gain margin alone cannot repair this boundary problem.

\begin{remark}[Condition \ref{con:density} and the weighting function]\label{rem:density_tension}
\Cref{con:density}(i) requires the conditional treatment density to be bounded below uniformly on $\mathcal{Z}$ within every subgroup of non-negligible probability.
Under this condition the density weighting \eqref{eq:weight} is bounded away from zero on the comparison interval, so the theory of \Cref{sec:theory} covers strict overlap and treats weighting as a choice of estimand emphasis rather than a proof of validity under positivity violations.
The finite-sample role of the weighting, and the empirical consequences of that role, are addressed in \Cref{s:application}.

Under comparable local conditional variances and bandwidths, sparse treatment regions tend to have greater curve-estimation variance, so the product density can downweight imprecisely estimated regions.
This motivation is not a general variance-reduction or optimal-weighting guarantee.

We note that the proposed framework accommodates other choices of $w(z)$: (i) uniform weights $w(z) = 1$ if the full treatment range is of scientific interest; (ii) target density weights $w(z) = f_{\text{target}}(z)$ to emphasize a particular population; or (iii) overlap weights $w(z) \propto \min\{f_L(z), f_R(z)\}$.
The deterministic objective bounds extend to bounded nonnegative weights with uniform estimation error; the stated stochastic rate uses the specific density bound in \Cref{lem:weight}.
As \Cref{rem:density_tension} explains, weighting does not remove the strict-overlap assumptions of the theory.
The choice is consequential and application-specific: \Cref{s:application} reports both uniform and density-weighted comparisons because they emphasize different dose regions.

\end{remark}

\section{Proofs}
\label{app:proofs}

\subsection{Marginal Pseudo-Outcomes with Estimated Nuisances}
\label{app:marginal_details}

\begin{lemma}[G-computation component determines the target]\label{lem:offset}
Assume $Y$ is integrable.
Let $w(Z, \bX)$ be any measurable weighting function of $(Z, \bX)$ for which $\E|w(Z,\bX)\{Y-\mu(Z,\bX)\}|<\infty$, and let $b(Z)$ be integrable; set $\varphi = w(Z,\bX)\{Y - \mu(Z,\bX)\} + b(Z)$.
Then for any measurable set $\mathcal{L}$ in the modifier space with $\Prob(\bV \in \mathcal{L}) > 0$,
\begin{equation}
\E[\varphi \mid Z = z, \bV \in \mathcal{L}] = b(z) \quad \text{for }P_{Z\mid\mathcal L}\text{-almost every }z.
\end{equation}
\end{lemma}

\begin{proof}[Proof of \Cref{lem:offset}]
By iterated expectation, conditioning first on $(Z, \bX)$,
$\E\big[w(Z,\bX)\{Y - \mu(Z,\bX)\} \mid Z = z, \bV \in \mathcal{L}\big]
= \E\big[w(z,\bX)\,\E\{Y - \mu(Z,\bX) \mid Z = z, \bX\} \mid Z = z, \bV \in \mathcal{L}\big] = 0$,
since $\E[Y \mid Z, \bX] = \mu(Z,\bX)$ by definition and $\bV$ is a function of $\bX$.
With the population quantities held fixed, the additive component depends only on $z$, so conditioning on $Z=z$ leaves it equal to $b(z)$; this does not assume that $b(z)$ is observed or known.
\end{proof}

\Cref{prop:insufficient} is stated with nuisances known exactly, which is the cleanest form, but the conclusion does not depend on that.

\begin{corollary}[Estimated nuisances]\label{cor:insufficient_est}
Let $\hat\varphi_i$ be \eqref{eq:pseudo_marginal} constructed from $(\hat\mu,\hat\pi,\hat\varpi,\hat\Prob)$ learned independently of evaluation unit $i$, with $\hat\pi$ truncated below at $\epsilon$, and take the following expectation conditional on these estimates.
Then
\begin{equation}\label{eq:cor_expansion}
\E[\hat\varphi_i \mid Z_i = z, \bV_i \in \mathcal{L}]
= \int \hat\mu(z, \bx)\, d\hat\Prob(\bx) + R_{\mathcal{L}}(z),
\end{equation}
where the remainder is
\begin{equation}\label{eq:cor_remainder}
R_{\mathcal{L}}(z) = \int \big\{\mu(z,\bx) - \hat\mu(z,\bx)\big\}\,
\frac{\pi(z \mid \bx) \,/\, \varpi(z \mid \mathcal{L})}{\hat\pi(z \mid \bx) \,/\, \hat\varpi(z)}
\; d\Prob(\bx \mid \bV \in \mathcal{L}).
\end{equation}
Provided the stabilizing ratios are uniformly bounded and $\sup_{\mathcal L,z}\|\hat\mu(z,\cdot)-\mu(z,\cdot)\|_{L^1(P_{\mathcal L})}=o_p(1)$, the remainder in \eqref{eq:cor_remainder} satisfies $\sup_{\mathcal L,z}|R_{\mathcal L}(z)|=o_p(1)$.
Integrated $L^2$ consistency alone does not imply this uniform-in-treatment conclusion.
If the empirical covariate law reuses evaluation observations rather than an independent sample, the displayed conditional expansion is not an exact finite-sample identity without additional conditioning; the relevant empirical g-computation dependence must be controlled separately as in \Cref{con:nuisance}.
\end{corollary}

Two features of \eqref{eq:cor_expansion} are worth drawing out.
First, uniform consistency of the outcome regression is sufficient for this failure to persist; neither the doubly robust product rate nor consistency of the treatment-density estimates is needed once the stabilizing ratio remains bounded.
Second, the leading term $\int \hat\mu(z,\bx)\, d\hat\Prob(\bx)$ is free of $\mathcal{L}$, and $\mathcal{L}$ enters \eqref{eq:cor_remainder} only through factors multiplying $\mu - \hat\mu$.
When the outcome regression is uniformly consistent, the only subgroup-dependent signal available to a tree built on \eqref{eq:pseudo_marginal} is the outcome-model error itself, which vanishes.
The marginal construction does not merely lose power at the margin; it has no population signal to lose.

Pointwise continuous-dose curves require a different construction from regular conditional influence-function targets; \Cref{rem:cif} explains the distinction.

Results on regression with estimated pseudo-outcomes \citepApp{kennedy2023towards} and orthogonal statistical learning \citepApp{foster2023orthogonal} motivate separating the error of an oracle second-stage regression from the error introduced by estimated nuisance functions.
Our setting differs from a fixed-target second-stage regression because each candidate subgroup induces a different pseudo-outcome, requiring uniform control over adaptively selected nodes and objective functions.
We use this decomposition for the specific local-linear subgroup estimator rather than developing oracle inequalities for arbitrary second-stage learners.

Replacing the empirical covariate law by its population counterpart contributes $\int\hat\mu(z,\bx)\,d(\hat\Prob-\Prob)(\bx)$.
An independent covariate sample gives a root-$n$ rate at a fixed dose under a bounded second moment; uniform control over doses and adaptive nodes requires the separate assumptions in \Cref{con:nuisance}.

Comparing \eqref{eq:subgroup_drc} with the g-computation component of \eqref{eq:pseudo_marginal} gives the required modification directly: the subgroup target averages $\mu(z,\bx)$ over the covariate distribution \emph{within} $\mathcal L$, rather than over the full population.
Thus the g-computation component must be subgroup-specific, and the residual correction must use the corresponding subgroup stabilizing density to retain double robustness.

We next explain why the construction does not require decomposing $\bX$ into candidate modifiers $\bV$ and the remaining adjustment covariates.
Writing $\bX=(\bV,\bW)$ for exposition, the subgroup curve is $\theta(z\mid\mathcal L)=\E\{\mu(z,\bV,\bW)\mid\bV\in\mathcal L\}$, which averages over the \emph{joint} distribution of all covariates among members of the node.
For a positive-probability node, the empirical g-computation component is obtained by averaging $\hat\mu(z,\bX_i)$ across observations with $\bV_i\in\mathcal L$, thereby accounting automatically for the observed distributions of both $\bV$ and $\bW$ in that subgroup.
There is therefore no need to estimate a separate distribution for $\bW\mid\bV$ or to assume that the effect modifiers and other confounders are independent.
For a fixed node, the empirical covariate average has a root-$n_{\mathcal L}$ sampling rate under appropriate regularity conditions, with uniform control over adaptively selected nodes addressed in \Cref{con:nuisance,lem:erf_uniform_sub}.
In contrast, estimating a point-conditional curve at $\bV=\bv$ when $\bV$ is continuous generally requires smoothing across values of $\bV$, introducing dimension-dependent estimation error.
Using positive-probability nodes simplifies estimation of the g-computation component, but does not remove the need to estimate $\mu$ or control its estimation error.

\Cref{cor:insufficient_est} shows the same conclusion survives nuisance estimation, so the failure cannot be attributed to an idealized construction.
This motivates the subgroup construction in \eqref{eq:pseudo_subgroup}.

\begin{remark}[The stabilizer is what buys double robustness]\label{rem:stabilizer}
\Cref{lem:offset} shows the stabilizer is irrelevant to the target when nuisances are known exactly, so \Cref{prop:insufficient} is entirely a statement about the g-computation component.
The role of the stabilizer appears only once $\hat\mu \neq \mu$.
Retaining the subgroup g-computation component but pairing it with the \emph{marginal} stabilizer $\hat\varpi(z)/\hat\pi(z \mid \bx)$ replaces the bracket in \eqref{eq:dr_remainder} by
$\{\pi(z \mid \bx)/\varpi(z \mid \mathcal{L})\} \cdot \{\hat\varpi(z)/\hat\pi(z \mid \bx)\} - 1$,
which at $(\hat\pi, \hat\varpi) = (\pi, \varpi)$ reduces to $\varpi(z)/\varpi(z \mid \mathcal{L}) - 1$.
The ratio need not equal one, so correct treatment modeling no longer guarantees cancellation for an arbitrary outcome-model error.
Equality at one dose does not imply independence, and particular errors can cancel after integration; independence of $Z$ and subgroup membership implies equality throughout the treatment support.
The two failure modes are therefore distinct: a marginal g-computation component destroys identification, while a marginal stabilizer preserves identification but sacrifices double robustness.
\Cref{app:po_sim} demonstrates both in a design where every quantity is available in closed form: the marginal construction collapses the estimated gap between two subgroups, while retaining the subgroup g-computation component but pairing it with the marginal stabilizer leaves a bias large enough to reverse the apparent direction of the dose-response at high treatment levels.
\end{remark}

\subsection{Helper Results}

\begin{lemma}[Weighting function convergence]\label{lem:weight}
Under \Cref{con:density}, suppose $\sup_{\mathcal L,z}|\hat f_{\mathcal L}(z)-f_{\mathcal L}(z)|=O_p(d_{w,n})$ simultaneously over all eligible children, where $d_{w,n}=\sqrt{\log n/(n_{\min}h_w)}+h_w^2$.
Then,
\begin{equation}
\sup_s\|\hat w_s-w_s\|_\infty=O_p(d_{w,n}).
\end{equation}
For uniform weighting, no density estimation is required and we set $d_{w,n}=0$.
\end{lemma}

The uniform convergence result below is stated for the subgroup-conditional pseudo-outcome \eqref{eq:pseudo_subgroup}, which is what \Cref{alg:general} evaluates.
It differs from the corresponding result for the marginal construction in three ways, each of which is a genuine cost of conditioning on the node rather than a bookkeeping change.
First, the stabilizing density $\varpi(\cdot\mid\mathcal L)$ is estimated using the training-fold observations belonging to the node, contributing a subgroup-density error rather than a full-sample marginal-density error.
Second, the g-computation component is evaluated as a within-node average, contributing the assumed uniform empirical-process term $t_n=\sqrt{\log n/n_{\min}}$.
Third, the pseudo-outcome varies with the candidate split.
The supremum in \Cref{prop:gain_uniform} therefore ranges over changing integrands and regions; \Cref{con:process} explicitly supplies the uniform bounds required for this class.

Define $r_n=a_{\mu,n}a_{\pi,n}$ and $s_n=a_{\mu,n}a_{\varpi,n}$ as the two components of the subgroup-specific doubly robust bias.
\begin{equation}\label{eq:sn}
\varepsilon_{\theta,n}=\sqrt{\frac{\log n}{n_{\min}h}}+h^2+t_n+r_n+s_n.
\end{equation}

\begin{lemma}[Uniform subgroup dose-response estimation]\label{lem:erf_uniform_sub}
Let $\hat\theta(z\mid\mathcal L)$ be the local-linear estimator in \eqref{eq:locpoly} formed from the subgroup pseudo-outcomes in \eqref{eq:pseudo_hat}.
Under \Crefrange{ass:consistency}{ass:overlap}, \Crefrange{con:smoothness}{con:splits}, and the explicit design and process bounds in \Cref{con:process},
\begin{equation}\label{eq:lem_sub_rate}
\sup_{\mathcal L\ \mathrm{eligible}}\sup_{z\in\mathcal Z}
|\hat\theta(z\mid\mathcal L)-\theta(z\mid\mathcal L)|=O_p(\varepsilon_{\theta,n}).
\end{equation}
The supremum is over candidate nodes reachable by at most $d_{\max}$ coordinate splits that meet the minimum-leaf and required training-fold constraints.
In particular, the right-hand side is $o_p(1)$ if the stated bandwidth conditions hold and the doubly robust products $r_n+s_n$ vanish.
\end{lemma}

The first two terms in \eqref{eq:sn} are the oracle local-linear stochastic error and bias, and $t_n$ accounts for replacing the subgroup covariate distribution by an empirical average.
The two products $r_n$ and $s_n$ arise from estimating, respectively, the conditional treatment density and the subgroup stabilizing density in combination with the outcome regression.
The centered nuisance-score fluctuation is included in the leading stochastic term through \Cref{con:process}(ii).
We use this coarser bound rather than assert an unproved localized refinement proportional to $a_{\mu,n}+a_{\pi,n}+a_{\varpi,n}$.
The rate $a_{\varpi,n}$ depends on how the subgroup density is estimated, and in particular need not equal the rate of a direct kernel density estimator when it is obtained by averaging fitted generalized propensity scores.

\begin{proof}[Proof of \Cref{lem:erf_uniform_sub}]
Write $\tilde\theta$ for the local-linear smoother applied to the oracle pseudo-outcomes, using the same observations, bandwidth, and equivalent weights as $\hat\theta$.
Then
\begin{equation}\label{eq:oracle_decomp}
\hat\theta(z\mid\mathcal L)-\theta(z\mid\mathcal L)
=\{\tilde\theta(z\mid\mathcal L)-\theta(z\mid\mathcal L)\}
+\{\hat\theta(z\mid\mathcal L)-\tilde\theta(z\mid\mathcal L)\}.
\end{equation}
On the matrix event in \Cref{con:process}, the equivalent weights reproduce affine functions:
$\sum_i\ell_{i,\mathcal L}(z)=1$ and $\sum_i\ell_{i,\mathcal L}(z)(Z_i-z)=0$.
Decomposing each oracle pseudo-outcome into $\theta_{\mathcal L}(Z_i)+\eta_i^{\mathcal L}$ therefore gives a uniform oracle error $O_p(h^2+a_n)$ by parts (i)--(ii) of that condition.

For an observation evaluated with fold-$k$ nuisance estimates, abbreviate $w_{\mathcal L}=\varpi_{\mathcal L}/\pi$ and $\hat w_{k,\mathcal L}=\hat\varpi_{k,\mathcal L}/\hat\pi_k$.
The exact estimated-minus-oracle decomposition is
\begin{align*}
R_i^{\mathcal L}
&=(\hat w_{k,\mathcal L}-w_{\mathcal L})(Z_i,\bX_i)\{Y_i-\mu(Z_i,\bX_i)\}\\
&\quad+\hat w_{k,\mathcal L}(Z_i,\bX_i)\{\mu(Z_i,\bX_i)-\hat\mu_k(Z_i,\bX_i)\}
+P_{\mathcal L}(\hat\mu_k-\mu)(Z_i,\cdot),
\end{align*}
\begin{equation}\label{eq:estimated_oracle_split}
\hat\varphi_i^{\mathcal L}-\varphi_i^{\mathcal L}=R_i^{\mathcal L}+e_{k,\mathcal L}(Z_i).
\end{equation}
For a fresh evaluation observation conditional on $\mathcal T_k$, the first weighted residual has conditional mean zero.
Bayes' identity in \eqref{eq:bayes_node} gives
\begin{equation}\label{eq:dr_bias_proof}
B_{k,\mathcal L}(t)=P_{\mathcal L}\left[(\mu-\hat\mu_k)(t,\cdot)
\left\{\frac{\pi(t\mid\cdot)\hat\varpi_{k,\mathcal L}(t)}{\varpi_{\mathcal L}(t)\hat\pi_k(t\mid\cdot)}-1\right\}\right].
\end{equation}
The exact ratio decomposition
\[
\frac{\pi\hat\varpi}{\varpi\hat\pi}-1
=\frac{(\pi-\hat\pi)\hat\varpi}{\varpi\hat\pi}
+\frac{\hat\varpi-\varpi}{\varpi}
\]
and Cauchy--Schwarz show that $\sup_{k,\mathcal L,t\in J}|B_{k,\mathcal L}(t)|=O_p(r_n+s_n)$ under the denominator bounds and \Cref{con:nuisance}.
Smoothing \eqref{eq:estimated_oracle_split}, subtract and add $B_{k(i),\mathcal L}(Z_i)$.
The centered sum is $O_p(a_n)$ by \Cref{con:process}(ii).
For observations with $Z_i\in J$, the sum of absolute equivalent weights transfers the uniform bias and g-computation component bounds to $O_p(r_n+s_n+t_n)$.
The remaining observations are controlled by \Cref{con:process}(iii).
Combining these terms with the oracle error proves \eqref{eq:lem_sub_rate}.
All bounds are simultaneous over the candidate class; no union bound over realized, data-dependent thresholds is used.
\end{proof}

% \falco{}

Uniform invertibility of the full local-linear design matrix, rather than a scalar kernel-denominator bound alone, is used above.
The resulting uniform rate and bounded target curves imply $\sup_{\mathcal L,z}|\hat\theta(z\mid\mathcal L)|=O_p(1)$.

\subsection{Proof of Lemma~\ref{lem:weight}}

We prove convergence of the weighting function estimator.

\textbf{Step 1: Simultaneous density bound.}
Use the premise of \Cref{lem:weight},
$\sup_{\mathcal L,z}|\hat f_{\mathcal L}(z)-f_{\mathcal L}(z)|=O_p(d_{w,n})$.
This is a simultaneous bound over eligible children, not a collection of pointwise-in-node statements.
Deriving it for a specified KDE and adaptive-node class is part of the empirical-process assumptions in \Cref{con:process}.
Every constant and bound in the following steps is uniform over the same children.

\textbf{Step 2: Product of densities.}
Define $\phi(z) = f_L(z) f_R(z)$ and $\hat{\phi}(z) = \hat{f}_L(z) \hat{f}_R(z)$.
Since $\|\hat{f}_g - f_g\|_\infty = \op(1)$ and $f_g$ is bounded above by $f_{\max}$ (Condition~\ref{con:density}), we have $\|\hat{f}_g\|_\infty \leq f_{\max} + \op(1)$, which is bounded with probability approaching one.

Therefore:
\begin{align}
|\hat{\phi}(z) - \phi(z)| &= |\hat{f}_L(z) \hat{f}_R(z) - f_L(z) f_R(z)| \\
&\leq |\hat{f}_L(z) - f_L(z)| \cdot |\hat{f}_R(z)| + |f_L(z)| \cdot |\hat{f}_R(z) - f_R(z)| \\
&\leq (f_{\max} + \op(1)) \|\hat{f}_L - f_L\|_\infty + f_{\max} \|\hat{f}_R - f_R\|_\infty \\
&= \Op\left(\sqrt{\frac{\log n}{n_{\min} h_w}} + h_w^2\right).
\end{align}

\textbf{Step 3: Normalizing constant.}
Let $c = \int_{\mathcal{Z}} \phi(z) dz$ and $\hat{c} = \int_{\mathcal{Z}} \hat{\phi}(z) dz$.
By Condition~\ref{con:density}, $f_L(z), f_R(z) \geq f_{\min} > 0$ for all $z \in \mathcal{Z}$.
Since $\mathcal{Z}$ is compact with length $|\mathcal{Z}| = z_{\max} - z_{\min}$:
\begin{equation}
c = \int_{\mathcal{Z}} f_L(z) f_R(z) dz \geq f_{\min}^2 \cdot |\mathcal{Z}| > 0.
\end{equation}

Moreover:
\begin{equation}
|\hat{c} - c| \leq \int_{\mathcal{Z}} |\hat{\phi}(z) - \phi(z)| dz \leq |\mathcal{Z}| \cdot \|\hat{\phi} - \phi\|_\infty = \Op\left(\sqrt{\frac{\log n}{n_{\min} h_w}} + h_w^2\right).
\end{equation}

\textbf{Step 4: Weighting function.}
Since $w_s(z) = \phi(z)/c$ and $\hat{w}_s(z) = \hat{\phi}(z)/\hat{c}$:
\begin{align}
|\hat{w}_s(z) - w_s(z)| &= \left|\frac{\hat{\phi}(z)}{\hat{c}} - \frac{\phi(z)}{c}\right| = \left|\frac{\hat{\phi}(z) c - \phi(z) \hat{c}}{\hat{c} \cdot c}\right| \\
&\leq \frac{1}{\hat{c} \cdot c} \left(|\hat{\phi}(z) - \phi(z)| \cdot c + |\phi(z)| \cdot |\hat{c} - c|\right).
\end{align}

Since $|\hat{c} - c| = \op(1)$ and $c \geq f_{\min}^2 |\mathcal{Z}| > 0$, we have $\hat{c} \geq c/2$ with probability approaching one for $n$ large.
Also, $|\phi(z)| \leq f_{\max}^2$.
Therefore:
\begin{align}
|\hat{w}_s(z) - w_s(z)| &\leq \frac{2}{c^2} \left(c \cdot \|\hat{\phi} - \phi\|_\infty + f_{\max}^2 \cdot |\hat{c} - c|\right) = \Op\left(\sqrt{\frac{\log n}{n_{\min} h_w}} + h_w^2\right).
\end{align}
\qed

\subsection{Proof of Proposition~\ref{prop:gain_uniform}}

Write $e_n=\sup_{\mathcal L,z}|\hat\theta(z\mid\mathcal L)-\theta(z\mid\mathcal L)|$ and $d_n=\sup_s\|\hat w_s-w_s\|_\infty$ over eligible nodes and splits.
By \Cref{lem:erf_uniform_sub,lem:weight}, $e_n=O_p(\varepsilon_{\theta,n})$ and $d_n=O_p(d_{w,n})$.
The boundedness conditions imply that the true and fitted subgroup curves and the weighting functions are uniformly bounded with probability approaching one.

\textbf{Step 1: Squared curve differences.}
For any parent and child curves, write $D(z)=\theta_C(z)-\theta_P(z)$ and $\hat D(z)=\hat\theta_C(z)-\hat\theta_P(z)$.
Then $\|\hat D-D\|_\infty\leq2e_n$, and
\begin{equation}
\left|\int_{\mathcal Z}\hat D(z)^2\hat w_s(z)\,dz-\int_{\mathcal Z}D(z)^2w_s(z)\,dz\right|
\leq C(e_n+d_n)
\end{equation}
for a constant $C$ uniform over candidate splits, by adding and subtracting $\int \hat D^2w_s$ and applying $|\hat D^2-D^2|\leq|\hat D-D|(|\hat D|+|D|)$.
The same argument with $D(z)=\theta_L(z)-\theta_R(z)$ gives the $L^2$ bound.

\textbf{Step 2: AUC and FANOVA.}
The inequality $||a|-|b||\leq|a-b|$ and boundedness of the weights give $|\hat{\mathcal G}_{\mathrm{AUC}}-\mathcal G_{\mathrm{AUC}}|\leq C(e_n+d_n)$.
For FANOVA, the two weighted squared-curve terms obey the bound in Step 1, and replacing population child proportions by sample proportions contributes $O_p(t_n)$ by \Cref{con:process}(iv).
This last term is included in $t_n$ and hence in $\varepsilon_{\theta,n}$.

\textbf{Step 3: Uniformity.}
The curve, density, and proportion bounds are simultaneous over the candidate rectangle class by \Cref{con:nuisance,con:process} and \Cref{lem:erf_uniform_sub,lem:weight}.
Thus, for each objective,
\begin{equation}
\sup_{s\in\mathcal S_n}|\hat{\mathcal G}(s)-\mathcal G(s)|
=O_p(\varepsilon_{\theta,n}+d_{w,n})=O_p(\varepsilon_n),
\end{equation}
including every nuisance, subgroup-density, empirical g-computation, and weighting term of \eqref{eq:sn}.
\qed

\subsection{Proof of \Cref{thm:tree_consistency}}

Let $E_n(\eta)=\{\sup_{s\in\mathcal S_n}|\hat{\mathcal G}(s)-\mathcal G(s)|\leq\eta\}$.
Uniform convergence and $\varepsilon_n/\lambda_n\to0$ imply $\Prob\{E_n(\lambda_n/2)\}\to1$.
Intersect this event with $A_n$ from \Cref{con:admissibility}; on the intersection every comparison uses the same empirical and population candidate set.
At a population node where the gain criterion calls for stopping, the maximum admissible gain is zero, so on this event $\max_s\hat{\mathcal G}(s)\leq\lambda_n/2<\lambda_n$, and the algorithm stops.
At a nonterminal population node with maximal gain at least $\gamma$, we have $\max_s\hat{\mathcal G}(s)\geq\gamma-\lambda_n/2>\lambda_n$ for all sufficiently large $n$, and the algorithm splits.

For any fixed node and the same admissible candidate set in both optimizations, optimality of $\hat s$ for $\hat{\mathcal G}$ gives
\begin{align}
0\leq\mathcal G(s^*)-\mathcal G(\hat s)
&\leq \{\mathcal G(s^*)-\hat{\mathcal G}(s^*)\}+\{\hat{\mathcal G}(\hat s)-\mathcal G(\hat s)\}\nonumber\\
&\leq2\sup_{s\in\mathcal S_n}|\hat{\mathcal G}(s)-\mathcal G(s)|,
\end{align}
which proves \eqref{eq:split_regret}.

Now restrict to a fixed finite candidate grid and impose \Cref{con:margin}.
On $A_n\cap E_n(\delta/3)$, the estimated maximizer at every population internal node equals the unique population maximizer, since every alternative has estimated gain at least $\delta/3$ smaller.
The stopping argument above also identifies terminal nodes correctly, so induction over the fixed depth yields the same branches and stopping decisions as the population greedy tree.
A union bound over at most $2^{d_{\max}}-1$ internal nodes gives $\Prob(\hat\psi=\psi^*)\to1$.

If thresholds form increasingly dense grids, adjacent split gains need not be separated by a fixed $\delta$, so the exact-recovery step does not apply.
The preceding regret inequality remains valid over each common candidate set, and ordinary argmax continuity yields threshold convergence if the population objective has a unique, well-separated maximizer away from every neighborhood of that maximizer and the candidate grids approximate the continuous domain.
This proves the stated distinction between gain consistency and exact tree recovery.
\qed

\subsection{Proof of \Cref{prop:subgroup}}

Fix a node $\mathcal{L}$ with $\Prob(\bV \in \mathcal{L}) > 0$ and write $\Prob_{\mathcal{L}}(\cdot) = \Prob(\cdot \mid \bV \in \mathcal{L})$.

\textbf{Step 1: identification.}
With nuisances known exactly, the residual term of \eqref{eq:pseudo_subgroup} has conditional mean zero by \Cref{lem:offset}, so
$\E[\varphi^{\mathcal{L}} \mid Z = z, \bV \in \mathcal{L}]$ equals the g-computation component, which is $\int \mu(z,\bx)\, d\Prob_{\mathcal{L}}(\bx) = \theta(z \mid \mathcal{L})$ by \eqref{eq:subgroup_drc}.

\textbf{Step 2: the conditional covariate law.}
By Bayes' theorem, the law of $\bX$ given $Z = z$ and $\bV \in \mathcal{L}$ is
\begin{equation}\label{eq:bayes_node}
d\Prob(\bx \mid Z = z, \bV \in \mathcal{L}) = \frac{\pi(z \mid \bx)}{\pi(z \mid \bV \in \mathcal{L})} \, d\Prob_{\mathcal{L}}(\bx),
\end{equation}
since $\pi(z \mid \bV \in \mathcal{L}) = \int \pi(z \mid \bx) \, d\Prob_{\mathcal{L}}(\bx)$ is the corresponding normalizing constant.

\textbf{Step 3: remainder with estimated nuisances.}
Condition on the fitted quantities and the empirical g-computation sample, which are independent of the fresh evaluation observation.
Put $v=\varpi(z\mid\mathcal L)$, $\hat v=\hat\varpi(z\mid\mathcal L)$, and
$\rho(z,\bx)=\pi(z\mid\bx)\hat v/\{v\hat\pi(z\mid\bx)\}$.
Using \eqref{eq:bayes_node} and retaining the estimated covariate law gives
\[
\E[\hat\varphi^{\mathcal L}\mid Z=z,\bV\in\mathcal L]
=P_{\mathcal L}[(\mu-\hat\mu)\rho](z)+\hat P_{\mathcal L}\hat\mu(z,\cdot).
\]
Add and subtract $P_{\mathcal L}\hat\mu(z,\cdot)$ and use
$P_{\mathcal L}(\mu-\hat\mu)+P_{\mathcal L}\hat\mu=P_{\mathcal L}\mu=\theta(z\mid\mathcal L)$.
This yields the two terms in \eqref{eq:dr_remainder} exactly.
Moreover,
\[
\rho-1=\frac{(\pi-\hat\pi)\hat v}{v\hat\pi}+\frac{\hat v-v}{v}.
\]
Thus, under positive denominator and upper-envelope bounds, the product term is bounded by a constant times
$\|\hat\mu-\mu\|_{L^2(P_{\mathcal L})}
\{\|\hat\pi-\pi\|_{L^2(P_{\mathcal L})}+|\hat v-v|\}$.
It vanishes if $\hat\mu=\mu$ or $\rho=1$, in particular if both density components are correct.
Correct $\hat\pi$ alone does not imply $\rho=1$ when $\hat v$ is separately estimated.
The empirical g-computation term remains even in these exact cancellation cases and requires its own convergence bound.
\qed

\subsection{Proof of the Claim in \Cref{rem:stabilizer}}

With the subgroup g-computation component and a marginal stabilizer, the product remainder becomes
\[
P_{\mathcal L}\left[(\mu-\hat\mu)
\left\{\frac{\pi(z\mid\cdot)}{\varpi(z\mid\mathcal L)}
\frac{\hat\varpi(z)}{\hat\pi(z\mid\cdot)}-1\right\}\right],
\]
in addition to the same empirical g-computation error.
When the propensity and marginal density are correct, this equals
$\{\varpi(z)/\varpi(z\mid\mathcal L)-1\}P_{\mathcal L}(\mu-\hat\mu)(z,\cdot)$.
It can vanish at particular doses or for particular outcome errors, but is not guaranteed to vanish across outcome-model errors.
For a nontrivial subgroup event, equality of the two densities almost everywhere on the entire treatment support is equivalent to independence of treatment and that event; equality on the comparison interval alone is insufficient for that conclusion.
\qed

\subsection{Modifier Selection: Separation Condition and Proof}
\label{app:modifier_selection}

Fix a version of the conditional dose-response surface and a covariate representation on its support.
Let $\mathcal M^*\subseteq\{1,\ldots,q\}$ denote its designated set of relevant coordinates, including level modifiers, and let $\mathcal M(\psi)$ denote the variables used by a tree.
With redundant coordinates this representation need not be unique; the result is relative to the chosen $\mathcal M^*$.
The notion of a modifier in this theorem is therefore broader than heterogeneity in additive dose contrasts, as discussed in \Cref{prop:l2_decomp}.
Let $\mathcal N_n(\mathcal M^*)$ contain all nodes reachable by up to $d_{\max}-1$ admissible splits on variables in $\mathcal M^*$, with population node probabilities at least $\rho_n=n_{\min}/(2n)$.

\begin{condition}[Modifier separation]\label{con:modifier_sep}
There is a constant $\delta_{\mathcal M}>0$ such that, at each $\mathcal L\in\mathcal N_n(\mathcal M^*)$ with positive maximal population gain, a split on a variable in $\mathcal M^*$ is admissible and
\begin{equation}
\max_{s\in\mathcal S(\mathcal L):j(s)\in\mathcal M^*}\mathcal G(s)
\geq\max_{s\in\mathcal S(\mathcal L):j(s)\notin\mathcal M^*}\mathcal G(s)+\delta_{\mathcal M}.
\end{equation}
Nodes with zero maximal population gain are governed by the stopping threshold and need not satisfy a separation inequality.
\end{condition}

The first conclusion of \Cref{thm:modifier_consistency} allows ties among genuinely relevant variables and does not require exact recovery of their thresholds.
It controls asymptotic inclusion of irrelevant variables under the stated separation condition, but does not supply finite-sample false discovery rate control.
The second conclusion additionally requires all true variables to appear in the population greedy tree, a stronger requirement than the absence of spurious splits.

\textit{Proof of \Cref{thm:modifier_consistency}.}

Set $E_n=\{\sup_{s\in\mathcal S_n}|\hat{\mathcal G}(s)-\mathcal G(s)|<\min(\delta_{\mathcal M}/3,\lambda_n/2)\}$.
By \Cref{prop:gain_uniform,con:stopping}, $\Prob(E_n)\to1$.
Intersect with $A_n$ from \Cref{con:admissibility} so that the separating relevant split is also empirically eligible.
By \Cref{con:process}(iv), every empirically eligible node has population mass at least $\rho_n=n_{\min}/(2n)$ on a further event with probability tending to one.
Use this intersection throughout the induction.

We prove by induction on depth that every split selected on $E_n$ is on a variable in $\mathcal M^*$.
The root is a node in $\mathcal N_n(\mathcal M^*)$.
Suppose all preceding splits use variables in $\mathcal M^*$, so the current eligible node is also in $\mathcal N_n(\mathcal M^*)$.
If its maximal population gain is zero, the stopping condition implies $\max_s\hat{\mathcal G}(s)<\lambda_n$ and no split is made.
Otherwise \Cref{con:modifier_sep} supplies a true-variable split $s^+$ whose population gain exceeds that of every null-variable split by at least $\delta_{\mathcal M}$.
On $E_n$, for any null-variable candidate $s^-$,
\begin{equation}
\hat{\mathcal G}(s^+)-\hat{\mathcal G}(s^-)
\geq\mathcal G(s^+)-\mathcal G(s^-)-2\sup_s|\hat{\mathcal G}(s)-\mathcal G(s)|
>\delta_{\mathcal M}/3>0.
\end{equation}
Thus the estimated maximizing split uses a variable in $\mathcal M^*$, completing the induction.
Because $d_{\max}$ is fixed, the tree contains finitely many internal nodes and a union bound establishes $\Prob\{\mathcal M(\hat\psi)\subseteq\mathcal M^*\}\to1$.

Under the additional fixed-grid margin condition, \Cref{thm:tree_consistency} gives exact recovery of $\psi^*$; when $\mathcal M(\psi^*)=\mathcal M^*$, equality of the modifier sets follows.
\qed

The argument separates relevant from irrelevant splitting variables, but never compares two relevant variables or two thresholds on the same relevant variable.
Consequently, ties among relevant splits do not prevent the first conclusion, even though they can prevent exact tree recovery.

\subsection{Proof of \Cref{prop:fanova_l2}}

Write $p = \Prob(\bV \in \mathcal{L}_L \mid \bV \in \mathcal{L})$ and abbreviate $\theta_L(z) = \theta(z \mid \mathcal{L}_L)$, $\theta_R(z) = \theta(z \mid \mathcal{L}_R)$, $\theta_P(z) = \theta(z \mid \mathcal{L})$.
By the law of total expectation applied to \eqref{eq:subgroup_drc},
$\theta_P(z) = p\,\theta_L(z) + (1-p)\,\theta_R(z)$, whence
\begin{equation}
\theta_L(z) - \theta_P(z) = (1-p)\{\theta_L(z) - \theta_R(z)\}, \qquad
\theta_R(z) - \theta_P(z) = -p\{\theta_L(z) - \theta_R(z)\}.
\end{equation}
Substituting into the population criterion \eqref{eq:fanova}, the integrand becomes
\begin{equation}
p(1-p)^2\{\theta_L - \theta_R\}^2 + (1-p)p^2\{\theta_L - \theta_R\}^2 = p(1-p)\{\theta_L - \theta_R\}^2,
\end{equation}
pointwise for the chosen curve versions.
Integrability permits integration against $w_s$, giving the stated identity.
For arbitrary fitted parent and child curves, completing the square instead gives
\begin{align}\label{eq:empirical_fanova}
\hat{\mathcal G}_{\mathrm{FANOVA}}(s)
&=\hat p(1-\hat p)\hat{\mathcal G}_{L^2}(s)\nonumber\\
&\quad+\int_{\mathcal Z}\{\hat\theta_P(z)-\hat p\hat\theta_L(z)-(1-\hat p)\hat\theta_R(z)\}^2\hat w_s(z)\,dz.
\end{align}
Thus separate parent fitting need not preserve the population identity exactly.

Without the proportional weights the corresponding factor is $(1-p)^2 + p^2 = 2p^2 - 2p + 1$, which attains its minimum $1/2$ at $p = 1/2$ and its maximum $1$ as $p \to 0$ or $p \to 1$; the proportional weighting therefore reverses the preference, from unbalanced to balanced splits.
\qed

\subsection{Objective Detection, Proofs, and Examples}\label{app:objective_detection}

Fix one split and write $d=d_s$, $\nu=\nu_s$, and $m=\int d\,d\nu$.
The same comparison measure must be used for all objectives at that split.
Since $\nu$ is a probability measure and $d\in L^2(\nu)$, $d\in L^1(\nu)$.
All almost-everywhere statements below concern $\nu$, not doses outside its support.

\begin{proof}[Proof of \Cref{prop:l2_decomp}]
Expand $(d-m)^2=d^2-2md+m^2$ and integrate to obtain $\int(d-m)^2\,d\nu=\int d^2\,d\nu-m^2$.
Since $\mathcal G_{\mathrm{AUC}}=|m|$, this proves the normalized identity.
For a nonnegative unnormalized weight with finite mass $M=\int_{\mathcal Z}w(z)\,dz>0$, use $m=M^{-1}\int dw$ instead; the same expansion gives
\[
\int d^2w=\frac{|\int dw|^2}{M}+\int(d-m)^2w.
\]
Taking $w=1$ gives $M=|\mathcal Z|$ and the unnormalized uniform-weight formula.
\end{proof}

\begin{proof}[Proof of \Cref{prop:detection}]
The AUC characterization follows from $|m|=0$ if and only if $m=0$.
A nonnegative integrable function has zero integral exactly when it is zero almost everywhere, so $\int d^2\,d\nu=0$ exactly when $d=0$ almost everywhere, and $\int(d-m)^2\,d\nu=0$ exactly when $d=m$ almost everywhere.
Since $p_s(1-p_s)>0$, \Cref{prop:fanova_l2} gives the identical zero-gain characterization for FANOVA.
The inequality $(d(z)-d(t))^2\leq2d(z)^2+2d(t)^2$ ensures integrability under $\nu\otimes\nu$.
Fubini's theorem and expansion then give
\[
\frac12\iint(d(z)-d(t))^2\,d\nu(z)\,d\nu(t)
=\int d^2\,d\nu-\left(\int d\,d\nu\right)^2
=\mathcal G_{\mathrm{shape}}.
\]
For subgroup curves, $d(z)-d(t)=\{\theta_L(z)-\theta_L(t)\}-\{\theta_R(z)-\theta_R(t)\}$, which proves the dose-contrast interpretation.
If $m=0$, the decomposition reduces to $\mathcal G_{L^2}=\mathcal G_{\mathrm{shape}}$; this common value is positive precisely when $d$ is not zero almost everywhere.
The FANOVA conclusion follows by multiplication by $p_s(1-p_s)$.
\end{proof}

\textit{Level and shape are distinct directions of heterogeneity.}
The decomposition is the orthogonal decomposition of $d\in L^2(\nu)$ into its constant part $m$ and the centered function $d-m$.
AUC measures only the constant component; its zero-gain set contains every nonzero mean-zero difference curve.
Shape removes every constant difference and detects the remaining variation in additive dose contrasts.
$L^2$ measures the sum of their squared magnitudes, and FANOVA multiplies that sum by the split's balance factor.
These are population detection statements, not finite-sample power or selection guarantees.
In particular, a greedy tree can miss modifiers whose candidate root splits all have zero gain, even if finer joint subgroups have different curves.

\textit{An analytic example family.}
On $\mathcal Z=[0,1]$ under uniform weighting, let
\[
\theta_L(z)=2+2z+\tfrac12\{c+a(2z-1)\},\qquad
\theta_R(z)=2+2z-\tfrac12\{c+a(2z-1)\}.
\]
Then $d(z)=c+a(2z-1)$, $\int_0^1(2z-1)\,dz=0$, and $\int_0^1(2z-1)^2\,dz=1/3$, giving
\begin{equation}\label{eq:detection_example}
\mathcal G_{\mathrm{AUC}}=|c|,\quad
\mathcal G_{\mathrm{shape}}=a^2/3,\quad
\mathcal G_{L^2}=c^2+a^2/3,\quad
\mathcal G_{\mathrm{FANOVA}}=p_s(1-p_s)(c^2+a^2/3).
\end{equation}
For $a=1,c=0$, both child curves increase but cross at $z=1/2$; the main-text example follows, and FANOVA equals $1/12$ when the split is balanced.
For $a=0,c\neq0$, the curves are parallel: every additive dose contrast agrees across children, yet all three primary objectives have positive gain.
When both coefficients are nonzero, level and shape heterogeneity coexist.
These curves can be realized causally by taking a binary subgroup indicator independent of a randomized continuous dose, setting $Y(z)=\theta_V(z)+\epsilon$ with independent mean-zero noise, and assigning positive probability to each subgroup.
A dose randomized uniformly over $[-1,2]$ makes $[0,1]$ an interior comparison interval and satisfies strict overlap there.
Thus the example is compatible with the identification setup and does not rely on confounding or estimation error.

\textit{Support and weighting.}
If a mean-zero $d$ is nonzero on a set of positive measure, both $\{d>0\}$ and $\{d<0\}$ must have positive measure; otherwise its integral could not vanish.
For continuous $d$ on an interval with a full-support comparison measure, this entails a crossing, but a crossing is not sufficient for zero AUC.
For example, the main-text difference $d(z)=2z-1$ has zero mean under $w=1$ but mean $1/3$ under the normalized weight $w(z)=2z$.
Under the latter weight, AUC is $1/3$, shape is $2/9$, and $L^2$ is $1/3$.
Changing weights therefore changes the question being asked, even for the same child curves.
If $w$ vanishes on a region, differences confined to that region are invisible to all of these weighted criteria.
With continuous curves and full-support weights, almost-everywhere equality or constancy extends to pointwise equality or constancy on the comparison interval.
For unnormalized weights of mass $M>0$, normalization divides AUC, $L^2$, shape, and FANOVA by $M$, leaving their zero-gain characterizations unchanged.
If this mass varies across candidate splits, normalization can change their ranking and must be part of the objective specification.

\section{Inference Within Terminal Nodes}
\label{app:honest}

An independent inference sample permits conditional pointwise inference once the oracle local-linear limit, nuisance remainder, and selection scope are specified.
We give a finite-family formulation, appropriate to a fixed finite split grid and fixed depth.

\begin{prop}[Inference within terminal nodes]\label{prop:honest}
Let $\mathcal D_n$ be an independent discovery sample, and let its selected terminal node $\widehat{\mathcal L}$ belong to a fixed finite family $\mathfrak L$ of measurable subgroups with probabilities bounded below by $p_0>0$.
Use an independent inference sample of $n$ i.i.d.\ units, with cross-fitted nuisances, and a deterministic bandwidth $h=h_n$.
For each possible node, the estimation routine uses only inference data; discovery data determine only which node is reported.
Fix an interior dose $z$.
Suppose \Crefrange{ass:consistency}{ass:overlap} hold, and for each $\mathcal L\in\mathfrak L$:
\begin{enumerate}
\item $f_{\mathcal L}(t)=\varpi(t\mid\mathcal L)$ is positive and continuous near $z$, and $\theta_{\mathcal L}$ is twice continuously differentiable there.
The specified conditional variance
$\nu_{\mathcal L}(t)=\Var(\varphi^{\mathcal L}\mid Z=t,\bV\in\mathcal L)$ is continuous at $z$ and satisfies $0<\nu_{\mathcal L}(z)<\infty$.
\item For some $\eta>0$, the conditional $(2+\eta)$-moment of $\varphi^{\mathcal L}-\theta_{\mathcal L}(Z)$ is bounded near $z$.
The kernel is a bounded symmetric compactly supported density or a Gaussian density, and $S=\int K(u)g(u)g(u)^\top du$, $g(u)=(1,u)^\top$, is positive definite.
For a Gaussian kernel, additionally assume the subgroup density, the curve, and this conditional residual moment are bounded over the treatment support, so kernel tails outside the neighborhood of $z$ are negligible.
\item $h\to0$, $nh\to\infty$, $nh^5\to0$, and, writing $\tilde\theta$ for the corresponding oracle smoother,
\[
\max_{\mathcal L\in\mathfrak L}\sqrt{n_{\mathcal L}h}
|\hat\theta(z\mid\mathcal L)-\tilde\theta(z\mid\mathcal L)|=o_p(1).
\]
Here $n_{\mathcal L}$ is the inference-sample node count.
The negligible-remainder condition is an assumption, not a consequence of double robustness alone.
\end{enumerate}
Then, with $T=\int K(u)^2g(u)g(u)^\top du$ and
\[
\sigma_{\mathcal L}^2(z)=\frac{\nu_{\mathcal L}(z)}{f_{\mathcal L}(z)}
e_1^\top S^{-1}TS^{-1}e_1,
\]
the conditional limit has the precise form
\begin{equation}\label{eq:honest_conditional}
\sup_{t\in\mathbb R}\left|
\Prob\left\{\left.
\frac{\sqrt{n_{\widehat{\mathcal L}}h}\{\hat\theta(z\mid\widehat{\mathcal L})-\theta(z\mid\widehat{\mathcal L})\}}
{\sigma_{\widehat{\mathcal L}}(z)}\leq t\,\right|\,\mathcal D_n\right\}-\Phi(t)
\right|\ \xrightarrow{p}\ 0.
\end{equation}
Replacing $\sigma_{\mathcal L}$ by an estimator additionally requires uniform consistency over $\mathfrak L$.
This is pointwise inference for the selected subgroup curve, not simultaneous coverage or inference on the selected tree.
\end{prop}

\begin{proof}
Fix $\mathcal L\in\mathfrak L$ and condition on its inference-sample count $m=n_{\mathcal L}$.
The observations within this deterministic subgroup have its conditional i.i.d.\ law.
Also $n_{\mathcal L}/n\to p_{\mathcal L}>0$, so $mh\to\infty$ and $mh^5\to0$ in probability.
Put $u_i=(Z_i-z)/h$, $\eta_i=\varphi_i^{\mathcal L}-\theta_{\mathcal L}(Z_i)$, and
\[
S_m=\frac1m\sum_{i=1}^m K_h(Z_i-z)g(u_i)g(u_i)^\top,
\qquad
U_m=\frac{\sqrt{mh}}m\sum_{i=1}^m K_h(Z_i-z)g(u_i)\eta_i.
\]
Local density continuity and the kernel moment bounds give
$S_m\to f_{\mathcal L}(z)S$ in probability.
For the Gaussian case the global envelope bounds control the complement of a fixed neighborhood of $z$.
Conditional residual means are zero, and change of variables gives
\[
\Var(U_m)=\int K(u)^2g(u)g(u)^\top
\nu_{\mathcal L}(z+hu)f_{\mathcal L}(z+hu)\,du
\ \longrightarrow\ f_{\mathcal L}(z)\nu_{\mathcal L}(z)T.
\]
The conditional $(2+\eta)$-moment bound and the finite integrals of
$|K(u)|^{2+\eta}\|g(u)\|^{2+\eta}$ make the Lyapunov sum $O((mh)^{-\eta/2})=o(1)$.
The triangular-array central limit theorem therefore yields the corresponding vector normal limit for $U_m$.
Local-linear affine reproduction and twice differentiability give an $O_p(h^2)$ intercept bias; the compact support or Gaussian tail bounds justify the Taylor remainder.
Hence
\[
\sqrt{mh}\{\tilde\theta(z\mid\mathcal L)-\theta_{\mathcal L}(z)\}
=e_1^\top S_m^{-1}U_m+O_p(\sqrt{mh}\,h^2),
\]
with limiting variance $\sigma_{\mathcal L}^2(z)$, since $mh^5\to0$.
The same limit holds with random $m=n_{\mathcal L}$: the conditional arguments apply uniformly when $m/n$ lies in a fixed neighborhood of $p_{\mathcal L}$, an event of probability tending to one.
Slutsky's theorem and the assumed remainder transfer it to $\hat\theta$.

Because $\mathfrak L$ is finite, convergence of each standardized distribution to the continuous standard normal law is uniform over this family and over its argument $t$.
Discovery/inference independence makes the inference law for each fixed subgroup unchanged after conditioning on $\mathcal D_n$; selection of one member therefore preserves this uniform bound, proving \eqref{eq:honest_conditional}.
Uniformly consistent variance estimates give the studentized version by the same argument.
\end{proof}

For data-dependent continuous thresholds with infinitely many possible selected nodes, independence alone does not supply \eqref{eq:honest_conditional}.
An extension requires uniform oracle central limit and remainder bounds over that node class, or an appropriate stabilization argument; it is not claimed here.

\begin{remark}[County-level dependence and interpretation of reported bands]\label{rem:bands}
The proposition above is an i.i.d.\ pointwise result and does not establish coverage for repeated county-year observations.
For county-year analyses, inference must account for within-county dependence and discovery/inference splitting must prevent the same county from contributing information to both phases when independence between phases is invoked.
A county-cluster sandwich or county-level bootstrap can address within-county correlation under additional conditions, but neither automatically accounts for spatial dependence across counties or removes smoothing bias.
\end{remark}

\section{Monte Carlo Simulations}
\label{app:simulations}

We evaluate the method's ability to recover the true effect modifiers and the true decision rules across varying signal strengths and data-generating conditions.

\subsection{Design}

We generate six confounders: $(X_1,\ldots,X_4)^\top\sim\mathcal N_4(0,I_4)$, $X_5 \sim \text{Uniform}\{-2, -1, 0, 1, 2\}$, and $X_6 \sim \text{Uniform}(-3, 3)$.
Ten candidate effect modifiers are generated as $E_1 \sim \text{Bernoulli}(0.5)$, $E_2 \sim \text{Bernoulli}(0.4)$, $E_3 \sim \text{Bernoulli}(0.3)$, $E_4 \sim \text{Bernoulli}(0.2)$, and $E_5,\ldots,E_{10} \sim \text{Bernoulli}(0.25)$.
Only $E_1$ and $E_2$ are true effect modifiers; the rest are null.
We consider two treatment mechanisms: (i) no confounding, $Z \sim \text{TruncNorm}(\mu = 3, \sigma = 5, \text{lb} = 0, \text{ub} = \infty)$, independent of covariates; and (ii) confounding, $Z \sim \text{TruncNorm}(\mu(\bX), \sigma = 5, \text{lb} = 0, \text{ub} = \infty)$ with $\mu(\bX) = 9(-0.8 + 0.1X_1 + 0.1X_2 - 0.1X_3 + 0.2X_4 + 0.1X_5 + 0.1X_6) + 17$.
The outcome follows
\begin{equation}
\label{eq:sim}
Y_i = -10 - \bX_i^\top \boldsymbol{\alpha} + 10 \cdot \tau(E_{1i}, E_{2i}) \cdot Z_i + \varepsilon_i, \quad \varepsilon_i \sim \mathcal{N}(0, \mu_i^2),
\end{equation}
where $\mu_i = -10 - \bX_i^\top \boldsymbol{\alpha} + 10 \cdot \tau(E_{1i}, E_{2i}) \cdot Z_i$ and $\boldsymbol{\alpha} = (2, 2, 3, -1, 2, 2)^\top$ and $\tau(E_1, E_2) = 1 + \beta(0.5 E_1 - 0.8 E_2)$ captures effect modification. Thus, the error standard deviation is $|\mu_i|$.

The parameter $\beta \in [0.01, 1.00]$ controls the strength of heterogeneity: $\beta = 0.01$ corresponds to almost no effect modification, while $\beta = 1.00$ yields maximal heterogeneity with four distinct dose-response curve slopes.

For Settings 1 and 2, we consider sample sizes $n \in \{5{,}000, 10{,}000\}$; Settings 3--5 use $n=5{,}000$.
We vary $\beta$ from $0.05$ to $1.00$ in increments of $0.05$, using $0.01$ in place of $0.00$ as the starting value.
For each configuration, we generate 200 datasets and apply the method with $d_{\max} = 3$ and $n_{\min} = 100$.
Performance is evaluated using precision (the fraction of discovered decision rules and effect modifiers that are true), recall (the fraction of true decision rules and effect modifiers that are discovered), and F1.
Given \eqref{eq:sim}, $E_1$ and $E_2$ modify the treatment slope additively; Setting 3 introduces an interaction between these modifiers.

Three further settings are considered: Setting 3 replaces the modification function with the interactive form $\tau = 1 + \beta(0.5 E_1 E_2 + 0.2 E_2)$, yielding three rather than four distinct dose-response curve groups; Setting 4 lets the treatment depend on the effect modifiers themselves, $\mu(\bX, E_1, E_2) = \mu(\bX) + 0.1E_1 + 0.1E_2$; and Setting 5 adds prognostic terms $20(E_{1i} + 2E_{2i})$ to the outcome model, so that both intercepts and exposure-response slopes vary across these covariates.

\subsection{Results}

\textit{Setting 1: no confounding.}
At $n = 5{,}000$, AUC and $L^2$ attain precision at or above $0.90$ for decision rules from $\beta \geq 0.40$ and recall at or above $0.90$ from $\beta \geq 0.30$, with FANOVA reaching the same thresholds at $\beta \geq 0.45$ and $\beta \geq 0.35$ respectively.
F1 exceeds $0.90$ from $\beta \geq 0.40$ for all three objectives and approaches one for $\beta \geq 0.60$.
At $\beta = 0.01$ all decision-rule metrics remain below $0.05$.
Increasing the sample size to $n = 10{,}000$ improves recovery at weaker signals: F1 exceeds $0.90$ from $\beta \geq 0.30$ for AUC and $L^2$ and from $\beta \geq 0.35$ for FANOVA, and exceeds $0.99$ for $\beta \geq 0.55$.

In Setting 1, the reported F1 scores for effect modifier recovery are at least as high as those for decision rule recovery.
At $n = 5{,}000$, F1 exceeds $0.90$ from $\beta \geq 0.30$ for AUC and $L^2$ and from $\beta \geq 0.35$ for FANOVA; at $n = 10{,}000$ the corresponding thresholds are $\beta \geq 0.25$ and $\beta \geq 0.30$.

\textit{Setting 2: with confounding.}
Performance remains strong under the simulated confounding mechanism, but degrades at weak heterogeneity at both sample sizes.
At $n = 5{,}000$ the largest F1 deficits relative to the unconfounded case are $0.196$ for decision rules and $0.132$ for effect modifiers, both attained by AUC at $\beta = 0.10$.
At $n = 10{,}000$ all three objectives exceed F1 of $0.95$ from $\beta \geq 0.40$ for both targets.

\textit{Comparison across objectives.}
For moderate to strong heterogeneity ($\beta \geq 0.40$) the three objectives perform comparably.
At weaker signal AUC and $L^2$ are often close, while FANOVA reaches some recovery thresholds at higher values of $\beta$; the differences vary by threshold, target, and sample size.
Differences can reflect both the balance factor and the separately fitted parent in the empirical FANOVA criterion (\Cref{eq:empirical_fanova}); attributing the observed performance gap to either component requires an implementation-matched comparison.
The similar performance of AUC and $L^2$ is consistent with the design of \eqref{eq:sim}: the four subgroup curves share an intercept at $Z=0$ and diverge without crossing over positive doses, so their weighted differences do not cancel.
AUC is itself a weighted level comparison; the crossing-curve example in \Cref{app:objective_detection} describes a different setting in which its population gain is zero despite nonzero shape heterogeneity.

\begin{remark}[Behavior under weak heterogeneity]\label{rem:null}
The original simulation summaries at $\beta=0.01$ show low rule-recovery rates but nonzero selection of effect modifiers, consistent with the tendency of a positive-gain rule to split under very weak heterogeneity.
These precision and recall summaries are not false discovery rates, and $\beta=0.01$ is a near-null rather than a null scenario.
The optional minimum-gain rule in \Cref{alg:general} targets asymptotically consistent stopping, but was not used in these simulations and does not by itself provide finite-sample false discovery rate control.
% \falco{}
\end{remark}

\subsection{Figures and Tables}

\begin{figure}
\centering
\includegraphics[width = 1.00\textwidth]{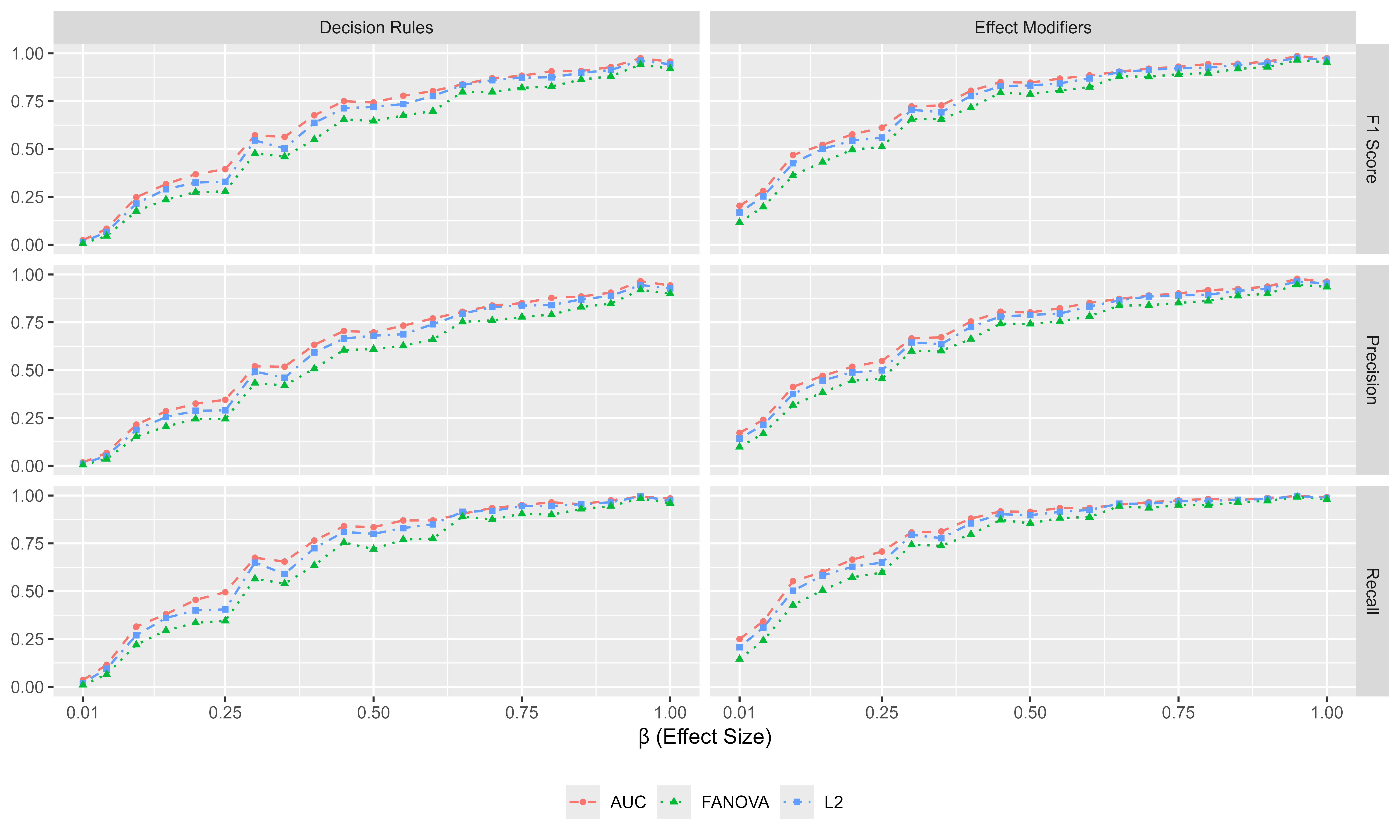}
\caption{Recall, Precision, and F1-score in \textbf{Setting 3} $(n = 5{,}000)$}
\label{fig:setting3}
\end{figure}

\begin{figure}
\centering
\includegraphics[width = 1.00\textwidth]{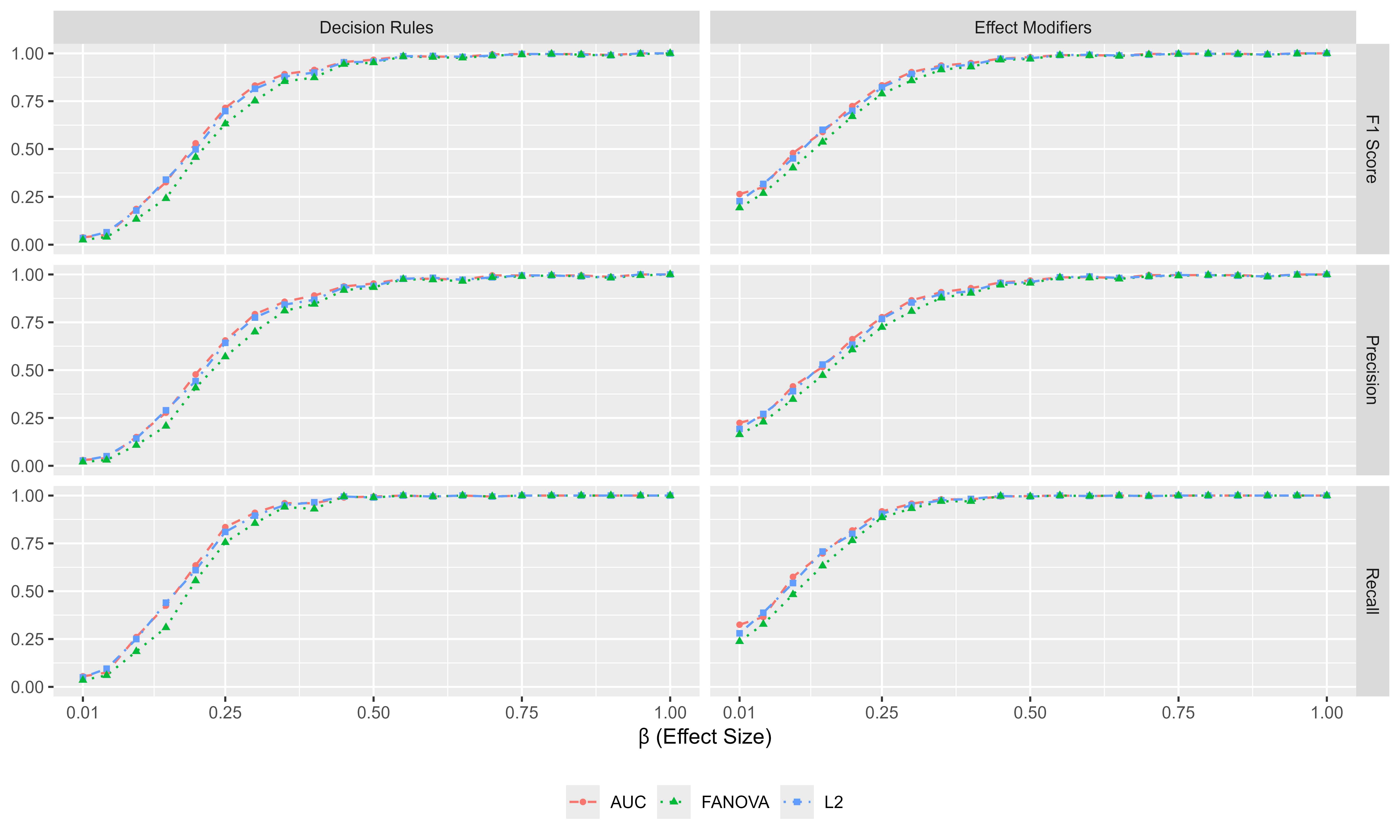}
\caption{Recall, Precision, and F1-score in \textbf{Setting 4} $(n = 5{,}000)$}
\label{fig:setting4}
\end{figure}

\begin{figure}
\centering
\includegraphics[width = 1.00\textwidth]{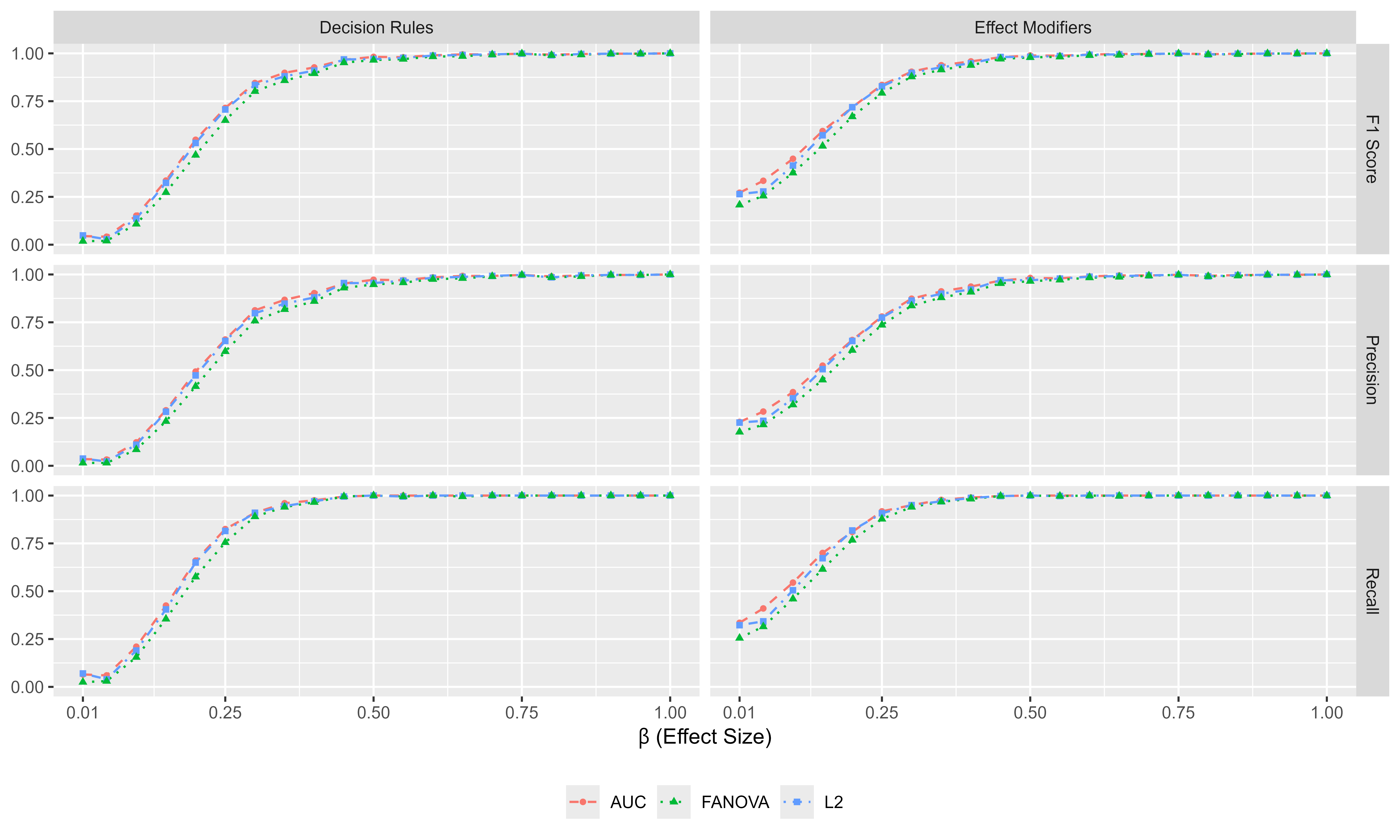}
\caption{Recall, Precision, and F1-score in \textbf{Setting 5} $(n = 5{,}000)$}
\label{fig:setting5}
\end{figure}

\Cref{fig:setting1_n5000,fig:setting1_n10000,fig:setting2_n5000,fig:setting2_n10000} plot precision, recall, and F1 against $\beta$ for Settings 1 and 2 at both sample sizes, and the corresponding numerical values are given in \Crefrange{table:setting1_n5000_dr}{table:setting2_n10000_em}.
\Cref{tab:sim_settings} summarizes all five settings, and \Cref{fig:setting3,fig:setting4,fig:setting5} report Settings 3 to 5.

\ifthenelse{\boolean{showfigures}}{
\begin{figure}[htbp]
\centering
\includegraphics[width=\textwidth]{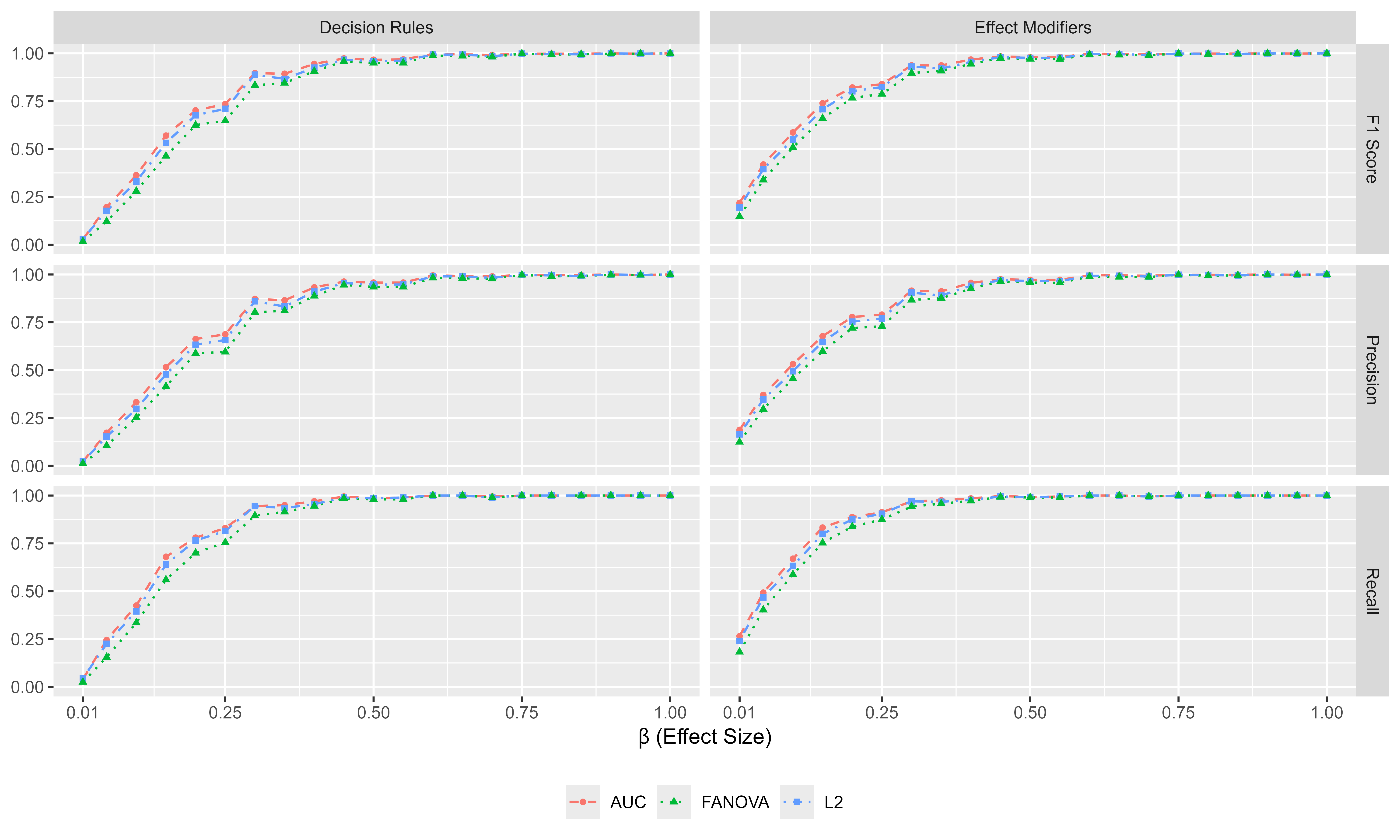}
\caption{Recall, precision, and F1 in \textbf{Setting 1} $(n = 5{,}000)$.}
\label{fig:setting1_n5000}
\end{figure}

\begin{figure}[htbp]
\centering
\includegraphics[width=\textwidth]{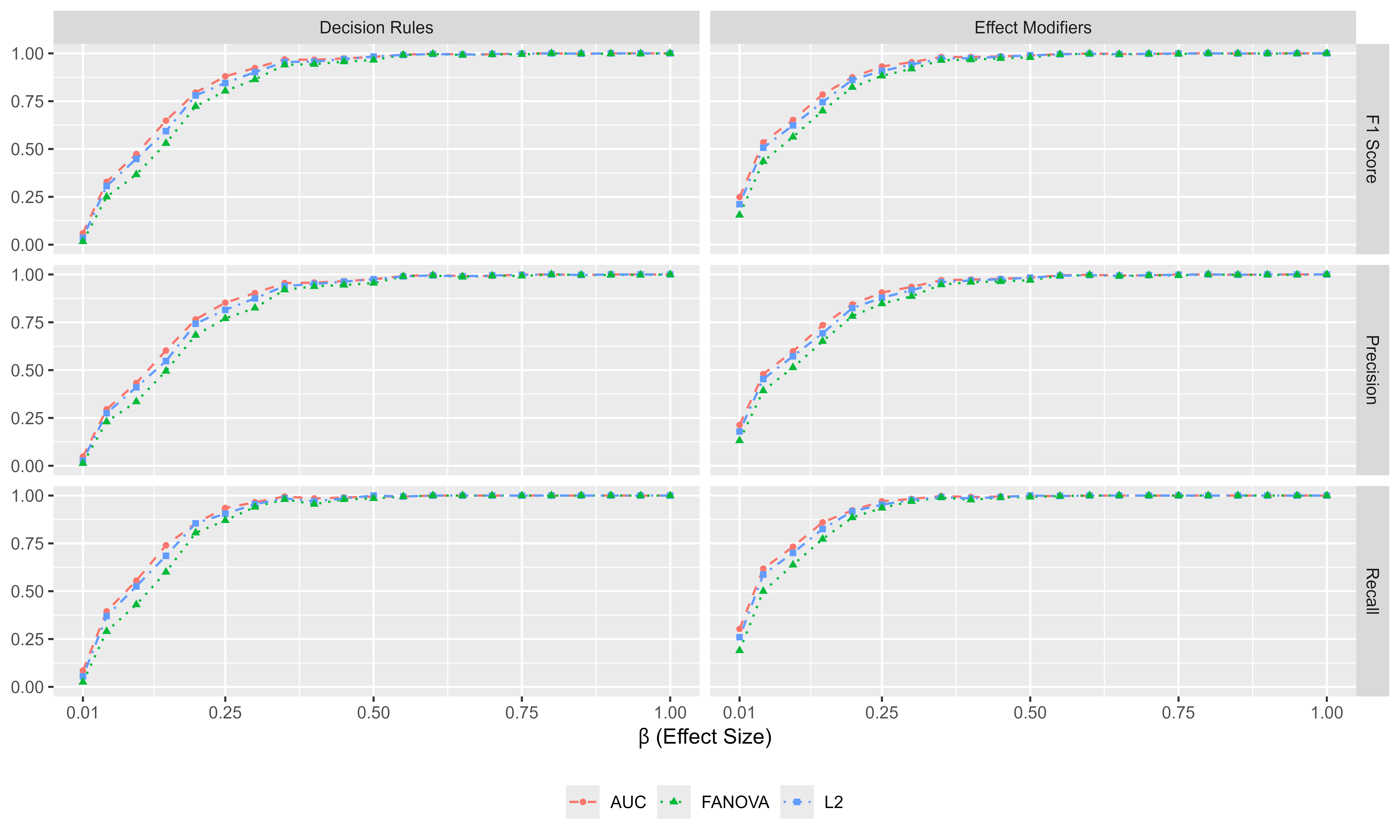}
\caption{Recall, precision, and F1 in \textbf{Setting 1} $(n = 10{,}000)$.}
\label{fig:setting1_n10000}
\end{figure}

\begin{figure}[htbp]
\centering
\includegraphics[width=\textwidth]{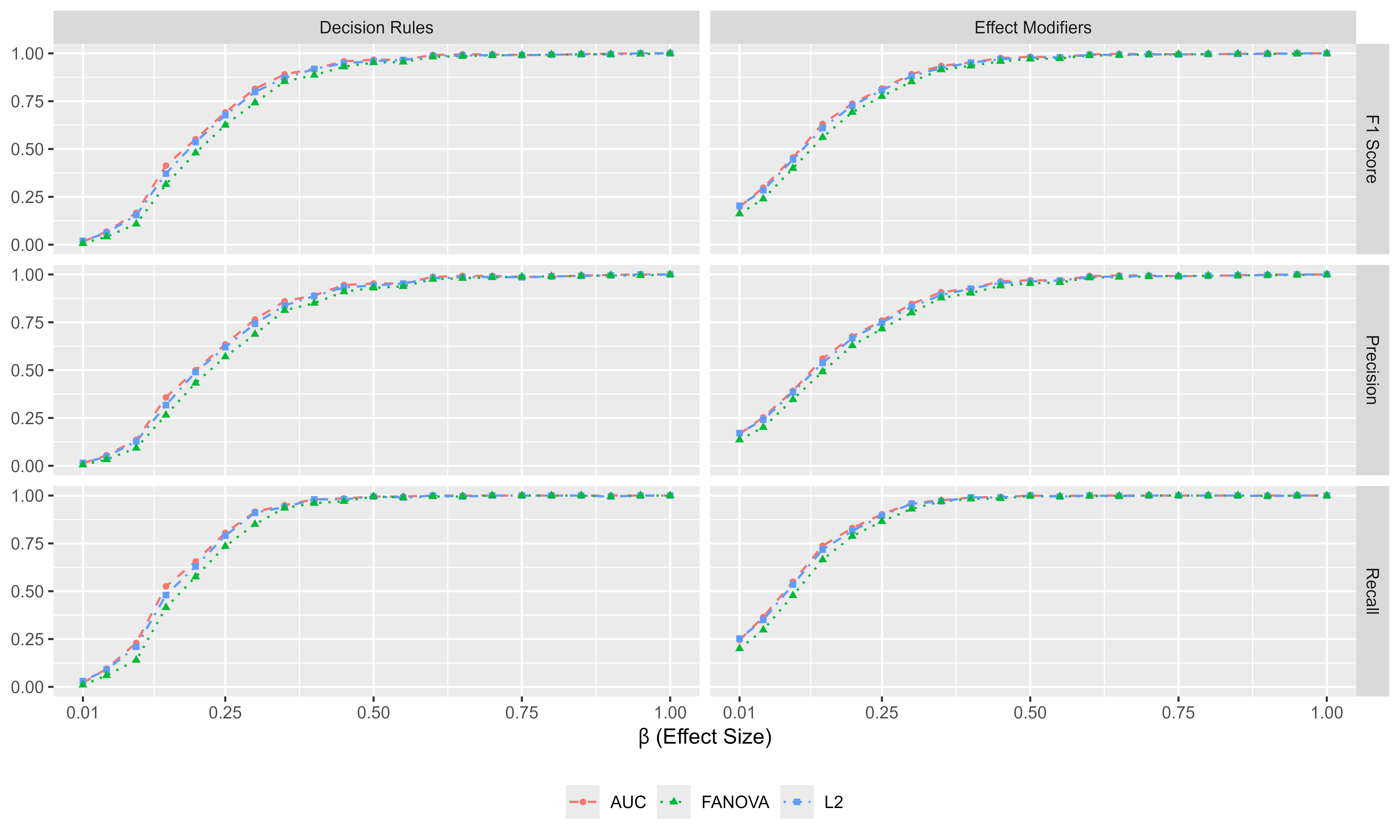}
\caption{Recall, precision, and F1 in \textbf{Setting 2} $(n = 5{,}000)$.}
\label{fig:setting2_n5000}
\end{figure}

\begin{figure}[htbp]
\centering
\includegraphics[width=\textwidth]{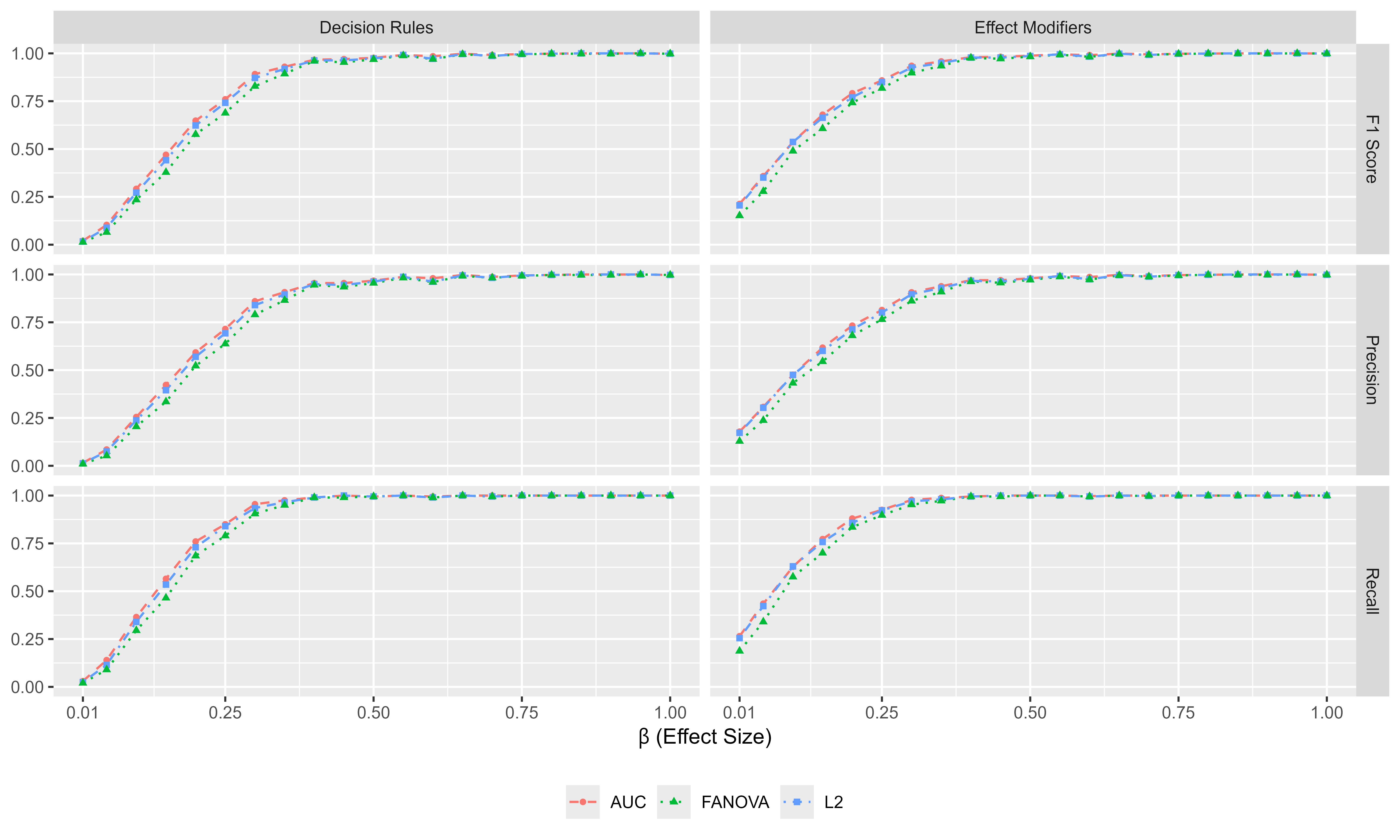}
\caption{Recall, precision, and F1 in \textbf{Setting 2} $(n = 10{,}000)$.}
\label{fig:setting2_n10000}
\end{figure}
}{}

\begin{table}[h!]
\centering
\footnotesize
\begin{tabular}{c|ccc|ccc|ccc}
\hline
\multirow{2}{*}{$\beta$}
& \multicolumn{3}{c}{\textbf{AUC}}
& \multicolumn{3}{c}{\textbf{FANOVA}}
& \multicolumn{3}{c}{$\mathbf{L^2}$} \\
\cline{2-10}
& F1 & Recall & Precision
& F1 & Recall & Precision
& F1 & Recall & Precision \\
\hline
0.01 & 0.028 & 0.040 & 0.022 & 0.017 & 0.025 & 0.013 & 0.030 & 0.045 & 0.022 \\
0.05 & 0.197 & 0.245 & 0.172 & 0.122 & 0.155 & 0.105 & 0.177 & 0.225 & 0.152 \\
0.10 & 0.363 & 0.425 & 0.333 & 0.280 & 0.335 & 0.253 & 0.330 & 0.395 & 0.297 \\
0.15 & 0.570 & 0.680 & 0.515 & 0.463 & 0.560 & 0.415 & 0.532 & 0.640 & 0.477 \\
0.20 & 0.702 & 0.780 & 0.662 & 0.625 & 0.700 & 0.588 & 0.677 & 0.765 & 0.632 \\
0.25 & 0.735 & 0.830 & 0.688 & 0.648 & 0.755 & 0.595 & 0.710 & 0.815 & 0.657 \\
0.30 & 0.897 & 0.945 & 0.873 & 0.833 & 0.895 & 0.802 & 0.888 & 0.945 & 0.860 \\
0.35 & 0.893 & 0.950 & 0.865 & 0.845 & 0.915 & 0.810 & 0.867 & 0.935 & 0.833 \\
0.40 & 0.945 & 0.970 & 0.932 & 0.907 & 0.945 & 0.887 & 0.925 & 0.955 & 0.910 \\
0.45 & 0.973 & 0.995 & 0.963 & 0.958 & 0.985 & 0.945 & 0.967 & 0.990 & 0.955 \\
0.50 & 0.967 & 0.985 & 0.958 & 0.950 & 0.980 & 0.935 & 0.960 & 0.985 & 0.948 \\
0.55 & 0.968 & 0.990 & 0.958 & 0.950 & 0.980 & 0.935 & 0.962 & 0.990 & 0.948 \\
0.60 & 0.997 & 1.000 & 0.995 & 0.988 & 1.000 & 0.983 & 0.993 & 1.000 & 0.990 \\
0.65 & 0.995 & 1.000 & 0.993 & 0.987 & 1.000 & 0.980 & 0.993 & 1.000 & 0.990 \\
0.70 & 0.992 & 0.995 & 0.990 & 0.982 & 0.990 & 0.978 & 0.985 & 0.990 & 0.983 \\
0.75 & 0.998 & 1.000 & 0.998 & 0.998 & 1.000 & 0.998 & 0.998 & 1.000 & 0.998 \\
0.80 & 0.998 & 1.000 & 0.998 & 0.993 & 1.000 & 0.990 & 0.997 & 1.000 & 0.995 \\
0.85 & 0.998 & 1.000 & 0.998 & 0.995 & 1.000 & 0.993 & 0.995 & 1.000 & 0.993 \\
0.90 & 1.000 & 1.000 & 1.000 & 0.998 & 1.000 & 0.998 & 1.000 & 1.000 & 1.000 \\
0.95 & 0.998 & 1.000 & 0.998 & 0.998 & 1.000 & 0.998 & 0.998 & 1.000 & 0.998 \\
1.00 & 1.000 & 1.000 & 1.000 & 1.000 & 1.000 & 1.000 & 1.000 & 1.000 & 1.000 \\
\hline
\end{tabular}
\caption{Recall, Precision, and F1-score for \textbf{Decision Rules} in \textbf{Setting 1} ($n = 5{,}000$)}
\label{table:setting1_n5000_dr}
\end{table}

\begin{table}[h!]
\centering
\footnotesize
\begin{tabular}{c|ccc|ccc|ccc}
\hline
\multirow{2}{*}{$\beta$}
& \multicolumn{3}{c}{\textbf{AUC}}
& \multicolumn{3}{c}{\textbf{FANOVA}}
& \multicolumn{3}{c}{$\mathbf{L^2}$} \\
\cline{2-10}
& F1 & Recall & Precision
& F1 & Recall & Precision
& F1 & Recall & Precision \\
\hline
0.01 & 0.218 & 0.265 & 0.188 & 0.147 & 0.182 & 0.124 & 0.195 & 0.240 & 0.164 \\
0.05 & 0.419 & 0.492 & 0.370 & 0.339 & 0.403 & 0.296 & 0.395 & 0.468 & 0.347 \\
0.10 & 0.587 & 0.670 & 0.531 & 0.508 & 0.588 & 0.456 & 0.549 & 0.632 & 0.494 \\
0.15 & 0.740 & 0.833 & 0.677 & 0.660 & 0.752 & 0.598 & 0.709 & 0.800 & 0.647 \\
0.20 & 0.822 & 0.887 & 0.777 & 0.767 & 0.838 & 0.720 & 0.802 & 0.875 & 0.753 \\
0.25 & 0.839 & 0.912 & 0.790 & 0.787 & 0.875 & 0.729 & 0.824 & 0.905 & 0.770 \\
0.30 & 0.937 & 0.970 & 0.915 & 0.897 & 0.943 & 0.867 & 0.931 & 0.970 & 0.905 \\
0.35 & 0.937 & 0.975 & 0.912 & 0.908 & 0.958 & 0.876 & 0.921 & 0.968 & 0.890 \\
0.40 & 0.968 & 0.985 & 0.956 & 0.945 & 0.973 & 0.926 & 0.956 & 0.978 & 0.941 \\
0.45 & 0.984 & 0.998 & 0.975 & 0.975 & 0.993 & 0.963 & 0.980 & 0.995 & 0.970 \\
0.50 & 0.980 & 0.993 & 0.972 & 0.971 & 0.990 & 0.958 & 0.977 & 0.993 & 0.966 \\
0.55 & 0.981 & 0.995 & 0.972 & 0.970 & 0.990 & 0.957 & 0.977 & 0.995 & 0.965 \\
0.60 & 0.998 & 1.000 & 0.997 & 0.993 & 1.000 & 0.988 & 0.996 & 1.000 & 0.993 \\
0.65 & 0.997 & 1.000 & 0.995 & 0.992 & 1.000 & 0.987 & 0.996 & 1.000 & 0.993 \\
0.70 & 0.995 & 0.998 & 0.993 & 0.990 & 0.995 & 0.987 & 0.992 & 0.995 & 0.989 \\
0.75 & 0.999 & 1.000 & 0.998 & 0.999 & 1.000 & 0.998 & 0.999 & 1.000 & 0.998 \\
0.80 & 0.999 & 1.000 & 0.998 & 0.996 & 1.000 & 0.993 & 0.998 & 1.000 & 0.997 \\
0.85 & 0.999 & 1.000 & 0.998 & 0.997 & 1.000 & 0.995 & 0.997 & 1.000 & 0.995 \\
0.90 & 1.000 & 1.000 & 1.000 & 0.999 & 1.000 & 0.998 & 1.000 & 1.000 & 1.000 \\
0.95 & 0.999 & 1.000 & 0.998 & 0.999 & 1.000 & 0.998 & 0.999 & 1.000 & 0.998 \\
1.00 & 1.000 & 1.000 & 1.000 & 1.000 & 1.000 & 1.000 & 1.000 & 1.000 & 1.000 \\
\hline
\end{tabular}
\caption{Recall, Precision, and F1-score for \textbf{Effect Modifiers} in \textbf{Setting 1} ($n = 5{,}000$)}
\label{table:setting1_n5000_em}
\end{table}

\begin{table}[h!]
\centering
\footnotesize
\begin{tabular}{c|ccc|ccc|ccc}
\hline
\multirow{2}{*}{$\beta$}
& \multicolumn{3}{c}{\textbf{AUC}}
& \multicolumn{3}{c}{\textbf{FANOVA}}
& \multicolumn{3}{c}{$\mathbf{L^2}$} \\
\cline{2-10}
& F1 & Recall & Precision
& F1 & Recall & Precision
& F1 & Recall & Precision \\
\hline
0.01 & 0.060 & 0.085 & 0.048 & 0.017 & 0.025 & 0.013 & 0.037 & 0.055 & 0.028 \\
0.05 & 0.328 & 0.395 & 0.295 & 0.250 & 0.290 & 0.230 & 0.307 & 0.370 & 0.275 \\
0.10 & 0.473 & 0.555 & 0.432 & 0.367 & 0.430 & 0.335 & 0.448 & 0.525 & 0.410 \\
0.15 & 0.648 & 0.740 & 0.603 & 0.530 & 0.600 & 0.495 & 0.593 & 0.685 & 0.547 \\
0.20 & 0.795 & 0.855 & 0.765 & 0.723 & 0.805 & 0.682 & 0.780 & 0.855 & 0.743 \\
0.25 & 0.880 & 0.935 & 0.853 & 0.803 & 0.870 & 0.770 & 0.845 & 0.905 & 0.815 \\
0.30 & 0.923 & 0.965 & 0.902 & 0.863 & 0.940 & 0.825 & 0.902 & 0.955 & 0.875 \\
0.35 & 0.968 & 0.995 & 0.955 & 0.940 & 0.980 & 0.920 & 0.955 & 0.985 & 0.940 \\
0.40 & 0.967 & 0.985 & 0.958 & 0.943 & 0.955 & 0.938 & 0.957 & 0.970 & 0.950 \\
0.45 & 0.973 & 0.990 & 0.965 & 0.957 & 0.980 & 0.945 & 0.970 & 0.985 & 0.963 \\
0.50 & 0.982 & 0.995 & 0.975 & 0.965 & 0.985 & 0.955 & 0.983 & 1.000 & 0.975 \\
0.55 & 0.993 & 0.995 & 0.993 & 0.992 & 0.995 & 0.990 & 0.992 & 0.995 & 0.990 \\
0.60 & 0.998 & 1.000 & 0.998 & 0.997 & 1.000 & 0.995 & 0.997 & 1.000 & 0.995 \\
0.65 & 0.995 & 1.000 & 0.993 & 0.992 & 1.000 & 0.988 & 0.993 & 1.000 & 0.990 \\
0.70 & 0.995 & 1.000 & 0.993 & 0.995 & 1.000 & 0.993 & 0.997 & 1.000 & 0.995 \\
0.75 & 1.000 & 1.000 & 1.000 & 0.995 & 1.000 & 0.993 & 0.998 & 1.000 & 0.998 \\
0.80 & 1.000 & 1.000 & 1.000 & 1.000 & 1.000 & 1.000 & 1.000 & 1.000 & 1.000 \\
0.85 & 0.998 & 1.000 & 0.998 & 0.998 & 1.000 & 0.998 & 0.998 & 1.000 & 0.998 \\
0.90 & 1.000 & 1.000 & 1.000 & 0.998 & 1.000 & 0.998 & 1.000 & 1.000 & 1.000 \\
0.95 & 1.000 & 1.000 & 1.000 & 0.998 & 1.000 & 0.998 & 1.000 & 1.000 & 1.000 \\
1.00 & 1.000 & 1.000 & 1.000 & 1.000 & 1.000 & 1.000 & 1.000 & 1.000 & 1.000 \\
\hline
\end{tabular}
\caption{Recall, Precision, and F1-score for \textbf{Decision Rules} in \textbf{Setting 1} $(n = 10{,}000)$}
\label{table:setting1_n10000_dr}
\end{table}

\begin{table}[h!]
\centering
\footnotesize
\begin{tabular}{c|ccc|ccc|ccc}
\hline
\multirow{2}{*}{$\beta$}
& \multicolumn{3}{c}{\textbf{AUC}}
& \multicolumn{3}{c}{\textbf{FANOVA}}
& \multicolumn{3}{c}{$\mathbf{L^2}$} \\
\cline{2-10}
& F1 & Recall & Precision
& F1 & Recall & Precision
& F1 & Recall & Precision \\
\hline
0.01 & 0.249 & 0.302 & 0.213 & 0.155 & 0.190 & 0.132 & 0.211 & 0.260 & 0.179 \\
0.05 & 0.534 & 0.618 & 0.479 & 0.435 & 0.500 & 0.393 & 0.507 & 0.588 & 0.453 \\
0.10 & 0.652 & 0.733 & 0.598 & 0.563 & 0.637 & 0.513 & 0.624 & 0.700 & 0.573 \\
0.15 & 0.785 & 0.860 & 0.735 & 0.700 & 0.772 & 0.651 & 0.746 & 0.825 & 0.693 \\
0.20 & 0.875 & 0.922 & 0.843 & 0.823 & 0.885 & 0.782 & 0.862 & 0.917 & 0.824 \\
0.25 & 0.931 & 0.970 & 0.906 & 0.882 & 0.935 & 0.848 & 0.908 & 0.953 & 0.878 \\
0.30 & 0.954 & 0.983 & 0.935 & 0.919 & 0.970 & 0.886 & 0.942 & 0.978 & 0.917 \\
0.35 & 0.981 & 0.998 & 0.970 & 0.964 & 0.990 & 0.947 & 0.973 & 0.993 & 0.960 \\
0.40 & 0.981 & 0.993 & 0.973 & 0.967 & 0.978 & 0.960 & 0.975 & 0.985 & 0.968 \\
0.45 & 0.985 & 0.995 & 0.978 & 0.974 & 0.990 & 0.963 & 0.982 & 0.993 & 0.975 \\
0.50 & 0.989 & 0.998 & 0.983 & 0.979 & 0.993 & 0.970 & 0.990 & 1.000 & 0.983 \\
0.55 & 0.997 & 0.998 & 0.996 & 0.995 & 0.998 & 0.993 & 0.995 & 0.998 & 0.993 \\
0.60 & 0.999 & 1.000 & 0.998 & 0.998 & 1.000 & 0.997 & 0.998 & 1.000 & 0.997 \\
0.65 & 0.997 & 1.000 & 0.995 & 0.995 & 1.000 & 0.992 & 0.996 & 1.000 & 0.993 \\
0.70 & 0.997 & 1.000 & 0.995 & 0.997 & 1.000 & 0.995 & 0.998 & 1.000 & 0.997 \\
0.75 & 1.000 & 1.000 & 1.000 & 0.997 & 1.000 & 0.995 & 0.999 & 1.000 & 0.998 \\
0.80 & 1.000 & 1.000 & 1.000 & 1.000 & 1.000 & 1.000 & 1.000 & 1.000 & 1.000 \\
0.85 & 0.999 & 1.000 & 0.998 & 0.999 & 1.000 & 0.998 & 0.999 & 1.000 & 0.998 \\
0.90 & 1.000 & 1.000 & 1.000 & 0.999 & 1.000 & 0.998 & 1.000 & 1.000 & 1.000 \\
0.95 & 1.000 & 1.000 & 1.000 & 0.999 & 1.000 & 0.998 & 1.000 & 1.000 & 1.000 \\
1.00 & 1.000 & 1.000 & 1.000 & 1.000 & 1.000 & 1.000 & 1.000 & 1.000 & 1.000 \\
\hline
\end{tabular}
\caption{Recall, Precision, and F1-score for \textbf{Effect Modifiers} in \textbf{Setting 1} $(n = 10{,}000)$}
\label{table:setting1_n10000_em}
\end{table}

\begin{table}[h!]
\centering
\footnotesize
\begin{tabular}{c|ccc|ccc|ccc}
\hline
\multirow{2}{*}{$\beta$}
& \multicolumn{3}{c}{\textbf{AUC}}
& \multicolumn{3}{c}{\textbf{FANOVA}}
& \multicolumn{3}{c}{$\mathbf{L^2}$} \\
\cline{2-10}
& F1 & Recall & Precision
& F1 & Recall & Precision
& F1 & Recall & Precision \\
\hline
0.01 & 0.015 & 0.020 & 0.013 & 0.007 & 0.010 & 0.005 & 0.020 & 0.030 & 0.015 \\
0.05 & 0.068 & 0.095 & 0.055 & 0.042 & 0.060 & 0.033 & 0.062 & 0.090 & 0.048 \\
0.10 & 0.167 & 0.230 & 0.135 & 0.108 & 0.140 & 0.092 & 0.155 & 0.210 & 0.128 \\
0.15 & 0.413 & 0.525 & 0.357 & 0.315 & 0.415 & 0.265 & 0.372 & 0.480 & 0.318 \\
0.20 & 0.552 & 0.655 & 0.500 & 0.480 & 0.575 & 0.432 & 0.537 & 0.630 & 0.490 \\
0.25 & 0.692 & 0.805 & 0.635 & 0.625 & 0.735 & 0.570 & 0.677 & 0.790 & 0.620 \\
0.30 & 0.815 & 0.915 & 0.765 & 0.742 & 0.850 & 0.688 & 0.798 & 0.910 & 0.743 \\
0.35 & 0.890 & 0.950 & 0.860 & 0.853 & 0.935 & 0.812 & 0.872 & 0.940 & 0.838 \\
0.40 & 0.920 & 0.980 & 0.890 & 0.887 & 0.960 & 0.850 & 0.918 & 0.980 & 0.887 \\
0.45 & 0.958 & 0.985 & 0.945 & 0.930 & 0.970 & 0.910 & 0.948 & 0.980 & 0.932 \\
0.50 & 0.967 & 0.995 & 0.953 & 0.952 & 0.995 & 0.930 & 0.960 & 0.995 & 0.943 \\
0.55 & 0.967 & 0.995 & 0.953 & 0.955 & 0.990 & 0.938 & 0.965 & 0.990 & 0.953 \\
0.60 & 0.992 & 1.000 & 0.988 & 0.982 & 0.995 & 0.975 & 0.987 & 1.000 & 0.980 \\
0.65 & 0.995 & 1.000 & 0.993 & 0.985 & 0.995 & 0.980 & 0.988 & 0.995 & 0.985 \\
0.70 & 0.995 & 1.000 & 0.993 & 0.990 & 1.000 & 0.985 & 0.992 & 1.000 & 0.988 \\
0.75 & 0.992 & 1.000 & 0.988 & 0.992 & 1.000 & 0.988 & 0.990 & 1.000 & 0.985 \\
0.80 & 0.992 & 1.000 & 0.988 & 0.993 & 1.000 & 0.990 & 0.993 & 1.000 & 0.990 \\
0.85 & 0.997 & 1.000 & 0.995 & 0.995 & 1.000 & 0.993 & 0.995 & 1.000 & 0.993 \\
0.90 & 0.998 & 1.000 & 0.998 & 0.995 & 0.995 & 0.995 & 0.995 & 0.995 & 0.995 \\
0.95 & 1.000 & 1.000 & 1.000 & 0.997 & 1.000 & 0.995 & 1.000 & 1.000 & 1.000 \\
1.00 & 1.000 & 1.000 & 1.000 & 1.000 & 1.000 & 1.000 & 1.000 & 1.000 & 1.000 \\
\hline
\end{tabular}
\caption{Recall, Precision, and F1-score for \textbf{Decision Rules} in \textbf{Setting 2} $(n = 5{,}000)$}
\label{table:setting2_n5000_dr}
\end{table}

\begin{table}[h!]
\centering
\footnotesize
\begin{tabular}{c|ccc|ccc|ccc}
\hline
\multirow{2}{*}{$\beta$}
& \multicolumn{3}{c}{\textbf{AUC}}
& \multicolumn{3}{c}{\textbf{FANOVA}}
& \multicolumn{3}{c}{$\mathbf{L^2}$} \\
\cline{2-10}
& F1 & Recall & Precision
& F1 & Recall & Precision
& F1 & Recall & Precision \\
\hline
0.01 & 0.198 & 0.245 & 0.166 & 0.162 & 0.200 & 0.136 & 0.203 & 0.253 & 0.171 \\
0.05 & 0.298 & 0.365 & 0.253 & 0.239 & 0.297 & 0.201 & 0.284 & 0.350 & 0.241 \\
0.10 & 0.455 & 0.550 & 0.392 & 0.399 & 0.477 & 0.347 & 0.445 & 0.535 & 0.385 \\
0.15 & 0.631 & 0.738 & 0.560 & 0.561 & 0.665 & 0.492 & 0.610 & 0.718 & 0.537 \\
0.20 & 0.738 & 0.830 & 0.676 & 0.692 & 0.787 & 0.628 & 0.727 & 0.815 & 0.667 \\
0.25 & 0.817 & 0.902 & 0.759 & 0.775 & 0.865 & 0.716 & 0.807 & 0.895 & 0.749 \\
0.30 & 0.891 & 0.960 & 0.846 & 0.852 & 0.930 & 0.800 & 0.881 & 0.958 & 0.831 \\
0.35 & 0.934 & 0.975 & 0.907 & 0.914 & 0.970 & 0.877 & 0.923 & 0.970 & 0.892 \\
0.40 & 0.952 & 0.990 & 0.927 & 0.934 & 0.983 & 0.902 & 0.952 & 0.990 & 0.926 \\
0.45 & 0.976 & 0.993 & 0.964 & 0.959 & 0.985 & 0.941 & 0.970 & 0.990 & 0.956 \\
0.50 & 0.982 & 1.000 & 0.970 & 0.971 & 0.998 & 0.953 & 0.978 & 1.000 & 0.963 \\
0.55 & 0.980 & 0.998 & 0.968 & 0.973 & 0.995 & 0.958 & 0.979 & 0.995 & 0.968 \\
0.60 & 0.995 & 1.000 & 0.992 & 0.989 & 0.998 & 0.983 & 0.992 & 1.000 & 0.987 \\
0.65 & 0.997 & 1.000 & 0.995 & 0.991 & 0.998 & 0.987 & 0.993 & 0.998 & 0.990 \\
0.70 & 0.997 & 1.000 & 0.995 & 0.994 & 1.000 & 0.990 & 0.995 & 1.000 & 0.992 \\
0.75 & 0.995 & 1.000 & 0.992 & 0.995 & 1.000 & 0.992 & 0.994 & 1.000 & 0.990 \\
0.80 & 0.995 & 1.000 & 0.992 & 0.996 & 1.000 & 0.993 & 0.996 & 1.000 & 0.993 \\
0.85 & 0.998 & 1.000 & 0.997 & 0.997 & 1.000 & 0.995 & 0.997 & 1.000 & 0.995 \\
0.90 & 0.999 & 1.000 & 0.998 & 0.997 & 0.998 & 0.997 & 0.997 & 0.998 & 0.997 \\
0.95 & 1.000 & 1.000 & 1.000 & 0.998 & 1.000 & 0.997 & 1.000 & 1.000 & 1.000 \\
1.00 & 1.000 & 1.000 & 1.000 & 1.000 & 1.000 & 1.000 & 1.000 & 1.000 & 1.000 \\
\hline
\end{tabular}
\caption{Recall, Precision, and F1-score for \textbf{Effect Modifiers} in \textbf{Setting 2} $(n = 5{,}000)$}
\label{table:setting2_n5000_em}
\end{table}

\begin{table}[h!]
\centering
\footnotesize
\begin{tabular}{c|ccc|ccc|ccc}
\hline
\multirow{2}{*}{$\beta$}
& \multicolumn{3}{c}{\textbf{AUC}}
& \multicolumn{3}{c}{\textbf{FANOVA}}
& \multicolumn{3}{c}{$\mathbf{L^2}$} \\
\cline{2-10}
& F1 & Recall & Precision
& F1 & Recall & Precision
& F1 & Recall & Precision \\
\hline
0.01 & 0.020 & 0.030 & 0.015 & 0.013 & 0.020 & 0.010 & 0.017 & 0.025 & 0.013 \\
0.05 & 0.103 & 0.140 & 0.085 & 0.065 & 0.090 & 0.052 & 0.088 & 0.115 & 0.075 \\
0.10 & 0.292 & 0.365 & 0.255 & 0.235 & 0.295 & 0.205 & 0.272 & 0.340 & 0.237 \\
0.15 & 0.470 & 0.565 & 0.422 & 0.378 & 0.465 & 0.335 & 0.442 & 0.535 & 0.395 \\
0.20 & 0.648 & 0.760 & 0.593 & 0.577 & 0.685 & 0.522 & 0.623 & 0.730 & 0.570 \\
0.25 & 0.760 & 0.850 & 0.715 & 0.688 & 0.790 & 0.637 & 0.742 & 0.840 & 0.693 \\
0.30 & 0.892 & 0.955 & 0.860 & 0.828 & 0.905 & 0.790 & 0.872 & 0.935 & 0.840 \\
0.35 & 0.930 & 0.975 & 0.907 & 0.893 & 0.950 & 0.865 & 0.918 & 0.965 & 0.895 \\
0.40 & 0.967 & 0.990 & 0.955 & 0.960 & 0.990 & 0.945 & 0.963 & 0.990 & 0.950 \\
0.45 & 0.970 & 1.000 & 0.955 & 0.953 & 0.990 & 0.935 & 0.963 & 1.000 & 0.945 \\
0.50 & 0.977 & 0.995 & 0.968 & 0.968 & 0.995 & 0.955 & 0.973 & 0.995 & 0.963 \\
0.55 & 0.992 & 1.000 & 0.988 & 0.988 & 1.000 & 0.983 & 0.992 & 1.000 & 0.988 \\
0.60 & 0.985 & 0.995 & 0.980 & 0.970 & 0.990 & 0.960 & 0.972 & 0.990 & 0.963 \\
0.65 & 0.998 & 1.000 & 0.998 & 0.995 & 1.000 & 0.993 & 0.997 & 1.000 & 0.995 \\
0.70 & 0.992 & 1.000 & 0.988 & 0.987 & 0.995 & 0.983 & 0.987 & 0.995 & 0.983 \\
0.75 & 0.997 & 1.000 & 0.995 & 0.995 & 1.000 & 0.993 & 0.997 & 1.000 & 0.995 \\
0.80 & 0.998 & 1.000 & 0.998 & 0.998 & 1.000 & 0.998 & 0.998 & 1.000 & 0.998 \\
0.85 & 1.000 & 1.000 & 1.000 & 0.998 & 1.000 & 0.998 & 1.000 & 1.000 & 1.000 \\
0.90 & 1.000 & 1.000 & 1.000 & 0.998 & 1.000 & 0.998 & 1.000 & 1.000 & 1.000 \\
0.95 & 1.000 & 1.000 & 1.000 & 1.000 & 1.000 & 1.000 & 1.000 & 1.000 & 1.000 \\
1.00 & 0.998 & 1.000 & 0.998 & 0.998 & 1.000 & 0.998 & 0.998 & 1.000 & 0.998 \\
\hline
\end{tabular}
\caption{Recall, Precision, and F1-score for \textbf{Decision Rules} in \textbf{Setting 2} $(n = 10{,}000)$}
\label{table:setting2_n10000_dr}
\end{table}

\begin{table}[h!]
\centering
\footnotesize
\begin{tabular}{c|ccc|ccc|ccc}
\hline
\multirow{2}{*}{$\beta$}
& \multicolumn{3}{c}{\textbf{AUC}}
& \multicolumn{3}{c}{\textbf{FANOVA}}
& \multicolumn{3}{c}{$\mathbf{L^2}$} \\
\cline{2-10}
& F1 & Recall & Precision
& F1 & Recall & Precision
& F1 & Recall & Precision \\
\hline
0.01 & 0.213 & 0.265 & 0.178 & 0.152 & 0.188 & 0.128 & 0.205 & 0.255 & 0.172 \\
0.05 & 0.358 & 0.435 & 0.307 & 0.279 & 0.340 & 0.237 & 0.351 & 0.422 & 0.303 \\
0.10 & 0.536 & 0.627 & 0.476 & 0.489 & 0.575 & 0.432 & 0.537 & 0.630 & 0.475 \\
0.15 & 0.679 & 0.772 & 0.618 & 0.607 & 0.700 & 0.545 & 0.663 & 0.757 & 0.601 \\
0.20 & 0.791 & 0.880 & 0.733 & 0.743 & 0.835 & 0.681 & 0.770 & 0.858 & 0.713 \\
0.25 & 0.859 & 0.925 & 0.814 & 0.818 & 0.897 & 0.765 & 0.850 & 0.922 & 0.801 \\
0.30 & 0.935 & 0.978 & 0.907 & 0.898 & 0.953 & 0.862 & 0.926 & 0.970 & 0.897 \\
0.35 & 0.958 & 0.988 & 0.938 & 0.934 & 0.973 & 0.909 & 0.950 & 0.980 & 0.929 \\
0.40 & 0.980 & 0.995 & 0.970 & 0.976 & 0.995 & 0.963 & 0.978 & 0.995 & 0.967 \\
0.45 & 0.982 & 1.000 & 0.970 & 0.972 & 0.995 & 0.957 & 0.978 & 1.000 & 0.963 \\
0.50 & 0.988 & 1.000 & 0.980 & 0.983 & 1.000 & 0.972 & 0.986 & 1.000 & 0.977 \\
0.55 & 0.995 & 1.000 & 0.992 & 0.993 & 1.000 & 0.988 & 0.995 & 1.000 & 0.992 \\
0.60 & 0.991 & 0.998 & 0.987 & 0.982 & 0.995 & 0.973 & 0.983 & 0.995 & 0.975 \\
0.65 & 0.999 & 1.000 & 0.998 & 0.997 & 1.000 & 0.995 & 0.998 & 1.000 & 0.997 \\
0.70 & 0.995 & 1.000 & 0.992 & 0.992 & 0.998 & 0.988 & 0.992 & 0.998 & 0.988 \\
0.75 & 0.998 & 1.000 & 0.997 & 0.997 & 1.000 & 0.995 & 0.998 & 1.000 & 0.997 \\
0.80 & 0.999 & 1.000 & 0.998 & 0.999 & 1.000 & 0.998 & 0.999 & 1.000 & 0.998 \\
0.85 & 1.000 & 1.000 & 1.000 & 0.999 & 1.000 & 0.998 & 1.000 & 1.000 & 1.000 \\
0.90 & 1.000 & 1.000 & 1.000 & 0.999 & 1.000 & 0.998 & 1.000 & 1.000 & 1.000 \\
0.95 & 1.000 & 1.000 & 1.000 & 1.000 & 1.000 & 1.000 & 1.000 & 1.000 & 1.000 \\
1.00 & 0.999 & 1.000 & 0.998 & 0.999 & 1.000 & 0.998 & 0.999 & 1.000 & 0.998 \\
\hline
\end{tabular}
\caption{Recall, Precision, and F1-score for \textbf{Effect Modifiers} in \textbf{Setting 2} $(n = 10{,}000)$}
\label{table:setting2_n10000_em}
\end{table}

\begin{table}[h!]
\footnotesize
\centering
\caption{Summary of simulation settings.}
\label{tab:sim_settings}
\begin{tabular}{clcccc}
\toprule
Setting & Description & $n$ & Effect modification & Confounding & Prognostic \\
\midrule
\multirow{2}{*}{1} & \multirow{2}{*}{Baseline (no confounding)} & 5,000 & Additive & No & No \\
 &  & 10,000 & Additive & No & No \\
\midrule
\multirow{2}{*}{2} & \multirow{2}{*}{With confounding} & 5,000 & Additive & Yes & No \\
 &  & 10,000 & Additive & Yes & No \\
\midrule
3 & Interactive modification & 5,000 & Interactive & No & No \\
4 & Confounded modifiers & 5,000 & Additive & Yes$^*$ & No \\
5 & Prognostic factors & 5,000 & Additive & Yes & Yes \\
\bottomrule
\multicolumn{6}{l}{\footnotesize $^*$Treatment additionally depends on effect modifiers $E_1$ and $E_2$.}
\end{tabular}
\end{table}

\subsection{Results for Settings 3--5}

\textit{Setting 3 (interactive modification).}

\begin{table}[h!]
\centering
\footnotesize
\begin{tabular}{c|ccc|ccc|ccc}
\hline
\multirow{2}{*}{$\beta$}
& \multicolumn{3}{c}{\textbf{AUC}}
& \multicolumn{3}{c}{\textbf{FANOVA}}
& \multicolumn{3}{c}{$\mathbf{L^2}$} \\
\cline{2-10}
& F1 & Recall & Precision
& F1 & Recall & Precision
& F1 & Recall & Precision \\
\hline
0.01 & 0.023 & 0.035 & 0.018 & 0.007 & 0.010 & 0.005 & 0.013 & 0.020 & 0.010 \\
0.05 & 0.083 & 0.115 & 0.068 & 0.045 & 0.065 & 0.035 & 0.067 & 0.095 & 0.052 \\
0.10 & 0.248 & 0.315 & 0.215 & 0.175 & 0.220 & 0.152 & 0.215 & 0.270 & 0.188 \\
0.15 & 0.317 & 0.380 & 0.285 & 0.235 & 0.295 & 0.205 & 0.290 & 0.360 & 0.255 \\
0.20 & 0.368 & 0.455 & 0.325 & 0.275 & 0.335 & 0.245 & 0.325 & 0.400 & 0.287 \\
0.25 & 0.395 & 0.495 & 0.345 & 0.278 & 0.345 & 0.245 & 0.328 & 0.405 & 0.290 \\
0.30 & 0.572 & 0.675 & 0.520 & 0.477 & 0.565 & 0.432 & 0.545 & 0.650 & 0.492 \\
0.35 & 0.563 & 0.655 & 0.517 & 0.460 & 0.540 & 0.420 & 0.503 & 0.590 & 0.460 \\
0.40 & 0.677 & 0.765 & 0.632 & 0.550 & 0.635 & 0.507 & 0.637 & 0.725 & 0.593 \\
0.45 & 0.750 & 0.840 & 0.705 & 0.655 & 0.755 & 0.605 & 0.713 & 0.810 & 0.665 \\
0.50 & 0.743 & 0.835 & 0.698 & 0.647 & 0.720 & 0.610 & 0.720 & 0.800 & 0.680 \\
0.55 & 0.778 & 0.870 & 0.733 & 0.675 & 0.770 & 0.627 & 0.735 & 0.830 & 0.688 \\
0.60 & 0.803 & 0.870 & 0.770 & 0.698 & 0.775 & 0.660 & 0.777 & 0.850 & 0.740 \\
0.65 & 0.838 & 0.905 & 0.805 & 0.798 & 0.890 & 0.752 & 0.835 & 0.915 & 0.795 \\
0.70 & 0.870 & 0.935 & 0.838 & 0.798 & 0.875 & 0.760 & 0.860 & 0.920 & 0.830 \\
0.75 & 0.883 & 0.950 & 0.850 & 0.820 & 0.905 & 0.777 & 0.873 & 0.945 & 0.838 \\
0.80 & 0.907 & 0.965 & 0.877 & 0.827 & 0.900 & 0.790 & 0.875 & 0.945 & 0.840 \\
0.85 & 0.908 & 0.955 & 0.885 & 0.863 & 0.930 & 0.830 & 0.898 & 0.955 & 0.870 \\
0.90 & 0.928 & 0.975 & 0.905 & 0.880 & 0.945 & 0.848 & 0.913 & 0.965 & 0.887 \\
0.95 & 0.975 & 0.995 & 0.965 & 0.942 & 0.985 & 0.920 & 0.962 & 0.995 & 0.945 \\
1.00 & 0.957 & 0.985 & 0.943 & 0.920 & 0.960 & 0.900 & 0.943 & 0.975 & 0.927 \\
\hline
\end{tabular}
\caption{Recall, Precision, and F1-score for \textbf{Decision Rules} in \textbf{Setting 3} $(n = 5{,}000)$}
\label{table:setting3_n5000_dr}
\end{table}

\begin{table}[h!]
\centering
\footnotesize
\begin{tabular}{c|ccc|ccc|ccc}
\hline
\multirow{2}{*}{$\beta$}
& \multicolumn{3}{c}{\textbf{AUC}}
& \multicolumn{3}{c}{\textbf{FANOVA}}
& \multicolumn{3}{c}{$\mathbf{L^2}$} \\
\cline{2-10}
& F1 & Recall & Precision
& F1 & Recall & Precision
& F1 & Recall & Precision \\
\hline
0.01 & 0.203 & 0.250 & 0.172 & 0.117 & 0.145 & 0.098 & 0.169 & 0.207 & 0.142 \\
0.05 & 0.281 & 0.343 & 0.240 & 0.198 & 0.242 & 0.168 & 0.253 & 0.310 & 0.214 \\
0.10 & 0.469 & 0.552 & 0.412 & 0.361 & 0.427 & 0.318 & 0.426 & 0.502 & 0.375 \\
0.15 & 0.522 & 0.600 & 0.470 & 0.432 & 0.505 & 0.383 & 0.500 & 0.583 & 0.446 \\
0.20 & 0.577 & 0.665 & 0.517 & 0.496 & 0.573 & 0.445 & 0.544 & 0.627 & 0.488 \\
0.25 & 0.612 & 0.708 & 0.547 & 0.512 & 0.598 & 0.455 & 0.559 & 0.650 & 0.499 \\
0.30 & 0.723 & 0.807 & 0.666 & 0.656 & 0.743 & 0.599 & 0.705 & 0.795 & 0.645 \\
0.35 & 0.728 & 0.812 & 0.672 & 0.655 & 0.738 & 0.601 & 0.693 & 0.777 & 0.636 \\
0.40 & 0.804 & 0.880 & 0.754 & 0.717 & 0.797 & 0.662 & 0.777 & 0.855 & 0.726 \\
0.45 & 0.850 & 0.917 & 0.805 & 0.794 & 0.873 & 0.743 & 0.829 & 0.902 & 0.780 \\
0.50 & 0.847 & 0.915 & 0.802 & 0.787 & 0.855 & 0.742 & 0.832 & 0.897 & 0.788 \\
0.55 & 0.868 & 0.935 & 0.823 & 0.805 & 0.882 & 0.754 & 0.844 & 0.915 & 0.796 \\
0.60 & 0.885 & 0.935 & 0.852 & 0.824 & 0.887 & 0.782 & 0.870 & 0.925 & 0.833 \\
0.65 & 0.904 & 0.953 & 0.873 & 0.880 & 0.945 & 0.838 & 0.902 & 0.958 & 0.865 \\
0.70 & 0.920 & 0.965 & 0.890 & 0.877 & 0.935 & 0.839 & 0.914 & 0.958 & 0.885 \\
0.75 & 0.930 & 0.975 & 0.901 & 0.891 & 0.950 & 0.852 & 0.922 & 0.970 & 0.891 \\
0.80 & 0.944 & 0.983 & 0.918 & 0.897 & 0.950 & 0.862 & 0.925 & 0.973 & 0.894 \\
0.85 & 0.946 & 0.978 & 0.924 & 0.919 & 0.965 & 0.888 & 0.940 & 0.978 & 0.914 \\
0.90 & 0.957 & 0.988 & 0.937 & 0.928 & 0.973 & 0.899 & 0.949 & 0.983 & 0.926 \\
0.95 & 0.986 & 0.998 & 0.978 & 0.965 & 0.993 & 0.947 & 0.977 & 0.998 & 0.963 \\
1.00 & 0.975 & 0.993 & 0.963 & 0.953 & 0.980 & 0.935 & 0.967 & 0.988 & 0.953 \\
\hline
\end{tabular}
\caption{Recall, Precision, and F1-score for \textbf{Effect Modifiers} in \textbf{Setting 3} $(n = 5{,}000)$}
\label{table:setting3_n5000_em}
\end{table}

\textit{Setting 4 (Confounded modifiers).}

\begin{table}[h!]
\centering
\footnotesize
\begin{tabular}{c|ccc|ccc|ccc}
\hline
\multirow{2}{*}{$\beta$}
& \multicolumn{3}{c}{\textbf{AUC}}
& \multicolumn{3}{c}{\textbf{FANOVA}}
& \multicolumn{3}{c}{$\mathbf{L^2}$} \\
\cline{2-10}
& F1 & Recall & Precision
& F1 & Recall & Precision
& F1 & Recall & Precision \\
\hline
0.01 & 0.038 & 0.055 & 0.030 & 0.025 & 0.035 & 0.020 & 0.035 & 0.050 & 0.028 \\
0.05 & 0.052 & 0.075 & 0.040 & 0.040 & 0.060 & 0.030 & 0.065 & 0.095 & 0.050 \\
0.10 & 0.187 & 0.260 & 0.150 & 0.133 & 0.185 & 0.107 & 0.178 & 0.250 & 0.142 \\
0.15 & 0.327 & 0.425 & 0.278 & 0.242 & 0.310 & 0.207 & 0.340 & 0.440 & 0.290 \\
0.20 & 0.530 & 0.635 & 0.477 & 0.457 & 0.555 & 0.407 & 0.498 & 0.610 & 0.443 \\
0.25 & 0.715 & 0.835 & 0.655 & 0.632 & 0.755 & 0.570 & 0.698 & 0.810 & 0.642 \\
0.30 & 0.832 & 0.910 & 0.792 & 0.752 & 0.855 & 0.700 & 0.815 & 0.895 & 0.775 \\
0.35 & 0.892 & 0.960 & 0.858 & 0.853 & 0.940 & 0.810 & 0.878 & 0.950 & 0.843 \\
0.40 & 0.913 & 0.960 & 0.890 & 0.873 & 0.930 & 0.845 & 0.900 & 0.965 & 0.868 \\
0.45 & 0.955 & 0.990 & 0.938 & 0.943 & 0.995 & 0.917 & 0.953 & 0.995 & 0.932 \\
0.50 & 0.967 & 0.995 & 0.953 & 0.952 & 0.990 & 0.932 & 0.957 & 0.990 & 0.940 \\
0.55 & 0.985 & 1.000 & 0.978 & 0.983 & 1.000 & 0.975 & 0.985 & 1.000 & 0.978 \\
0.60 & 0.983 & 0.995 & 0.978 & 0.980 & 0.995 & 0.973 & 0.987 & 0.995 & 0.983 \\
0.65 & 0.982 & 1.000 & 0.973 & 0.977 & 1.000 & 0.965 & 0.982 & 1.000 & 0.973 \\
0.70 & 0.995 & 0.995 & 0.995 & 0.988 & 0.995 & 0.985 & 0.988 & 0.995 & 0.985 \\
0.75 & 0.997 & 1.000 & 0.995 & 0.993 & 1.000 & 0.990 & 0.997 & 1.000 & 0.995 \\
0.80 & 0.997 & 1.000 & 0.995 & 0.997 & 1.000 & 0.995 & 0.997 & 1.000 & 0.995 \\
0.85 & 0.997 & 1.000 & 0.995 & 0.993 & 1.000 & 0.990 & 0.993 & 1.000 & 0.990 \\
0.90 & 0.992 & 1.000 & 0.988 & 0.988 & 1.000 & 0.983 & 0.990 & 1.000 & 0.985 \\
0.95 & 1.000 & 1.000 & 1.000 & 0.997 & 1.000 & 0.995 & 1.000 & 1.000 & 1.000 \\
1.00 & 1.000 & 1.000 & 1.000 & 1.000 & 1.000 & 1.000 & 1.000 & 1.000 & 1.000 \\
\hline
\end{tabular}
\caption{Recall, Precision, and F1-score for \textbf{Decision Rules} in \textbf{Setting 4} $(n = 5{,}000)$}
\label{table:setting4_n5000_dr}
\end{table}

\begin{table}[h!]
\centering
\footnotesize
\begin{tabular}{c|ccc|ccc|ccc}
\hline
\multirow{2}{*}{$\beta$}
& \multicolumn{3}{c}{\textbf{AUC}}
& \multicolumn{3}{c}{\textbf{FANOVA}}
& \multicolumn{3}{c}{$\mathbf{L^2}$} \\
\cline{2-10}
& F1 & Recall & Precision
& F1 & Recall & Precision
& F1 & Recall & Precision \\
\hline
0.01 & 0.265 & 0.325 & 0.224 & 0.193 & 0.237 & 0.163 & 0.228 & 0.280 & 0.193 \\
0.05 & 0.300 & 0.365 & 0.258 & 0.269 & 0.328 & 0.229 & 0.318 & 0.388 & 0.271 \\
0.10 & 0.479 & 0.575 & 0.415 & 0.402 & 0.482 & 0.347 & 0.451 & 0.542 & 0.390 \\
0.15 & 0.589 & 0.698 & 0.517 & 0.536 & 0.632 & 0.472 & 0.601 & 0.708 & 0.529 \\
0.20 & 0.724 & 0.818 & 0.662 & 0.670 & 0.765 & 0.607 & 0.700 & 0.800 & 0.633 \\
0.25 & 0.833 & 0.917 & 0.777 & 0.788 & 0.885 & 0.724 & 0.823 & 0.905 & 0.767 \\
0.30 & 0.902 & 0.958 & 0.865 & 0.858 & 0.932 & 0.807 & 0.892 & 0.950 & 0.853 \\
0.35 & 0.936 & 0.980 & 0.907 & 0.914 & 0.970 & 0.877 & 0.928 & 0.975 & 0.897 \\
0.40 & 0.949 & 0.980 & 0.928 & 0.929 & 0.970 & 0.902 & 0.941 & 0.983 & 0.912 \\
0.45 & 0.973 & 0.995 & 0.958 & 0.966 & 0.998 & 0.945 & 0.972 & 0.998 & 0.955 \\
0.50 & 0.980 & 0.998 & 0.968 & 0.971 & 0.995 & 0.955 & 0.974 & 0.995 & 0.960 \\
0.55 & 0.991 & 1.000 & 0.985 & 0.990 & 1.000 & 0.983 & 0.991 & 1.000 & 0.985 \\
0.60 & 0.991 & 0.998 & 0.986 & 0.989 & 0.998 & 0.983 & 0.993 & 0.998 & 0.989 \\
0.65 & 0.989 & 1.000 & 0.982 & 0.986 & 1.000 & 0.977 & 0.989 & 1.000 & 0.982 \\
0.70 & 0.997 & 0.998 & 0.997 & 0.993 & 0.998 & 0.990 & 0.993 & 0.998 & 0.990 \\
0.75 & 0.998 & 1.000 & 0.997 & 0.996 & 1.000 & 0.993 & 0.998 & 1.000 & 0.997 \\
0.80 & 0.998 & 1.000 & 0.997 & 0.998 & 1.000 & 0.997 & 0.998 & 1.000 & 0.997 \\
0.85 & 0.998 & 1.000 & 0.997 & 0.996 & 1.000 & 0.993 & 0.996 & 1.000 & 0.993 \\
0.90 & 0.995 & 1.000 & 0.992 & 0.993 & 1.000 & 0.988 & 0.994 & 1.000 & 0.990 \\
0.95 & 1.000 & 1.000 & 1.000 & 0.998 & 1.000 & 0.997 & 1.000 & 1.000 & 1.000 \\
1.00 & 1.000 & 1.000 & 1.000 & 1.000 & 1.000 & 1.000 & 1.000 & 1.000 & 1.000 \\
\hline
\end{tabular}
\caption{Recall, Precision, and F1-score for \textbf{Effect Modifiers} in \textbf{Setting 4} $(n = 5{,}000)$}
\label{table:setting4_n5000_em}
\end{table}

\textit{Setting 5 (Prognostic factors).}

\begin{table}[h!]
\centering
\footnotesize
\begin{tabular}{c|ccc|ccc|ccc}
\hline
\multirow{2}{*}{$\beta$}
& \multicolumn{3}{c}{\textbf{AUC}}
& \multicolumn{3}{c}{\textbf{FANOVA}}
& \multicolumn{3}{c}{$\mathbf{L^2}$} \\
\cline{2-10}
& F1 & Recall & Precision
& F1 & Recall & Precision
& F1 & Recall & Precision \\
\hline
0.01 & 0.045 & 0.065 & 0.035 & 0.018 & 0.025 & 0.015 & 0.048 & 0.070 & 0.037 \\
0.05 & 0.042 & 0.060 & 0.033 & 0.020 & 0.030 & 0.015 & 0.028 & 0.040 & 0.022 \\
0.10 & 0.152 & 0.210 & 0.122 & 0.108 & 0.155 & 0.085 & 0.137 & 0.190 & 0.110 \\
0.15 & 0.335 & 0.425 & 0.290 & 0.273 & 0.355 & 0.233 & 0.323 & 0.405 & 0.282 \\
0.20 & 0.548 & 0.660 & 0.492 & 0.468 & 0.575 & 0.415 & 0.532 & 0.650 & 0.472 \\
0.25 & 0.715 & 0.825 & 0.660 & 0.650 & 0.755 & 0.598 & 0.707 & 0.815 & 0.652 \\
0.30 & 0.845 & 0.910 & 0.812 & 0.802 & 0.890 & 0.757 & 0.835 & 0.910 & 0.797 \\
0.35 & 0.898 & 0.960 & 0.868 & 0.858 & 0.940 & 0.818 & 0.880 & 0.945 & 0.848 \\
0.40 & 0.927 & 0.975 & 0.902 & 0.895 & 0.965 & 0.860 & 0.910 & 0.970 & 0.880 \\
0.45 & 0.967 & 0.995 & 0.953 & 0.952 & 0.995 & 0.930 & 0.968 & 0.995 & 0.955 \\
0.50 & 0.982 & 1.000 & 0.973 & 0.965 & 1.000 & 0.948 & 0.970 & 1.000 & 0.955 \\
0.55 & 0.980 & 1.000 & 0.970 & 0.970 & 0.995 & 0.958 & 0.977 & 0.995 & 0.968 \\
0.60 & 0.990 & 1.000 & 0.985 & 0.983 & 1.000 & 0.975 & 0.988 & 1.000 & 0.983 \\
0.65 & 0.995 & 1.000 & 0.993 & 0.985 & 0.995 & 0.980 & 0.992 & 1.000 & 0.988 \\
0.70 & 0.995 & 1.000 & 0.993 & 0.993 & 1.000 & 0.990 & 0.995 & 1.000 & 0.993 \\
0.75 & 0.998 & 1.000 & 0.998 & 0.998 & 1.000 & 0.998 & 0.998 & 1.000 & 0.998 \\
0.80 & 0.993 & 1.000 & 0.990 & 0.990 & 1.000 & 0.985 & 0.990 & 1.000 & 0.985 \\
0.85 & 0.997 & 1.000 & 0.995 & 0.993 & 1.000 & 0.990 & 0.997 & 1.000 & 0.995 \\
0.90 & 0.998 & 1.000 & 0.998 & 0.998 & 1.000 & 0.998 & 0.998 & 1.000 & 0.998 \\
0.95 & 0.998 & 1.000 & 0.998 & 0.998 & 1.000 & 0.998 & 0.998 & 1.000 & 0.998 \\
1.00 & 1.000 & 1.000 & 1.000 & 1.000 & 1.000 & 1.000 & 1.000 & 1.000 & 1.000 \\
\hline
\end{tabular}
\caption{Recall, Precision, and F1-score for \textbf{Decision Rules} in \textbf{Setting 5} $(n = 5{,}000)$}
\label{table:setting5_n5000_dr}
\end{table}

\begin{table}[h!]
\centering
\footnotesize
\begin{tabular}{c|ccc|ccc|ccc}
\hline
\multirow{2}{*}{$\beta$}
& \multicolumn{3}{c}{\textbf{AUC}}
& \multicolumn{3}{c}{\textbf{FANOVA}}
& \multicolumn{3}{c}{$\mathbf{L^2}$} \\
\cline{2-10}
& F1 & Recall & Precision
& F1 & Recall & Precision
& F1 & Recall & Precision \\
\hline
0.01 & 0.272 & 0.335 & 0.229 & 0.208 & 0.255 & 0.177 & 0.265 & 0.323 & 0.226 \\
0.05 & 0.334 & 0.410 & 0.283 & 0.255 & 0.315 & 0.215 & 0.278 & 0.343 & 0.235 \\
0.10 & 0.449 & 0.545 & 0.385 & 0.376 & 0.460 & 0.319 & 0.413 & 0.505 & 0.352 \\
0.15 & 0.594 & 0.700 & 0.523 & 0.515 & 0.615 & 0.449 & 0.572 & 0.672 & 0.505 \\
0.20 & 0.719 & 0.810 & 0.657 & 0.669 & 0.767 & 0.604 & 0.719 & 0.818 & 0.652 \\
0.25 & 0.835 & 0.917 & 0.780 & 0.792 & 0.877 & 0.736 & 0.828 & 0.907 & 0.774 \\
0.30 & 0.904 & 0.950 & 0.873 & 0.878 & 0.940 & 0.837 & 0.898 & 0.950 & 0.863 \\
0.35 & 0.938 & 0.978 & 0.912 & 0.914 & 0.968 & 0.879 & 0.927 & 0.970 & 0.899 \\
0.40 & 0.958 & 0.990 & 0.937 & 0.938 & 0.983 & 0.908 & 0.948 & 0.988 & 0.922 \\
0.45 & 0.980 & 0.998 & 0.968 & 0.971 & 0.998 & 0.953 & 0.981 & 0.998 & 0.970 \\
0.50 & 0.989 & 1.000 & 0.982 & 0.979 & 1.000 & 0.965 & 0.982 & 1.000 & 0.970 \\
0.55 & 0.988 & 1.000 & 0.980 & 0.982 & 0.998 & 0.972 & 0.986 & 0.998 & 0.978 \\
0.60 & 0.994 & 1.000 & 0.990 & 0.990 & 1.000 & 0.983 & 0.993 & 1.000 & 0.988 \\
0.65 & 0.997 & 1.000 & 0.995 & 0.991 & 0.998 & 0.987 & 0.995 & 1.000 & 0.992 \\
0.70 & 0.997 & 1.000 & 0.995 & 0.996 & 1.000 & 0.993 & 0.997 & 1.000 & 0.995 \\
0.75 & 0.999 & 1.000 & 0.998 & 0.999 & 1.000 & 0.998 & 0.999 & 1.000 & 0.998 \\
0.80 & 0.996 & 1.000 & 0.993 & 0.994 & 1.000 & 0.990 & 0.994 & 1.000 & 0.990 \\
0.85 & 0.998 & 1.000 & 0.997 & 0.996 & 1.000 & 0.993 & 0.998 & 1.000 & 0.997 \\
0.90 & 0.999 & 1.000 & 0.998 & 0.999 & 1.000 & 0.998 & 0.999 & 1.000 & 0.998 \\
0.95 & 0.999 & 1.000 & 0.998 & 0.999 & 1.000 & 0.998 & 0.999 & 1.000 & 0.998 \\
1.00 & 1.000 & 1.000 & 1.000 & 1.000 & 1.000 & 1.000 & 1.000 & 1.000 & 1.000 \\
\hline
\end{tabular}
\caption{Recall, Precision, and F1-score for \textbf{Effect Modifiers} in \textbf{Setting 5} $(n = 5{,}000)$}
\label{table:setting5_n5000_em}
\end{table}

% Flush all long simulation tables before placing supplementary application figures.
\clearpage
\section{Additional Application Results}
\label{app:app_results}

\subsection{Discovered Trees and Subgroup Estimates with Binarized Covariates}
\label{app:additional_disc}

\ifthenelse{\boolean{showfigures}}{
\begin{figure}[H]
\centering
\includegraphics[width=\textwidth]{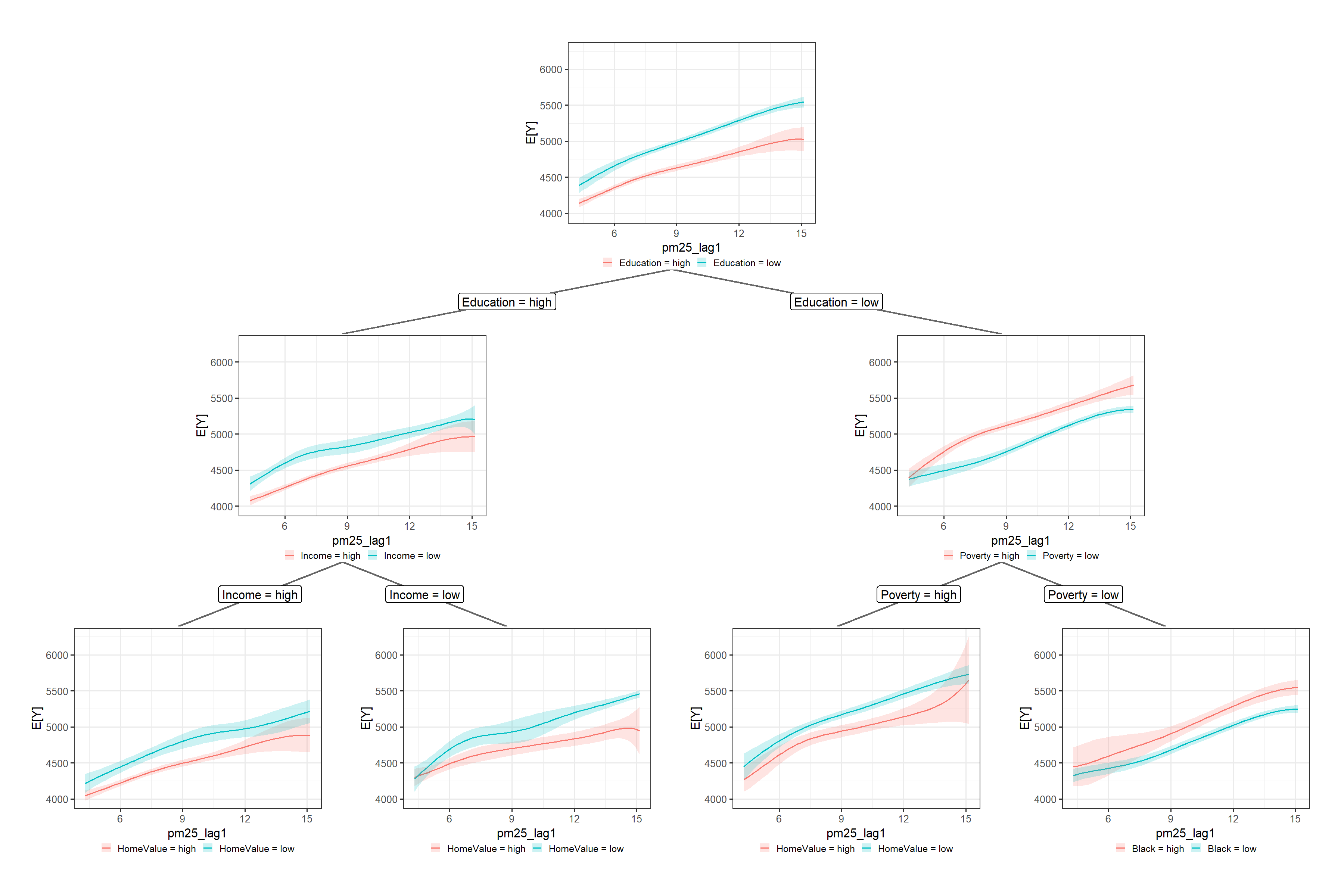}
\caption{Discovered honest tree under the $L^2$, AUC, and FANOVA objectives with density weighting.
Each panel shows the estimated dose-response curves of the two children of the corresponding split, with pointwise confidence bands based on cluster-robust standard errors.}
\label{fig:tree_l2_density_pm25}
\end{figure}
}{}

\begin{table}[H]
\centering
\footnotesize
\setlength{\tabcolsep}{2pt}
\begin{tabular}{>{\raggedright\arraybackslash}p{0.30\textwidth}cc|rrr|rrr|rrr}
\hline
\multirow{2}{*}{Subgroup}
& \multirow{2}{*}{$n$}
& \multirow{2}{*}{Counties}
& \multicolumn{3}{c|}{$5\ \mu$g/m$^3$}
& \multicolumn{3}{c|}{$9\ \mu$g/m$^3$}
& \multicolumn{3}{c}{$12\ \mu$g/m$^3$} \\
\cline{4-12}
& & 
& Est. & SE$_{\mathrm{CR}}$ & SE$_{\mathrm{C}}$
& Est. & SE$_{\mathrm{CR}}$ & SE$_{\mathrm{C}}$
& Est. & SE$_{\mathrm{CR}}$ & SE$_{\mathrm{C}}$ \\
\hline
(1) Population & 26751 & 1408 & 4368 & 21.3 & 29.1 & 4803 & 17.0 & 38.6 & 5067 & 19.1 & 36.5 \\
\quad$\vdash$ (2) Education = high & 13746 & 1125 & 4229 & 22.8 & 31.4 & 4632 & 22.6 & 36.6 & 4852 & 34.7 & 33.6 \\
\quad\quad$\vdash$ (3) HomeValue = high & 9750 & 902 & 4165 & 25.5 & 30.8 & 4533 & 22.8 & 34.6 & 4739 & 40.6 & 33.6 \\
\quad\quad\quad$\vdash$ (4) Income = high & 7943 & 847 & 4116 & 27.4 & 33.5 & 4495 & 23.6 & 35.9 & 4720 & 50.1 & 37.7 \\
\quad\quad\quad$\vdash$ (5) Income = low & 1807 & 328 & 4373 & 43.2 & 46.1 & 4699 & 48.6 & 62.4 & 4834 & 49.3 & 53.3 \\
\quad\quad$\vdash$ (6) HomeValue = low & 3996 & 556 & 4385 & 39.6 & 51.7 & 4863 & 54.0 & 69.8 & 5115 & 40.0 & 47.6 \\
\quad\quad\quad$\vdash$ (7) Black = high & 1209 & 256 & 4575 & 59.1 & 62.1 & 4974 & 86.3 & 87.3 & 5190 & 46.3 & 53.5 \\
\quad\quad\quad$\vdash$ (8) Black = low & 2787 & 357 & 4319 & 49.6 & 59.4 & 4816 & 51.8 & 70.3 & 5095 & 86.6 & 76.7 \\
\quad$\vdash$ (9) Education = low & 13005 & 975 & 4506 & 46.4 & 51.9 & 4984 & 20.7 & 41.0 & 5289 & 23.1 & 43.6 \\
\quad\quad$\vdash$ (10) Poverty = high & 8189 & 700 & 4552 & 53.1 & 58.7 & 5119 & 24.8 & 42.3 & 5390 & 31.7 & 48.8 \\
\quad\quad\quad$\vdash$ (11) HomeValue = high & 1759 & 345 & 4407 & 84.2 & 86.7 & 4939 & 42.5 & 51.9 & 5140 & 60.6 & 63.5 \\
\quad\quad\quad$\vdash$ (12) HomeValue = low & 6430 & 599 & 4608 & 69.0 & 69.8 & 5170 & 27.9 & 45.4 & 5458 & 33.6 & 50.3 \\
\quad\quad$\vdash$ (13) Poverty = low & 4816 & 611 & 4423 & 51.1 & 61.5 & 4754 & 26.2 & 38.9 & 5117 & 22.4 & 34.8 \\
\quad\quad\quad$\vdash$ (14) Black = high & 1660 & 266 & 4493 & 150.7 & 173.2 & 4904 & 53.1 & 55.4 & 5286 & 39.5 & 45.0 \\
\quad\quad\quad$\vdash$ (15) Black = low & 3156 & 397 & 4373 & 42.3 & 45.7 & 4678 & 28.8 & 40.9 & 5022 & 24.3 & 32.5 \\
\hline
\end{tabular}
\caption{Number of county-years ($n$) and counties, estimated age-standardized mortality rates among adults aged 65 and older (deaths per 100,000 person-years) at exposures of 5, 9, and 12 $\mu$g/m$^3$, and standard errors within each subgroup on the discovered tree under the $L^2$ and AUC objectives with uniform weighting. SE$_{\mathrm{CR}}$: cluster-robust standard error, clustered by county. SE$_{\mathrm{C}}$: Conley spatial standard error with a Bartlett kernel and a 200 km cutoff.}
\label{table:tab_tree_l2_unweighted}
\end{table}

\begin{table}[H]
\centering
\footnotesize
\setlength{\tabcolsep}{2pt}
\begin{tabular}{>{\raggedright\arraybackslash}p{0.30\textwidth}cc|rrr|rrr|rrr}
\hline
\multirow{2}{*}{Subgroup}
& \multirow{2}{*}{$n$}
& \multirow{2}{*}{Counties}
& \multicolumn{3}{c|}{$5\ \mu$g/m$^3$}
& \multicolumn{3}{c|}{$9\ \mu$g/m$^3$}
& \multicolumn{3}{c}{$12\ \mu$g/m$^3$} \\
\cline{4-12}
& & 
& Est. & SE$_{\mathrm{CR}}$ & SE$_{\mathrm{C}}$
& Est. & SE$_{\mathrm{CR}}$ & SE$_{\mathrm{C}}$
& Est. & SE$_{\mathrm{CR}}$ & SE$_{\mathrm{C}}$ \\
\hline
(1) Population & 26751 & 1408 & 4368 & 21.3 & 29.1 & 4803 & 17.0 & 38.6 & 5067 & 19.1 & 36.5 \\
\quad$\vdash$ (2) Education = high & 13746 & 1125 & 4229 & 22.8 & 31.4 & 4632 & 22.6 & 36.6 & 4852 & 34.7 & 33.6 \\
\quad\quad$\vdash$ (3) Income = high & 9862 & 1012 & 4149 & 24.9 & 34.2 & 4555 & 22.3 & 36.4 & 4786 & 46.0 & 37.9 \\
\quad\quad\quad$\vdash$ (4) HomeValue = high & 7943 & 847 & 4116 & 27.4 & 33.5 & 4495 & 23.6 & 35.9 & 4720 & 50.1 & 37.7 \\
\quad\quad\quad$\vdash$ (5) HomeValue = low & 1919 & 371 & 4313 & 51.3 & 66.5 & 4799 & 52.3 & 73.6 & 4973 & 59.9 & 70.1 \\
\quad\quad$\vdash$ (6) Income = low & 3884 & 615 & 4430 & 35.4 & 41.0 & 4826 & 49.9 & 59.1 & 5022 & 38.5 & 43.1 \\
\quad\quad\quad$\vdash$ (7) HomeValue = high & 1807 & 328 & 4373 & 43.2 & 46.1 & 4699 & 48.6 & 62.4 & 4834 & 49.3 & 53.3 \\
\quad\quad\quad$\vdash$ (8) HomeValue = low & 2077 & 408 & 4462 & 56.9 & 66.5 & 4929 & 81.3 & 90.4 & 5195 & 48.5 & 52.7 \\
\quad$\vdash$ (9) Education = low & 13005 & 975 & 4506 & 46.4 & 51.9 & 4984 & 20.7 & 41.0 & 5289 & 23.1 & 43.6 \\
\quad\quad$\vdash$ (10) Poverty = high & 8189 & 700 & 4552 & 53.1 & 58.7 & 5119 & 24.8 & 42.3 & 5390 & 31.7 & 48.8 \\
\quad\quad\quad$\vdash$ (11) HomeValue = high & 1759 & 345 & 4407 & 84.2 & 86.7 & 4939 & 42.5 & 51.9 & 5140 & 60.6 & 63.5 \\
\quad\quad\quad$\vdash$ (12) HomeValue = low & 6430 & 599 & 4608 & 69.0 & 69.8 & 5170 & 27.9 & 45.4 & 5458 & 33.6 & 50.3 \\
\quad\quad$\vdash$ (13) Poverty = low & 4816 & 611 & 4423 & 51.1 & 61.5 & 4754 & 26.2 & 38.9 & 5117 & 22.4 & 34.8 \\
\quad\quad\quad$\vdash$ (14) Black = high & 1660 & 266 & 4493 & 150.7 & 173.2 & 4904 & 53.1 & 55.4 & 5286 & 39.5 & 45.0 \\
\quad\quad\quad$\vdash$ (15) Black = low & 3156 & 397 & 4373 & 42.3 & 45.7 & 4678 & 28.8 & 40.9 & 5022 & 24.3 & 32.5 \\
\hline
\end{tabular}
\caption{Number of county-years ($n$) and counties, estimated age-standardized mortality rates among adults aged 65 and older (deaths per 100,000 person-years) at exposures of 5, 9, and 12 $\mu$g/m$^3$, and standard errors within each subgroup on the discovered tree under the $L^2$, AUC, and FANOVA objectives with density weighting. SE$_{\mathrm{CR}}$: cluster-robust standard error, clustered by county. SE$_{\mathrm{C}}$: Conley spatial standard error with a Bartlett kernel and a 200 km cutoff.}
\label{table:tab_tree_l2_density}
\end{table}

\begin{table}[H]
\centering
\footnotesize
\setlength{\tabcolsep}{2pt}
\begin{tabular}{>{\raggedright\arraybackslash}p{0.30\textwidth}cc|rrr|rrr|rrr}
\hline
\multirow{2}{*}{Subgroup}
& \multirow{2}{*}{$n$}
& \multirow{2}{*}{Counties}
& \multicolumn{3}{c|}{$5\ \mu$g/m$^3$}
& \multicolumn{3}{c|}{$9\ \mu$g/m$^3$}
& \multicolumn{3}{c}{$12\ \mu$g/m$^3$} \\
\cline{4-12}
& & 
& Est. & SE$_{\mathrm{CR}}$ & SE$_{\mathrm{C}}$
& Est. & SE$_{\mathrm{CR}}$ & SE$_{\mathrm{C}}$
& Est. & SE$_{\mathrm{CR}}$ & SE$_{\mathrm{C}}$ \\
\hline
(1) Population & 26751 & 1408 & 4368 & 21.3 & 29.1 & 4803 & 17.0 & 38.6 & 5067 & 19.1 & 36.5 \\
\quad$\vdash$ (2) Education = high & 13746 & 1125 & 4229 & 22.8 & 31.4 & 4632 & 22.6 & 36.6 & 4852 & 34.7 & 33.6 \\
\quad\quad$\vdash$ (3) HomeValue = high & 9750 & 902 & 4165 & 25.5 & 30.8 & 4533 & 22.8 & 34.6 & 4739 & 40.6 & 33.6 \\
\quad\quad\quad$\vdash$ (4) Poverty = high & 3080 & 478 & 4255 & 37.6 & 41.6 & 4627 & 39.6 & 48.2 & 4815 & 68.3 & 71.1 \\
\quad\quad\quad$\vdash$ (5) Poverty = low & 6670 & 712 & 4126 & 32.2 & 37.2 & 4489 & 24.8 & 35.5 & 4697 & 45.1 & 34.6 \\
\quad\quad$\vdash$ (6) HomeValue = low & 3996 & 556 & 4385 & 39.6 & 51.7 & 4863 & 54.0 & 69.8 & 5115 & 40.0 & 47.6 \\
\quad\quad\quad$\vdash$ (7) Black = high & 1209 & 256 & 4575 & 59.1 & 62.1 & 4974 & 86.3 & 87.3 & 5190 & 46.3 & 53.5 \\
\quad\quad\quad$\vdash$ (8) Black = low & 2787 & 357 & 4319 & 49.6 & 59.4 & 4816 & 51.8 & 70.3 & 5095 & 86.6 & 76.7 \\
\quad$\vdash$ (9) Education = low & 13005 & 975 & 4506 & 46.4 & 51.9 & 4984 & 20.7 & 41.0 & 5289 & 23.1 & 43.6 \\
\quad\quad$\vdash$ (10) Poverty = high & 8189 & 700 & 4552 & 53.1 & 58.7 & 5119 & 24.8 & 42.3 & 5390 & 31.7 & 48.8 \\
\quad\quad\quad$\vdash$ (11) HomeValue = high & 1759 & 345 & 4407 & 84.2 & 86.7 & 4939 & 42.5 & 51.9 & 5140 & 60.6 & 63.5 \\
\quad\quad\quad$\vdash$ (12) HomeValue = low & 6430 & 599 & 4608 & 69.0 & 69.8 & 5170 & 27.9 & 45.4 & 5458 & 33.6 & 50.3 \\
\quad\quad$\vdash$ (13) Poverty = low & 4816 & 611 & 4423 & 51.1 & 61.5 & 4754 & 26.2 & 38.9 & 5117 & 22.4 & 34.8 \\
\quad\quad\quad$\vdash$ (14) Black = high & 1660 & 266 & 4493 & 150.7 & 173.2 & 4904 & 53.1 & 55.4 & 5286 & 39.5 & 45.0 \\
\quad\quad\quad$\vdash$ (15) Black = low & 3156 & 397 & 4373 & 42.3 & 45.7 & 4678 & 28.8 & 40.9 & 5022 & 24.3 & 32.5 \\
\hline
\end{tabular}
\caption{Number of county-years ($n$) and counties, estimated age-standardized mortality rates among adults aged 65 and older (deaths per 100,000 person-years) at exposures of 5, 9, and 12 $\mu$g/m$^3$, and standard errors within each subgroup on the discovered tree under the FANOVA objective with uniform weighting. SE$_{\mathrm{CR}}$: cluster-robust standard error, clustered by county. SE$_{\mathrm{C}}$: Conley spatial standard error with a Bartlett kernel and a 200 km cutoff.}
\label{table:tab_tree_fanova_unweighted}
\end{table}

\ifthenelse{\boolean{showfigures}}{
\begin{figure}[H]
\centering
\includegraphics[width=\textwidth]{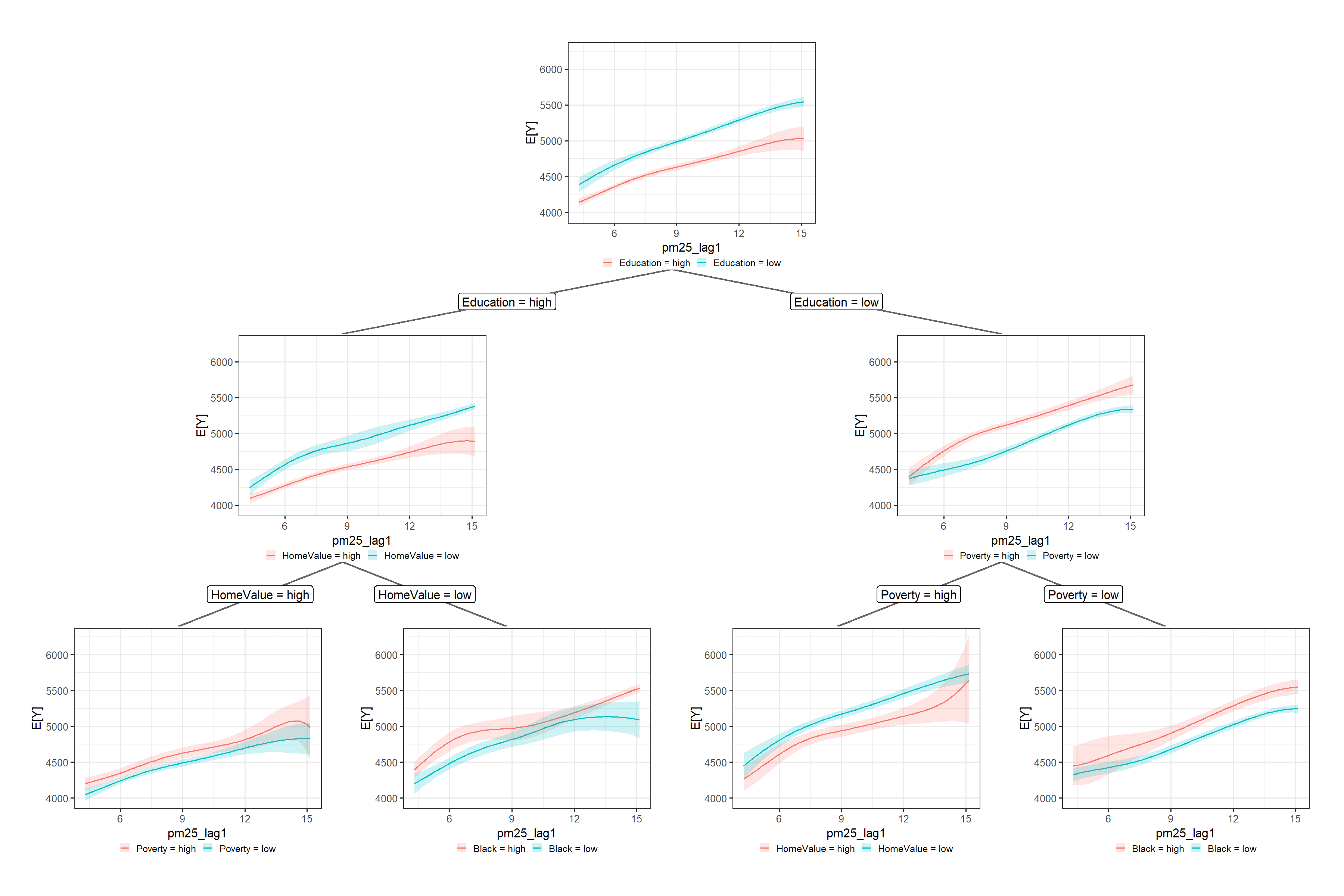}
\caption{Discovered honest tree under the FANOVA objective with uniform weighting.
Each panel shows the estimated dose-response curves of the two children of the corresponding split, with pointwise confidence bands based on cluster-robust standard errors.}
\label{fig:tree_fanova_unweighted_pm25}
\end{figure}
}{}

In the discovery sample, splitting higher-education, higher-home-value counties on poverty yields groups of 3{,}032 and 6{,}378 county-years, compared with 7{,}602 and 1{,}808 for income.
The more balanced poverty split is selected by FANOVA (\Cref{fig:tree_fanova_unweighted_pm25}).

\subsection{Discovered Trees and Subgroup Estimates with Continuous Covariates}
\label{app:additional_cont}

In this analysis we retain continuous covariates and compute candidate median splits within each discovery node. The binarized analysis uses medians fixed across all 53{,}469 county-years.

\ifthenelse{\boolean{showfigures}}{
\begin{figure}[H]
\centering
\includegraphics[width=\textwidth]{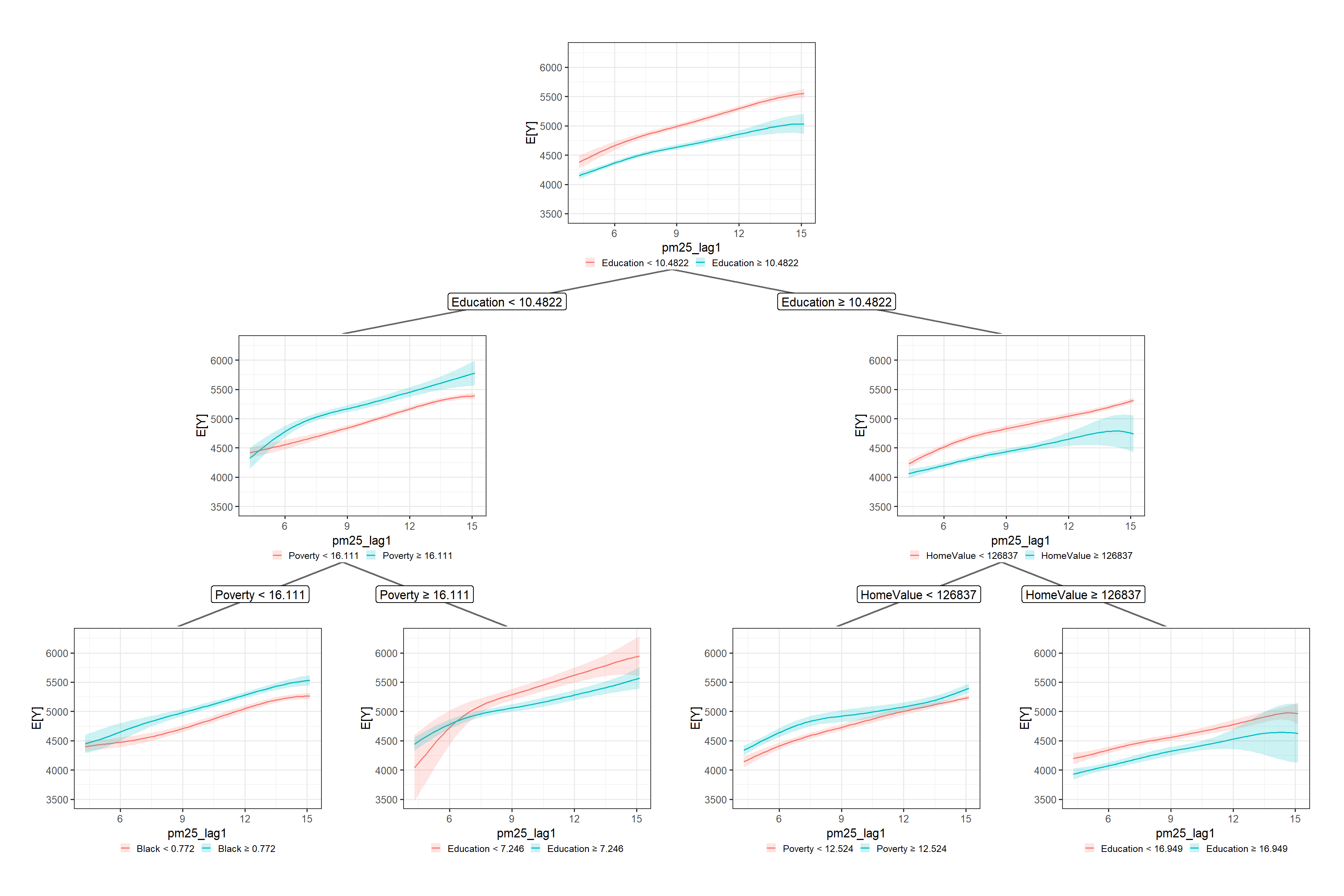}
\caption{Discovered honest tree under the $L^2$ objective with uniform weighting and continuous covariates. Each panel shows the estimated dose-response curves of the two children of the corresponding split, with pointwise confidence bands based on cluster-robust standard errors.}
\label{fig:tree_L2_unweighted_cont_pm25}
\end{figure}
}{}

\begin{table}[H]
\centering
\footnotesize
\setlength{\tabcolsep}{2pt}
\begin{tabular}{>{\raggedright\arraybackslash}p{0.30\textwidth}cc|rrr|rrr|rrr}
\hline
\multirow{2}{*}{Subgroup}
& \multirow{2}{*}{$n$}
& \multirow{2}{*}{Counties}
& \multicolumn{3}{c|}{$5\ \mu$g/m$^3$}
& \multicolumn{3}{c|}{$9\ \mu$g/m$^3$}
& \multicolumn{3}{c}{$12\ \mu$g/m$^3$} \\
\cline{4-12}
& & 
& Est. & SE$_{\mathrm{CR}}$ & SE$_{\mathrm{C}}$
& Est. & SE$_{\mathrm{CR}}$ & SE$_{\mathrm{C}}$
& Est. & SE$_{\mathrm{CR}}$ & SE$_{\mathrm{C}}$ \\
\hline
(1) Population & 26751 & 1408 & 4368 & 21.3 & 29.1 & 4803 & 17.0 & 38.6 & 5067 & 19.1 & 36.5 \\
\quad$\vdash$ (2) Education $<$ 10.48 & 12589 & 963 & 4502 & 49.1 & 54.4 & 4990 & 21.1 & 41.5 & 5297 & 23.5 & 43.9 \\
\quad\quad$\vdash$ (3) Poverty $<$ 16.11 & 6900 & 771 & 4474 & 43.2 & 53.0 & 4842 & 23.1 & 38.1 & 5167 & 20.2 & 31.5 \\
\quad\quad\quad$\vdash$ (4) Black $<$ 0.77 & 3479 & 431 & 4433 & 41.7 & 42.7 & 4709 & 29.2 & 42.3 & 5049 & 28.5 & 36.3 \\
\quad\quad\quad$\vdash$ (5) Black $\geq$ 0.77 & 3421 & 450 & 4521 & 86.1 & 105.9 & 4982 & 32.8 & 42.3 & 5279 & 26.6 & 34.7 \\
\quad\quad$\vdash$ (6) Poverty $\geq$ 16.11 & 5689 & 536 & 4526 & 74.1 & 78.1 & 5168 & 30.7 & 49.8 & 5453 & 42.2 & 59.7 \\
\quad\quad\quad$\vdash$ (7) Education $<$ 7.25 & 2691 & 302 & 4335 & 234.7 & 235.8 & 5288 & 49.6 & 72.6 & 5619 & 65.9 & 82.4 \\
\quad\quad\quad$\vdash$ (8) Education $\geq$ 7.25 & 2998 & 492 & 4599 & 50.7 & 53.7 & 5059 & 32.4 & 39.2 & 5280 & 44.2 & 48.2 \\
\quad$\vdash$ (9) Education $\geq$ 10.48 & 14162 & 1144 & 4238 & 22.7 & 31.4 & 4637 & 22.1 & 36.2 & 4857 & 34.1 & 33.4 \\
\quad\quad$\vdash$ (10) HomeValue $<$ \$126{.}8K & 7326 & 907 & 4354 & 27.9 & 38.4 & 4826 & 26.8 & 41.2 & 5045 & 24.7 & 33.9 \\
\quad\quad\quad$\vdash$ (11) Poverty $<$ 12.52 & 3821 & 601 & 4258 & 38.1 & 50.7 & 4728 & 28.4 & 46.6 & 5002 & 32.7 & 47.0 \\
\quad\quad\quad$\vdash$ (12) Poverty $\geq$ 12.52 & 3505 & 597 & 4464 & 34.0 & 38.8 & 4921 & 51.6 & 58.5 & 5079 & 40.0 & 45.2 \\
\quad\quad$\vdash$ (13) HomeValue $\geq$ \$126{.}8K & 6836 & 634 & 4118 & 30.2 & 35.8 & 4435 & 26.8 & 36.2 & 4650 & 59.6 & 42.7 \\
\quad\quad\quad$\vdash$ (14) Education $<$ 16.95 & 3319 & 496 & 4252 & 36.5 & 41.2 & 4558 & 26.9 & 33.9 & 4776 & 41.7 & 39.1 \\
\quad\quad\quad$\vdash$ (15) Education $\geq$ 16.95 & 3517 & 345 & 3991 & 35.8 & 41.3 & 4319 & 38.4 & 50.2 & 4528 & 88.4 & 63.1 \\
\hline
\end{tabular}
\caption{Number of county-years ($n$) and counties, estimated age-standardized mortality rates among adults aged 65 and older (deaths per 100,000 person-years) at exposures of 5, 9, and 12 $\mu$g/m$^3$, and standard errors within each subgroup on the tree discovered with continuous modifiers under the $L^2$ objective with uniform weighting. SE$_{\mathrm{CR}}$: cluster-robust standard error, clustered by county. SE$_{\mathrm{C}}$: Conley spatial standard error with a Bartlett kernel and a 200 km cutoff.}
\label{table:table:tab_tree_l2_unweighted_cont}
\end{table}

\ifthenelse{\boolean{showfigures}}{
\begin{figure}[H]
\centering
\includegraphics[width=\textwidth]{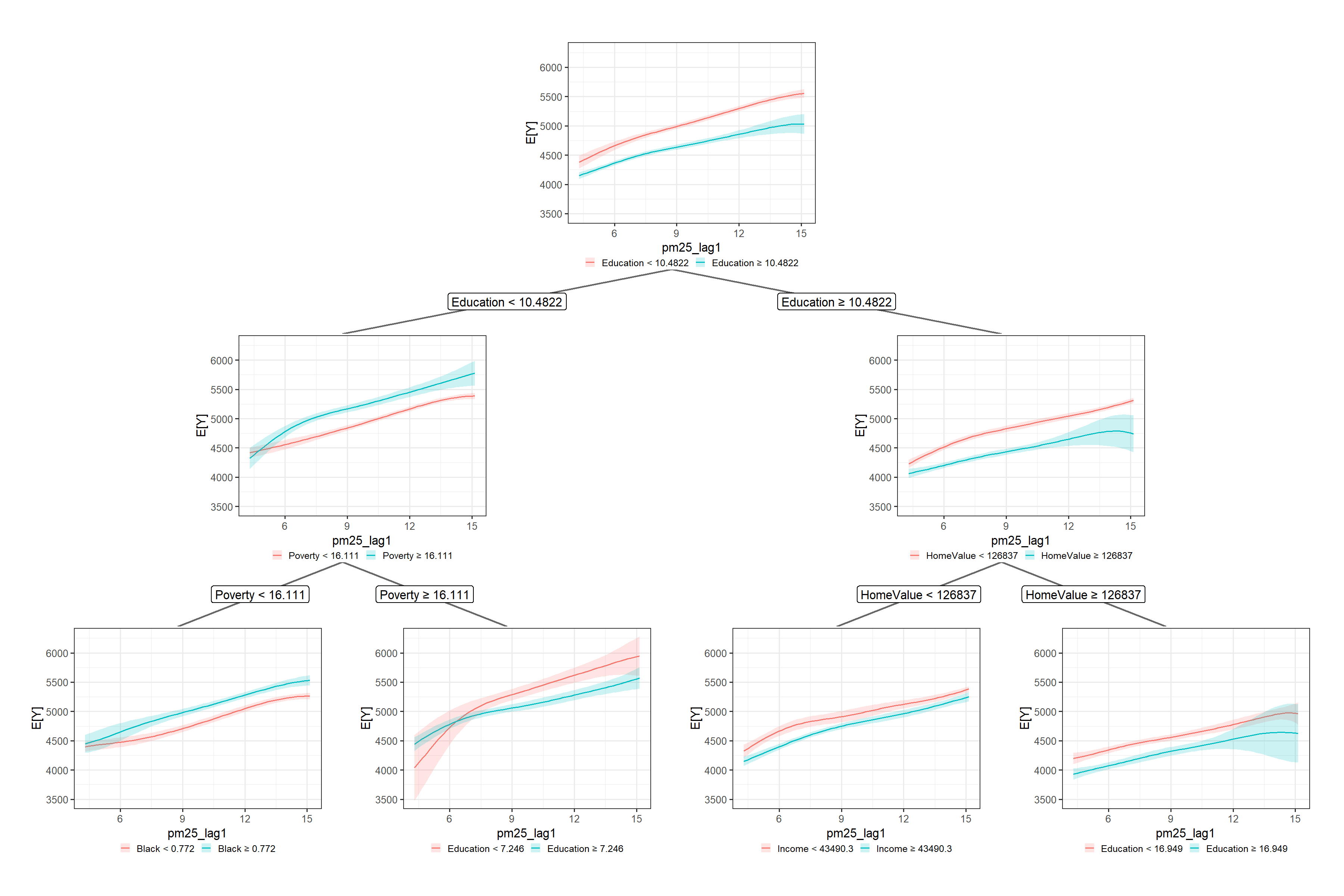}
\caption{Discovered honest tree under the AUC objective with uniform weighting and continuous covariates. Each panel shows the estimated dose-response curves of the two children of the corresponding split, with pointwise confidence bands based on cluster-robust standard errors.}
\label{fig:tree_auc_unweighted_cont_pm25}
\end{figure}
}{}

\begin{table}[H]
\centering
\footnotesize
\setlength{\tabcolsep}{2pt}
\begin{tabular}{>{\raggedright\arraybackslash}p{0.30\textwidth}cc|rrr|rrr|rrr}
\hline
\multirow{2}{*}{Subgroup}
& \multirow{2}{*}{$n$}
& \multirow{2}{*}{Counties}
& \multicolumn{3}{c|}{$5\ \mu$g/m$^3$}
& \multicolumn{3}{c|}{$9\ \mu$g/m$^3$}
& \multicolumn{3}{c}{$12\ \mu$g/m$^3$} \\
\cline{4-12}
& & 
& Est. & SE$_{\mathrm{CR}}$ & SE$_{\mathrm{C}}$
& Est. & SE$_{\mathrm{CR}}$ & SE$_{\mathrm{C}}$
& Est. & SE$_{\mathrm{CR}}$ & SE$_{\mathrm{C}}$ \\
\hline
(1) Population & 26751 & 1408 & 4368 & 21.3 & 29.1 & 4803 & 17.0 & 38.6 & 5067 & 19.1 & 36.5 \\
\quad$\vdash$ (2) Education $<$ 10.48 & 12589 & 963 & 4502 & 49.1 & 54.4 & 4990 & 21.1 & 41.5 & 5297 & 23.5 & 43.9 \\
\quad\quad$\vdash$ (3) Poverty $<$ 16.11 & 6900 & 771 & 4474 & 43.2 & 53.0 & 4842 & 23.1 & 38.1 & 5167 & 20.2 & 31.5 \\
\quad\quad\quad$\vdash$ (4) Black $<$ 0.77 & 3479 & 431 & 4433 & 41.7 & 42.7 & 4709 & 29.2 & 42.3 & 5049 & 28.5 & 36.3 \\
\quad\quad\quad$\vdash$ (5) Black $\geq$ 0.77 & 3421 & 450 & 4521 & 86.1 & 105.9 & 4982 & 32.8 & 42.3 & 5279 & 26.6 & 34.7 \\
\quad\quad$\vdash$ (6) Poverty $\geq$ 16.11 & 5689 & 536 & 4526 & 74.1 & 78.1 & 5168 & 30.7 & 49.8 & 5453 & 42.2 & 59.7 \\
\quad\quad\quad$\vdash$ (7) Education $<$ 7.25 & 2691 & 302 & 4335 & 234.7 & 235.8 & 5288 & 49.6 & 72.6 & 5619 & 65.9 & 82.4 \\
\quad\quad\quad$\vdash$ (8) Education $\geq$ 7.25 & 2998 & 492 & 4599 & 50.7 & 53.7 & 5059 & 32.4 & 39.2 & 5280 & 44.2 & 48.2 \\
\quad$\vdash$ (9) Education $\geq$ 10.48 & 14162 & 1144 & 4238 & 22.7 & 31.4 & 4637 & 22.1 & 36.2 & 4857 & 34.1 & 33.4 \\
\quad\quad$\vdash$ (10) HomeValue $<$ \$126.8K & 7326 & 907 & 4354 & 27.9 & 38.4 & 4826 & 26.8 & 41.2 & 5045 & 24.7 & 33.9 \\
\quad\quad\quad$\vdash$ (11) Income $<$ \$43.5K & 3416 & 610 & 4479 & 41.6 & 51.6 & 4910 & 49.0 & 57.0 & 5124 & 35.4 & 39.6 \\
\quad\quad\quad$\vdash$ (12) Income $\geq$ \$43.5K & 3910 & 673 & 4252 & 32.8 & 44.4 & 4749 & 25.5 & 43.2 & 4964 & 33.0 & 48.8 \\
\quad\quad$\vdash$ (13) HomeValue $\geq$ \$126.8K & 6836 & 634 & 4118 & 30.2 & 35.8 & 4435 & 26.8 & 36.2 & 4650 & 59.6 & 42.7 \\
\quad\quad\quad$\vdash$ (14) Education $<$ 16.95 & 3319 & 496 & 4252 & 36.5 & 41.2 & 4558 & 26.9 & 33.9 & 4776 & 41.7 & 39.1 \\
\quad\quad\quad$\vdash$ (15) Education $\geq$ 16.95 & 3517 & 345 & 3991 & 35.8 & 41.3 & 4319 & 38.4 & 50.2 & 4528 & 88.4 & 63.1 \\
\hline
\end{tabular}
\caption{Number of county-years ($n$) and counties, estimated age-standardized mortality rates among adults aged 65 and older (deaths per 100,000 person-years) at exposures of 5, 9, and 12 $\mu$g/m$^3$, and standard errors within each subgroup on the tree discovered with continuous modifiers under the AUC objective with uniform weighting. SE$_{\mathrm{CR}}$: cluster-robust standard error, clustered by county. SE$_{\mathrm{C}}$: Conley spatial standard error with a Bartlett kernel and a 200 km cutoff.}
\label{table:table:tab_tree_auc_unweighted_cont}
\end{table}

\ifthenelse{\boolean{showfigures}}{
\begin{figure}[H]
\centering
\includegraphics[width=\textwidth]{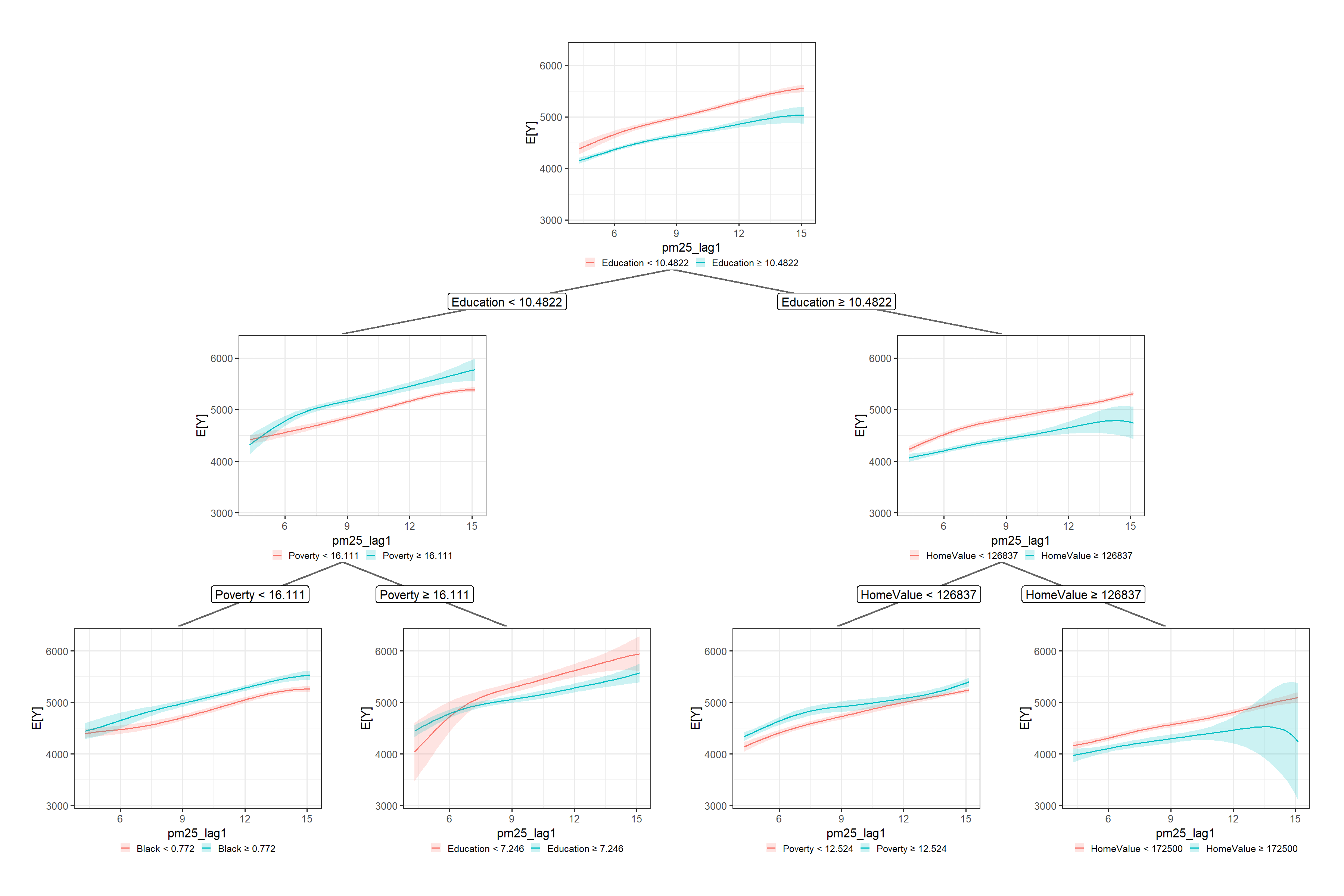}
\caption{Discovered honest tree under the FANOVA objective with uniform weighting and continuous covariates. Each panel shows the estimated dose-response curves of the two children of the corresponding split, with pointwise confidence bands based on cluster-robust standard errors.}
\label{fig:tree_fanova_unweighted_cont_pm25}
\end{figure}
}{}

\begin{table}[H]
\centering
\footnotesize
\setlength{\tabcolsep}{2pt}
\begin{tabular}{>{\raggedright\arraybackslash}p{0.30\textwidth}cc|rrr|rrr|rrr}
\hline
\multirow{2}{*}{Subgroup}
& \multirow{2}{*}{$n$}
& \multirow{2}{*}{Counties}
& \multicolumn{3}{c|}{$5\ \mu$g/m$^3$}
& \multicolumn{3}{c|}{$9\ \mu$g/m$^3$}
& \multicolumn{3}{c}{$12\ \mu$g/m$^3$} \\
\cline{4-12}
& & 
& Est. & SE$_{\mathrm{CR}}$ & SE$_{\mathrm{C}}$
& Est. & SE$_{\mathrm{CR}}$ & SE$_{\mathrm{C}}$
& Est. & SE$_{\mathrm{CR}}$ & SE$_{\mathrm{C}}$ \\
\hline
(1) Population & 26751 & 1408 & 4368 & 21.3 & 29.1 & 4803 & 17.0 & 38.6 & 5067 & 19.1 & 36.5 \\
\quad$\vdash$ (2) Education $<$ 10.48 & 12589 & 963 & 4502 & 49.1 & 54.4 & 4990 & 21.1 & 41.5 & 5297 & 23.5 & 43.9 \\
\quad\quad$\vdash$ (3) Poverty $<$ 16.11 & 6900 & 771 & 4474 & 43.2 & 53.0 & 4842 & 23.1 & 38.1 & 5167 & 20.2 & 31.5 \\
\quad\quad\quad$\vdash$ (4) Black $<$ 0.77 & 3479 & 431 & 4433 & 41.7 & 42.7 & 4709 & 29.2 & 42.3 & 5049 & 28.5 & 36.3 \\
\quad\quad\quad$\vdash$ (5) Black $\geq$ 0.77 & 3421 & 450 & 4521 & 86.1 & 105.9 & 4982 & 32.8 & 42.3 & 5279 & 26.6 & 34.7 \\
\quad\quad$\vdash$ (6) Poverty $\geq$ 16.11 & 5689 & 536 & 4526 & 74.1 & 78.1 & 5168 & 30.7 & 49.8 & 5453 & 42.2 & 59.7 \\
\quad\quad\quad$\vdash$ (7) Education $<$ 7.25 & 2691 & 302 & 4335 & 234.7 & 235.8 & 5288 & 49.6 & 72.6 & 5619 & 65.9 & 82.4 \\
\quad\quad\quad$\vdash$ (8) Education $\geq$ 7.25 & 2998 & 492 & 4599 & 50.7 & 53.7 & 5059 & 32.4 & 39.2 & 5280 & 44.2 & 48.2 \\
\quad$\vdash$ (9) Education $\geq$ 10.48 & 14162 & 1144 & 4238 & 22.7 & 31.4 & 4637 & 22.1 & 36.2 & 4857 & 34.1 & 33.4 \\
\quad\quad$\vdash$ (10) HomeValue $<$ \$126.8K & 7326 & 907 & 4354 & 27.9 & 38.4 & 4826 & 26.8 & 41.2 & 5045 & 24.7 & 33.9 \\
\quad\quad\quad$\vdash$ (11) Poverty $<$ 12.52 & 3821 & 601 & 4258 & 38.1 & 50.7 & 4728 & 28.4 & 46.6 & 5002 & 32.7 & 47.0 \\
\quad\quad\quad$\vdash$ (12) Poverty $\geq$ 12.52 & 3505 & 597 & 4464 & 34.0 & 38.8 & 4921 & 51.6 & 58.5 & 5079 & 40.0 & 45.2 \\
\quad\quad$\vdash$ (13) HomeValue $\geq$ \$126.8K & 6836 & 634 & 4118 & 30.2 & 35.8 & 4435 & 26.8 & 36.2 & 4650 & 59.6 & 42.7 \\
\quad\quad\quad$\vdash$ (14) HomeValue $<$ \$172.5K & 3361 & 574 & 4218 & 29.9 & 36.1 & 4570 & 26.3 & 37.9 & 4806 & 28.8 & 32.7 \\
\quad\quad\quad$\vdash$ (15) HomeValue $\geq$ \$172.5K & 3475 & 348 & 4027 & 47.2 & 49.7 & 4296 & 42.3 & 48.9 & 4463 & 129.5 & 95.0 \\
\hline
\end{tabular}
\caption{Number of county-years ($n$) and counties, estimated age-standardized mortality rates among adults aged 65 and older (deaths per 100,000 person-years) at exposures of 5, 9, and 12 $\mu$g/m$^3$, and standard errors within each subgroup on the tree discovered with continuous modifiers under the FANOVA objective with uniform weighting. SE$_{\mathrm{CR}}$: cluster-robust standard error, clustered by county. SE$_{\mathrm{C}}$: Conley spatial standard error with a Bartlett kernel and a 200 km cutoff.}
\label{table:tab_tree_fanova_unweighted_cont}
\end{table}

\ifthenelse{\boolean{showfigures}}{
\begin{figure}[H]
\centering
\includegraphics[width=\textwidth]{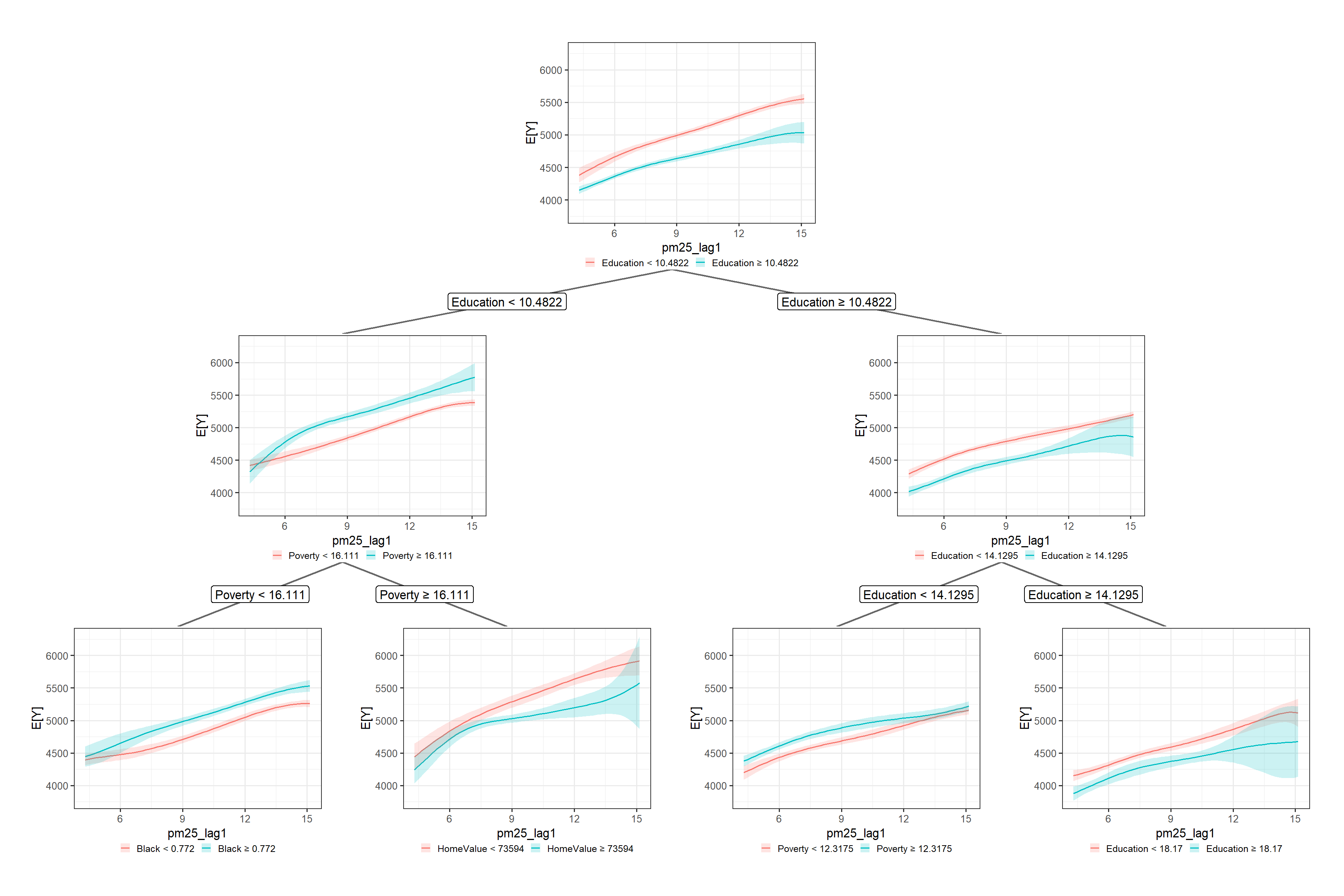}
\caption{Discovered honest tree under the $L^2$, AUC, and FANOVA objectives with density weighting and continuous covariates. Each panel shows the estimated dose-response curves of the two children of the corresponding split, with pointwise confidence bands based on cluster-robust standard errors.}
\label{fig:tree_l2_density_cont_pm25}
\end{figure}
}{}

\begin{table}[H]
\centering
\footnotesize
\setlength{\tabcolsep}{2pt}
\begin{tabular}{>{\raggedright\arraybackslash}p{0.30\textwidth}cc|rrr|rrr|rrr}
\hline
\multirow{2}{*}{Subgroup}
& \multirow{2}{*}{$n$}
& \multirow{2}{*}{Counties}
& \multicolumn{3}{c|}{$5\ \mu$g/m$^3$}
& \multicolumn{3}{c|}{$9\ \mu$g/m$^3$}
& \multicolumn{3}{c}{$12\ \mu$g/m$^3$} \\
\cline{4-12}
& & 
& Est. & SE$_{\mathrm{CR}}$ & SE$_{\mathrm{C}}$
& Est. & SE$_{\mathrm{CR}}$ & SE$_{\mathrm{C}}$
& Est. & SE$_{\mathrm{CR}}$ & SE$_{\mathrm{C}}$ \\
\hline
(1) Population & 26751 & 1408 & 4368 & 21.3 & 29.1 & 4803 & 17.0 & 38.6 & 5067 & 19.1 & 36.5 \\
\quad$\vdash$ (2) Education $<$ 10.48 & 12589 & 963 & 4502 & 49.1 & 54.4 & 4990 & 21.1 & 41.5 & 5297 & 23.5 & 43.9 \\
\quad\quad$\vdash$ (3) Poverty $<$ 16.11 & 6900 & 771 & 4474 & 43.2 & 53.0 & 4842 & 23.1 & 38.1 & 5167 & 20.2 & 31.5 \\
\quad\quad\quad$\vdash$ (4) Black $<$ 0.77 & 3479 & 431 & 4433 & 41.7 & 42.7 & 4709 & 29.2 & 42.3 & 5049 & 28.5 & 36.3 \\
\quad\quad\quad$\vdash$ (5) Black $\geq$ 0.77 & 3421 & 450 & 4521 & 86.1 & 105.9 & 4982 & 32.8 & 42.3 & 5279 & 26.6 & 34.7 \\
\quad\quad$\vdash$ (6) Poverty $\geq$ 16.11 & 5689 & 536 & 4526 & 74.1 & 78.1 & 5168 & 30.7 & 49.8 & 5453 & 42.2 & 59.7 \\
\quad\quad\quad$\vdash$ (7) HomeValue $<$ \$73.6K & 2963 & 346 & 4617 & 89.6 & 91.2 & 5289 & 47.3 & 67.7 & 5635 & 45.3 & 65.2 \\
\quad\quad\quad$\vdash$ (8) HomeValue $\geq$ \$73.6K & 2726 & 427 & 4448 & 92.4 & 94.3 & 5029 & 31.6 & 40.2 & 5197 & 75.7 & 77.2 \\
\quad$\vdash$ (9) Education $\geq$ 10.48 & 14162 & 1144 & 4238 & 22.7 & 31.4 & 4637 & 22.1 & 36.2 & 4857 & 34.1 & 33.4 \\
\quad\quad$\vdash$ (10) Education $<$ 14.13 & 7031 & 932 & 4389 & 28.0 & 37.3 & 4789 & 23.8 & 37.9 & 4985 & 26.1 & 32.7 \\
\quad\quad\quad$\vdash$ (11) Poverty $<$ 12.32 & 3488 & 585 & 4301 & 42.6 & 54.2 & 4686 & 28.4 & 44.3 & 4923 & 33.3 & 40.0 \\
\quad\quad\quad$\vdash$ (12) Poverty $\geq$ 12.32 & 3543 & 645 & 4475 & 31.8 & 36.9 & 4886 & 37.1 & 44.3 & 5035 & 40.1 & 45.2 \\
\quad\quad$\vdash$ (13) Education $\geq$ 14.13 & 7131 & 708 & 4093 & 28.3 & 34.8 & 4489 & 30.1 & 41.8 & 4717 & 61.1 & 45.6 \\
\quad\quad\quad$\vdash$ (14) Education $<$ 18.17 & 3784 & 615 & 4211 & 32.3 & 35.9 & 4592 & 30.2 & 36.8 & 4865 & 48.9 & 38.0 \\
\quad\quad\quad$\vdash$ (15) Education $\geq$ 18.17 & 3347 & 368 & 3974 & 41.1 & 46.1 & 4371 & 44.7 & 56.0 & 4554 & 102.5 & 76.8 \\
\hline
\end{tabular}
\caption{Number of county-years ($n$) and counties, estimated age-standardized mortality rates among adults aged 65 and older (deaths per 100,000 person-years) at exposures of 5, 9, and 12 $\mu$g/m$^3$, and standard errors within each subgroup on the tree discovered with continuous modifiers and density weighting under the $L^2$, AUC, and FANOVA objectives with density weighting. SE$_{\mathrm{CR}}$: cluster-robust standard error, clustered by county. SE$_{\mathrm{C}}$: Conley spatial standard error with a Bartlett kernel and a 200 km cutoff.}
\label{table:tab_tree_l2_density_cont}
\end{table}

With continuous covariates and uniform weighting, the three objectives select different trees but share the branch below 10.5\% educational attainment (\Cref{fig:tree_L2_unweighted_cont_pm25,fig:tree_auc_unweighted_cont_pm25,fig:tree_fanova_unweighted_cont_pm25}).
This branch splits on poverty, then on Black population share below 16.11\% poverty and again on education above that threshold.
Counties with poverty above 16.11\% and educational attainment below 7.25\% have the highest fitted mortality from 9 $\mu$g/m$^3$ upward.
With density weighting all objectives select the same tree (\Cref{fig:tree_l2_density_cont_pm25}). The higher-poverty branch splits on home value, and repeated splits on education yield the lowest fitted rates in counties with educational attainment of at least 18.2\%.
The selected modifiers across trees are the same as in the binarized analysis.
\subsection{Pointwise Confidence Bands with Conley Spatial Standard Errors}
\label{app:additional_conley}

The following figures use Conley spatial standard errors with a Bartlett kernel and a 200 km cutoff. The corresponding tables report both county-clustered and spatial standard errors.

\ifthenelse{\boolean{showfigures}}{
\begin{figure}[htbp]
\centering
\includegraphics[width=0.75\textwidth]{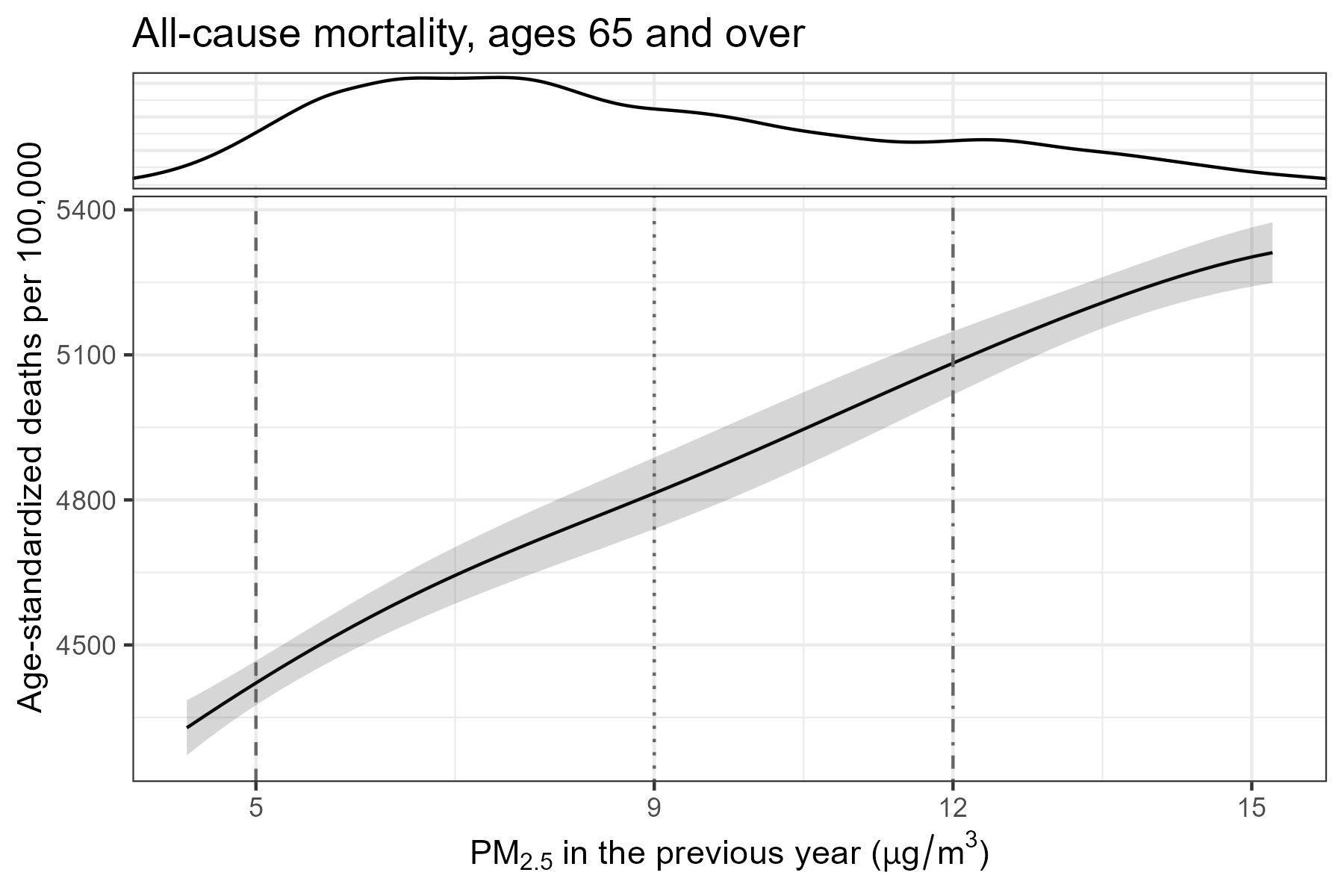}
\caption{\footnotesize Estimated population-average dose-response curve relating one-year-lagged fine particulate matter (PM$_{2.5}$) exposure, on the horizontal axis, to county-level mortality (deaths per 100,000 person-years) with pointwise confidence bands based on the Conley spatial standard errors with a Bartlett kernel and a 200 km cutoff.}
\label{fig:erf_pm25_conley}
\end{figure}
}{}

\ifthenelse{\boolean{showfigures}}{
\begin{figure}[H]
\centering
\includegraphics[width=\textwidth]{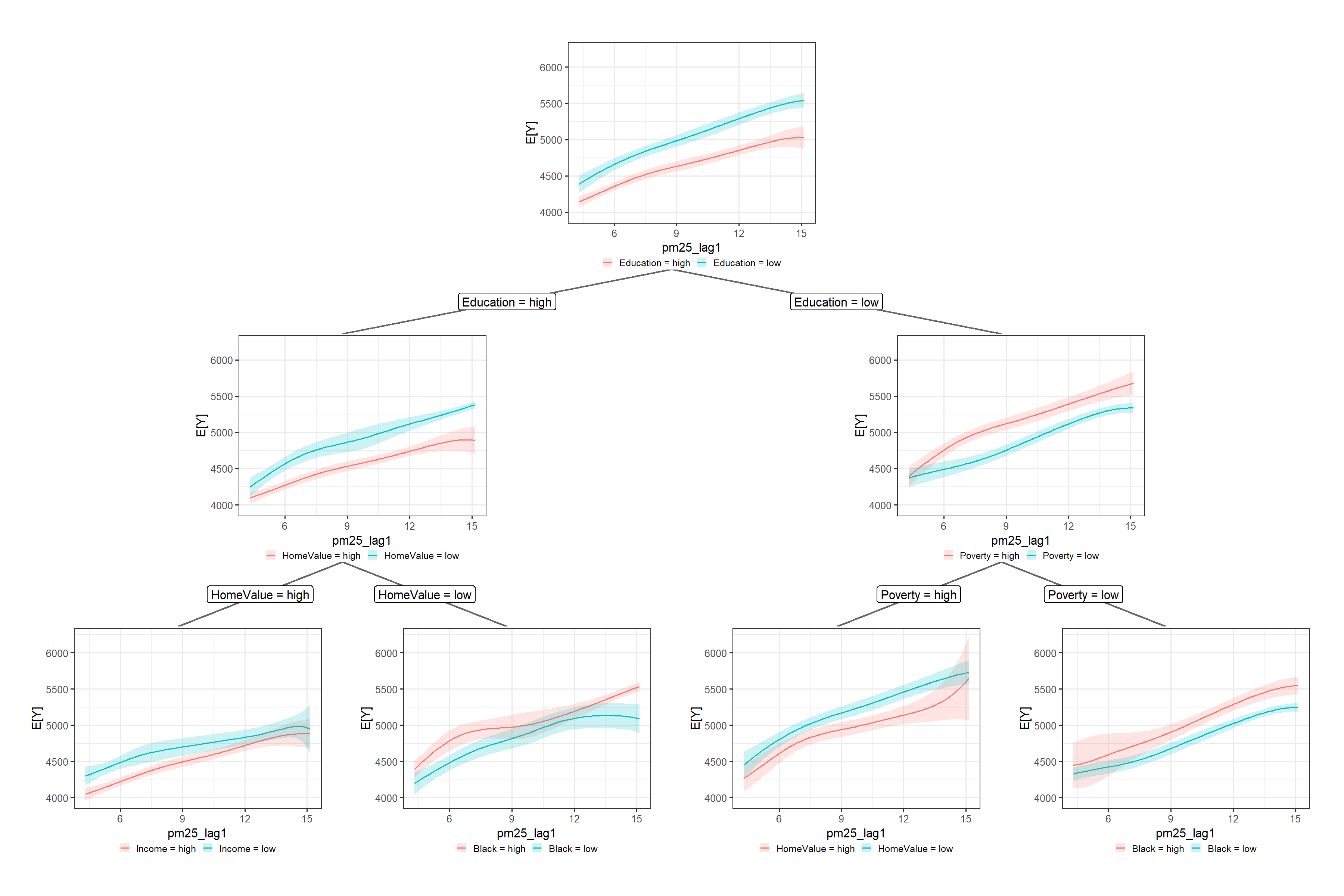}
\caption{Discovered honest tree under the $L^2$ and AUC objectives with uniform weighting.
Each panel shows the estimated dose-response curves of the two children of the corresponding split, with pointwise confidence bands based on the Conley spatial standard errors with a Bartlett kernel and a 200 km cutoff.}
\label{fig:tree_l2_unweighted_pm25_conley}
\end{figure}
}{}

\ifthenelse{\boolean{showfigures}}{
\begin{figure}[H]
\centering
\includegraphics[width=\textwidth]{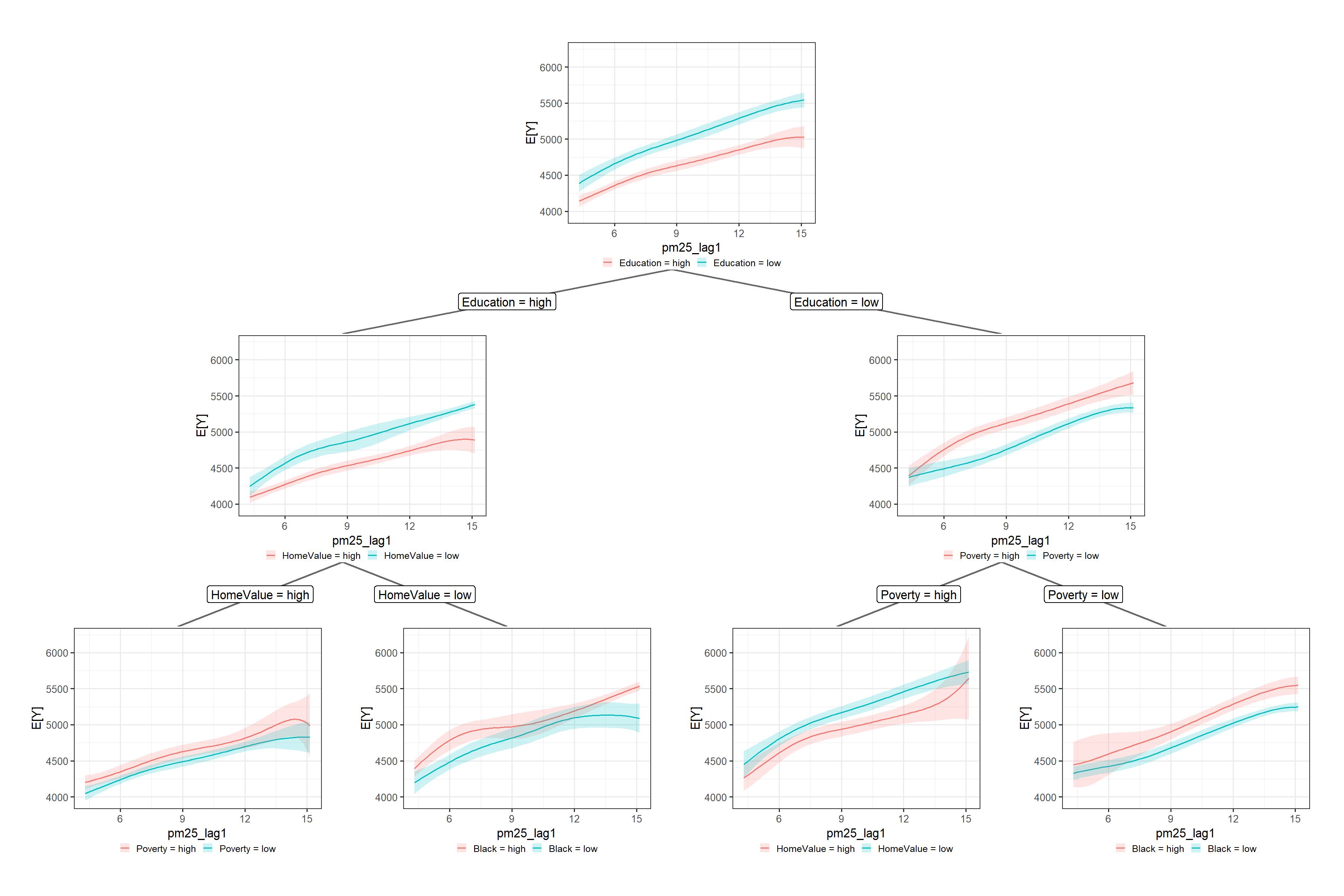}
\caption{Discovered honest tree under the FANOVA objective with uniform weighting.
Each panel shows the estimated dose-response curves of the two children of the corresponding split, with pointwise confidence bands based on the Conley spatial standard errors with a Bartlett kernel and a 200 km cutoff.}
\label{fig:tree_fanova_unweighted_pm25_conley}
\end{figure}
}{}

\ifthenelse{\boolean{showfigures}}{
\begin{figure}[H]
\centering
\includegraphics[width=\textwidth]{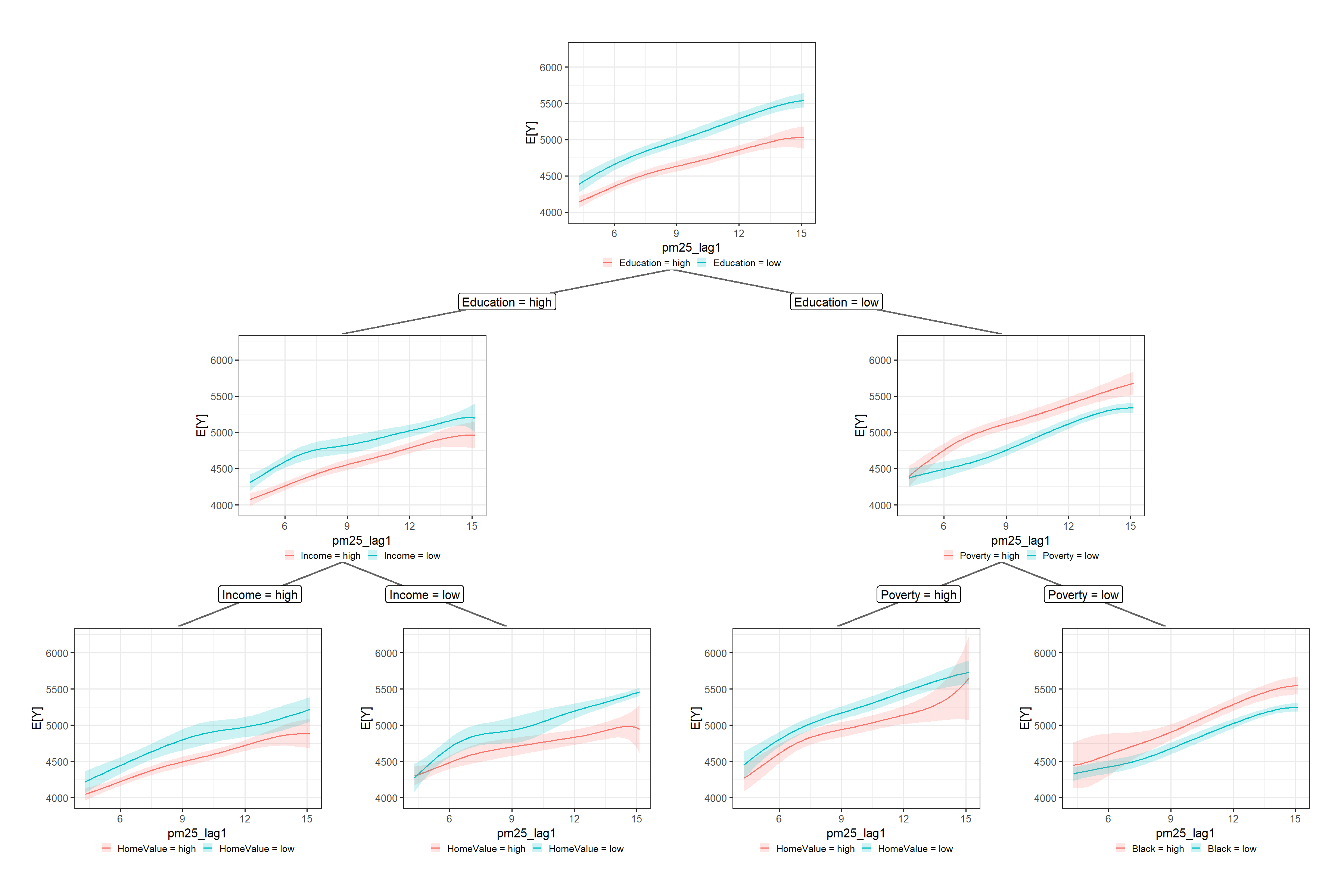}
\caption{Discovered honest tree under the $L^2$, AUC, and FANOVA objective with density weighting.
Each panel shows the estimated dose-response curves of the two children of the corresponding split, with pointwise confidence bands based on the Conley spatial standard errors with a Bartlett kernel and a 200 km cutoff.}
\label{fig:tree_l2_density_pm25_conley}
\end{figure}
}{}

\ifthenelse{\boolean{showfigures}}{
\begin{figure}[H]
\centering
\includegraphics[width=\textwidth]{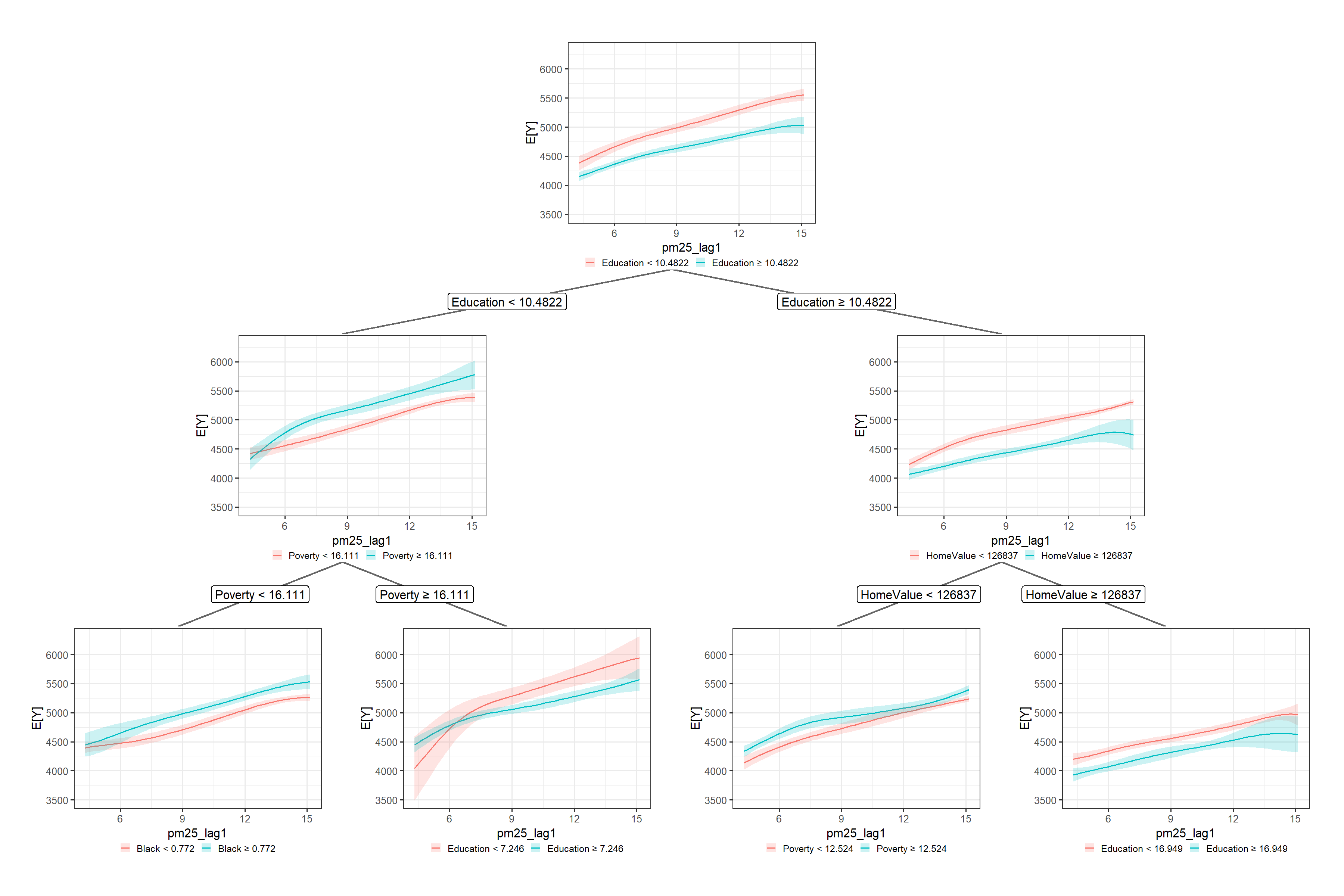}
\caption{Discovered honest tree under the $L^2$ objective with uniform weighting and continuous covariates. Each panel shows the estimated dose-response curves of the two children of the corresponding split, with pointwise confidence bands based on the Conley spatial standard errors with a Bartlett kernel and a 200 km cutoff.}
\label{fig:tree_L2_unweighted_cont_conley_pm25}
\end{figure}
}{}

\ifthenelse{\boolean{showfigures}}{
\begin{figure}[H]
\centering
\includegraphics[width=\textwidth]{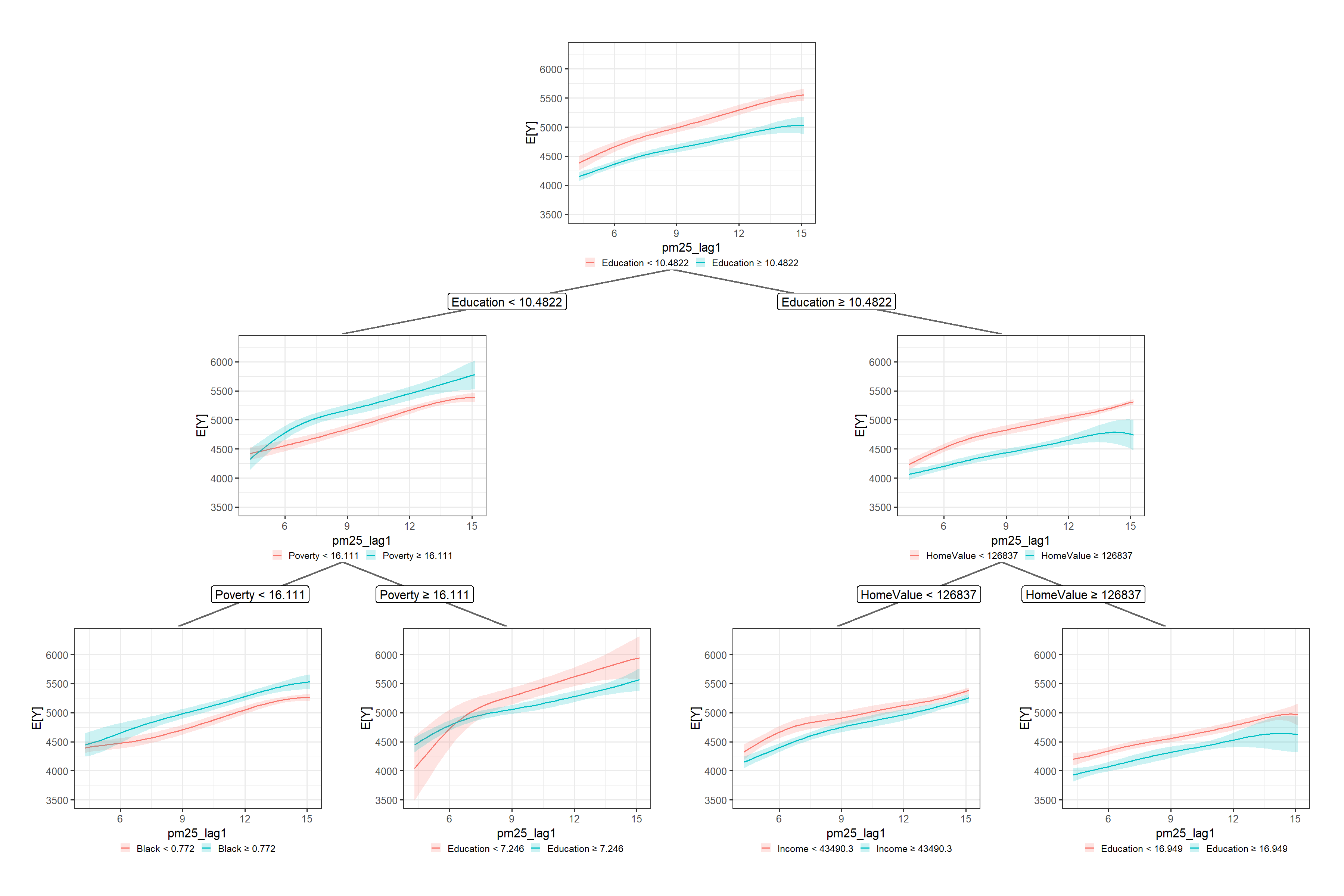}
\caption{Discovered honest tree under the AUC objective with uniform weighting and continuous covariates. Each panel shows the estimated dose-response curves of the two children of the corresponding split, with pointwise confidence bands based on the Conley spatial standard errors with a Bartlett kernel and a 200 km cutoff.}
\label{fig:tree_auc_unweighted_cont_conley_pm25}
\end{figure}
}{}

\ifthenelse{\boolean{showfigures}}{
\begin{figure}[H]
\centering
\includegraphics[width=\textwidth]{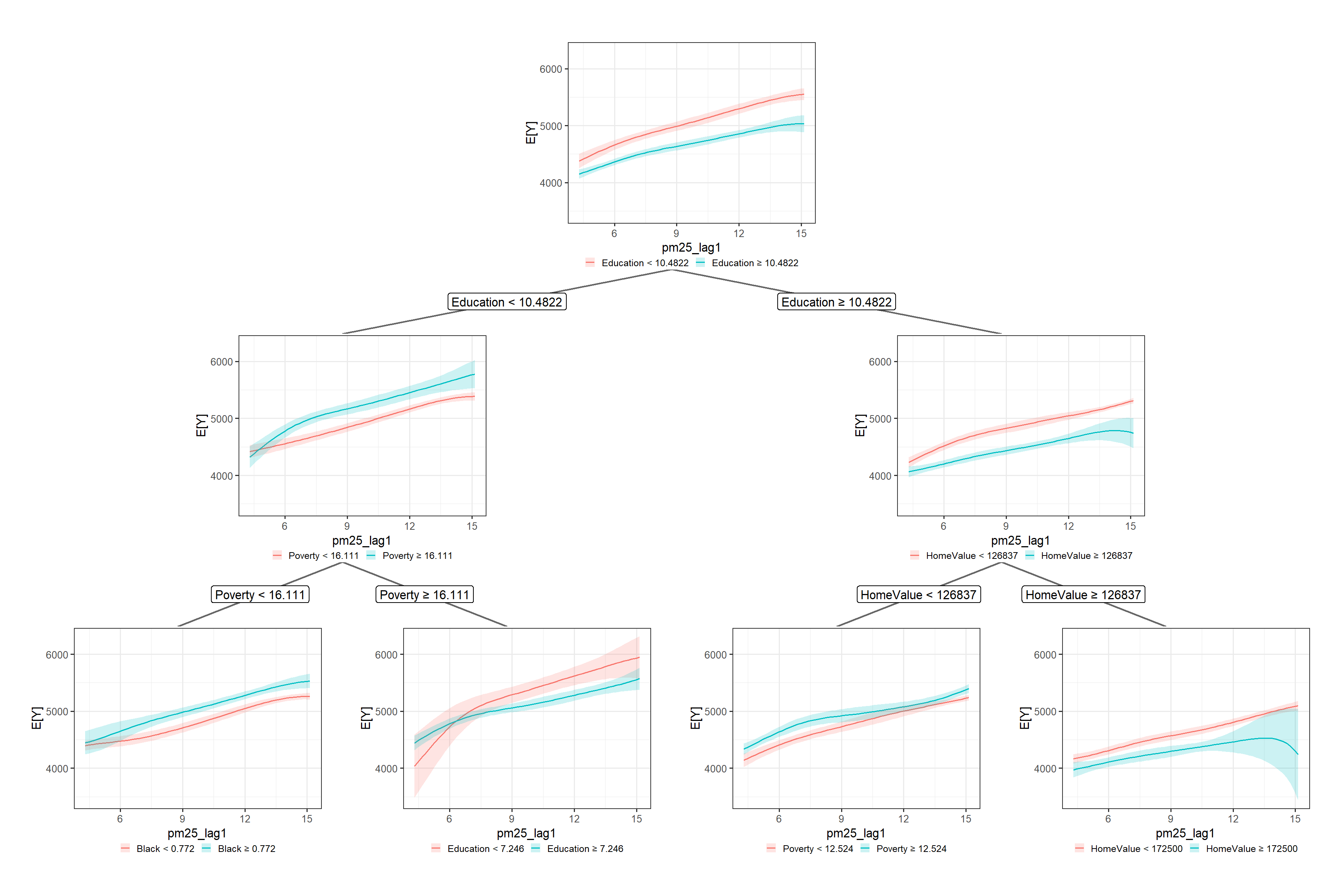}
\caption{Discovered honest tree under the FANOVA objective with uniform weighting and continuous covariates. Each panel shows the estimated dose-response curves of the two children of the corresponding split, with pointwise confidence bands based on the Conley spatial standard errors with a Bartlett kernel and a 200 km cutoff.}
\label{fig:tree_fanova_unweighted_cont_conley_pm25}
\end{figure}
}{}

\ifthenelse{\boolean{showfigures}}{
\begin{figure}[H]
\centering
\includegraphics[width=\textwidth]{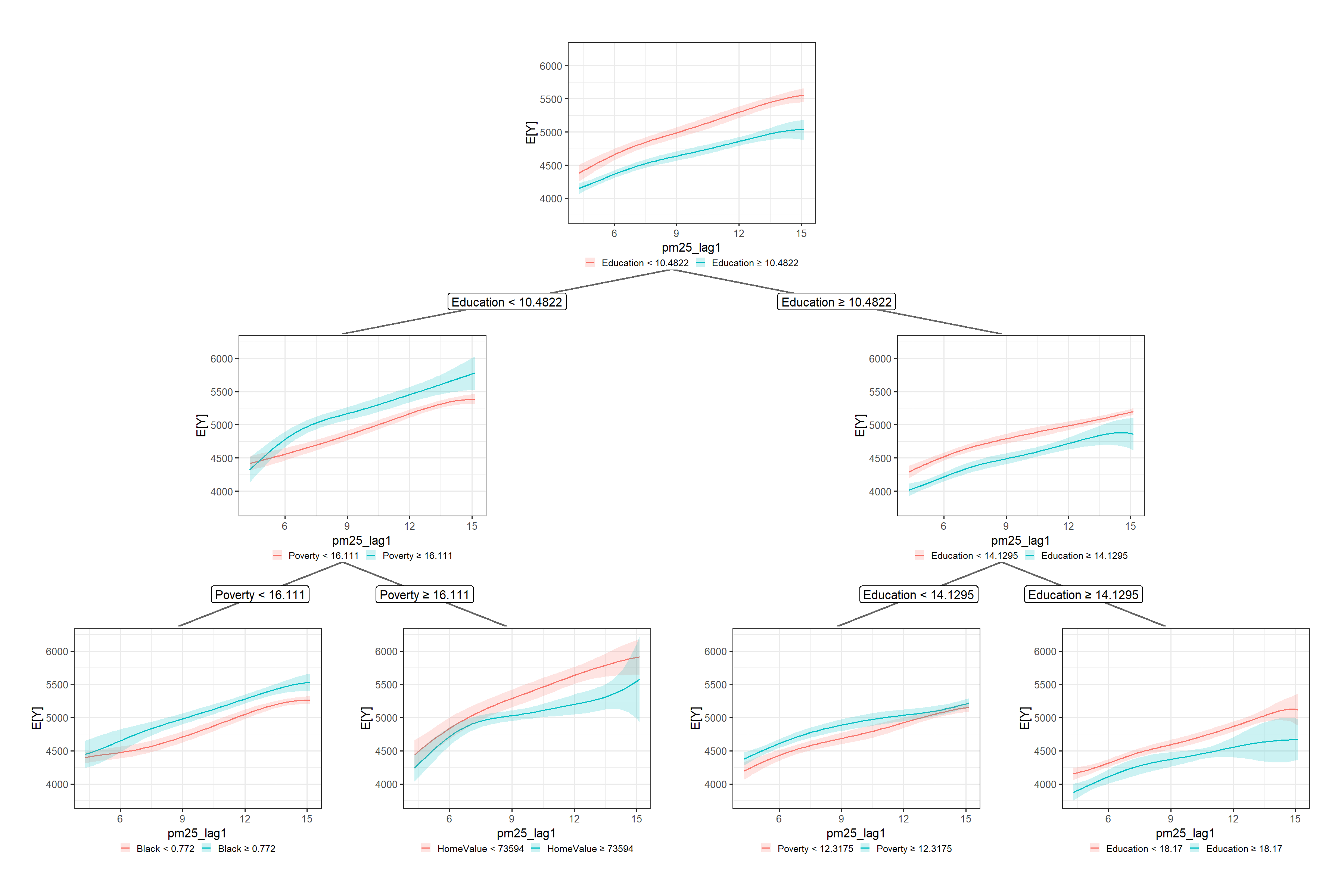}
\caption{Discovered honest tree under the $L^2$, AUC, and FANOVA objectives with density weighting and continuous covariates. Each panel shows the estimated dose-response curves of the two children of the corresponding split, with pointwise confidence bands based on the Conley spatial standard errors with a Bartlett kernel and a 200 km cutoff.}
\label{fig:tree_l2_density_cont_conley_pm25}
\end{figure}
}{}

\section{Further Comparison with Existing Approaches}
\label{app:related}

Because several established literatures bear on this problem, we distinguish their statistical targets from ours: direct discovery of an interpretable, covariate-defined partition together with an estimated dose-response curve for each subgroup.

\textit{Discretizing the exposure.}
A natural shortcut is to replace the continuous exposure $Z$ by the binary indicator $D=\1\{Z>c\}$ and apply methods for binary treatments.
However, intervening on the indicator alone does not specify the continuous dose assigned, so a causal contrast additionally requires defining the exposure regimes above and below the threshold, for example two stochastic policies $g_1(z\mid\bx)$ and $g_0(z\mid\bx)$ supported on their respective ranges.
Such a comparison targets a scalar contrast between the potential outcomes under those regimes, rather than the full dose-response curve $z\mapsto\E\{Y(z)\mid\bV\}$.
As \citeApp{lee2026bridging} show, binarization can yield a legitimate causal estimand, and this estimand is well suited to settings where a policy-relevant threshold already exists and the estimand of interest compares regimes above and below it \citeApp[see, e.g.,][]{lee2021discovering,
bargagli2024causal}.
A contrast between exposure regimes is, nonetheless, a different estimand from a dose-response curve, yielding a scalar rather than a function of $z$.
Relatedly, \citeApp{lundberg2026causal} advocate for an additive shift estimand as an alternative to the dose-response curve when exposure is strongly stratified by confounders, thereby avoiding counterfactual comparisons at treatment values rarely observed within certain subgroups.
Our application uses the full curve to compare estimated mortality changes below and above the regulatory standard (\Cref{sec:q2}).

\textit{Dose-response curve estimation.}
A substantial literature estimates $\theta(z) = \E[Y(z)]$ itself \citepApp{Kennedy2017,ren2021bayesian,  giffin2023generalized,wu2024matching,colangelo2026double}.
Where heterogeneity is of interest, the standard practice is to fix subgroups in advance and estimate $\theta(z \mid \bV) = \E[Y(z) \mid \bV]$ separately within each one \citeApp[see, e.g.,][]{josey2023air}.
General frameworks for debiased estimation of conditional objects \citepApp{semenova2021debiased, chernozhukov2024conditional} do not close the gap either: as we show in \Cref{rem:cif}, the conditional dose-response curve at a treatment level fails the regularity condition on which those constructions rest.

\textit{Flexible modeling of the full surface.}
For example, \citetApp{Woody2020} use Bayesian tree ensembles for covariate-dependent treatment slopes, retaining a linear-in-exposure working model; \citetApp{ren2021bayesian} develop Gaussian-process estimation of a causal exposure-response curve.
Such models can represent variation that a small partition does not capture, although predictive performance depends on the model and data.
The fitted object, however, does not provide a single interpretable partition: the model summarizes the surface pointwise rather than delivering a small set of rules.
These rules are what would make the output usable for targeted decision-making of the type described in our motivating application in \Cref{sec:rq}.

\textit{Variable importance and change-plane models.}
A third approach assigns to each covariate a scalar measuring its contribution to effect heterogeneity \citepApp{shin2025treatment, hines2025variable,bargagli2026causal}.
A variable-importance summary does not by itself specify an interpretable partition or its subgroup curves, and its population relevance target need not coincide with the split-based $\mathcal M^*$ in \Cref{thm:modifier_consistency}.
Related work discovers modifiers of stochastic shift interventions \citepApp{mccoy2024data}, a different causal estimand from the full dose-response curves considered here.
Closest to our aim, \citeApp{jin2023change} develop a semiparametric change-plane model and a doubly robust test for the existence of two subgroups with differential treatment effects under a continuous treatment.
This targets a hypothesis about heterogeneity rather than a partition: the number of subgroups is fixed at two and the boundary between them is a hyperplane in the covariates, which is estimated from the data.
Our aim is complementary, in that we discover a partition up to a chosen depth limit defined by interpretable conjunctions and attach an estimated dose-response curve to each cell, at the cost of the formal test that their approach provides.

\textit{Continuous-treatment subgroup discovery.}
\citetApp{tran2019triggers} propose trigger-based causal trees that seek individualized thresholds in continuous treatment intensity, rather than entire subgroup dose-response curves.
\citetApp{cerqua2024clustered} estimate cluster-specific dose-response curves by data-driven clustering, whereas our target is a small partition defined by explicit covariate thresholds and furnished with a subgroup-specific doubly robust curve in every node.

\textit{Further methodological directions.}
Finally, sample splitting reduces effective sample size for both discovery and inference, and alternative splitting ratios or repeated sample splitting could improve efficiency.
Generic oracle inequalities for alternative second-stage learners, as well as formal extensions to clustered or spatial observations, remain useful theoretical directions.
When the scientific question concerns marginal exposure effects, derivative-based objectives could compare $\int_{\mathcal{Z}} [\theta'(z \mid \mathcal{L}_L) - \theta'(z \mid \mathcal{L}_R)]^2 w_s(z)\, dz$, connecting to work on incremental effects for continuous treatments \citepApp{hines2023optimally, schindl2024incremental, mccoy2024data}.
Derivative estimation would introduce additional variance and stronger smoothness requirements.
The shape component of \Cref{prop:l2_decomp} is a derivative-free alternative that is invariant to level shifts, and we leave derivative-based objectives to future work.

%\bibliographystyleApp{plainnat}
\bibliographystyleApp{abbrvnat}
\bibliographyApp{biblio}

\end{document}